\documentclass[11pt,a4paper]{article} 

\usepackage{mathtools}

        \usepackage[utf8]{inputenc}
        \usepackage{fontenc}
	    \usepackage[italian,english]{babel}
	    \usepackage{amsmath}
	    \usepackage{geometry}
	    \usepackage{amsthm}
	    \usepackage{mathrsfs}
	    \usepackage{bbold}
	    \usepackage{mathtools}
	    \usepackage{amsfonts}
	    \usepackage{graphicx}
	    \usepackage{setspace}
	    \usepackage{physics}
	    \usepackage{comment}
	    \usepackage{layout}
	    \usepackage{xcolor}
        \usepackage[big]{layaureo}
	    \usepackage{dsfont}	
	    \usepackage{slashed}
 	    \usepackage{amssymb}
	    \usepackage{tensor}
	    \usepackage{fancyhdr} 
 	    \usepackage{tikz-cd}
 	    \usepackage{braket}
 	    \usepackage{tikz}
 	    \usepackage[compat=1.1.0]{tikz-feynman}
 	    \usepackage{subfigure}
	    \usepackage{fancyhdr} 
	    \usepackage{booktabs}
	    \usepackage{braket}
	    \usepackage{caption}
	    \usepackage{wrapfig}

\usepackage{jheppub} 
\usepackage{natbib}
\bibstyle{JHEP}

\author[a]{Vladimir Bashmakov} 
\author[b]{Jacopo Sisti}

\affiliation[a]{Institute for Molecules and Materials, Radboud University, Heijendaalseweg 135,6525AJ\\ Nijmegen, The Netherlands}
\affiliation[b]{Department of Physics and Astronomy, Uppsala University, Box 516, SE-75120 Uppsala, Sweden}

\emailAdd{vladimir.bashmakov@ru.nl, jacopo.sisti@physics.uu.se}   

\abstract{Defects in conformal field theories are interesting objects to study from both formal and applied points of view. In this paper, we construct conformal defects in free scalar field CFTs in diverse dimensions. After discussing the possible quadratic defects, we explore interacting setups. These are realized by coupling the bulk free scalar to lower-dimensional theories, including the unitary family of minimal models $\mathcal{M}(p, p+1)$. Another example involves coupling to a two-dimensional free scalar field, from which we construct surface defects for the bulk dimensions three and five. Additionally, we consider monodromy defects associated with a global $U(1)$ flavour symmetry. In these theories, we study both self-defect interactions and couplings to Minimal Models, finding new IR defect fixed points. For all our examples, we provide results for correlation functions, such as those involving the bulk stress tensor and the displacement operator, and for the defect central charges.}

	\newcommand{\beq}{\begin{equation}}
	\newcommand{\bea}{\begin{eqnarray}}
	\newcommand{\eea}{\end{eqnarray}}
	\newcommand{\eeq}{\end{equation}}

	\newcommand{\e}{\mathrm{e}}

	\renewcommand{\a}{\alpha}

	\renewcommand{\e}{\epsilon}

	\makeatletter
	\newcommand{\aextp}{\@ifnextchar^\@aextp{\@aextp^{\,}}}
	\def\@aextp^#1{\mathop{\bigwedge\nolimits^{\!#1}}}
	\makeatother
	
	\makeatletter
	\newcommand{\extp}{\@ifnextchar_\@extp{\@extp_{\,}}}
	\def\@extp_#1{\mathop{\aextp\nolimits_{\!#1}}}
	\makeatother

	\theoremstyle{definition}

\preprint{}

\preprint{UUITP-28/24}

\title{Exploring Defects with Degrees of Freedom in Free Scalar CFTs}

\begin{document}

\maketitle

\section{Introduction}
\label{sec:intro}

Conformal invariance is a powerful condition that significantly constrains the dynamics of a physical theory. In the realm of Quantum Field Theories (QFTs), Conformal Field Theories (or CFTs) are much more constrained than their more general counterparts, and hence are considerably more tractable \cite{DiFrancesco:1997nk, Rychkov:2016iqz, Simmons-Duffin:2016gjk}. The usage of constraints imposed by conformal symmetry culminated in the Conformal Bootstrap Program \cite{Poland:2018epd}.

Given the advances in the understanding of CFTs, it is natural to ask whether we can deform a theory in such a way that the $d$-dimensional conformal group $SO(d+1, 1)$\footnote{In this paper we are going to consider Euclidean CFTs.} is broken, but not completely, such that some features of the original conformal invariance are retained. A way to do it, relevant for both theoretical reasons and experimental needs, is to introduce boundaries or defects. Our focus in this paper is going to be on the latter ones, even though boundaries and defects share many common properties.

A defect, broadly speaking, is a manifold $\Sigma_p$ of dimension $p$ ($q=d-p$, is the codimension of the defect), embedded into the bulk space $\mathcal{M}_d$ of dimension $d$ (such that $\Sigma_p \subset \mathcal{M}_d$), on which the theory has some special properties. Such distinguished manifolds evidently break certain symmetries of the original theory. In particular, a planar $p$-dimensional defect, embedded in the Euclidean $d$-dimensional bulk, breaks the original conformal group $SO(d+1,1)$, with the maximal possible preserved subgroup being $SO(p+1,1)\times SO(d-p)$. The first factor here corresponds to the conformal group along the defect, and the second factor represents the rotations around the defect. A defect preserving the above-mentioned maximal subgroup is called a conformal defect, and the whole system is referred to as a Defect Conformal Field Theory (DCFT).

In a DCFT, it is convenient to split the $d$ coordinates into coordinates along the defect $\sigma^i$, $i=1,...,p$, and $x_\perp^{a}$, $a=1,...,d-p$, of the orthogonal direction. Throughout this work, we will choose the location of the defect to be $x_\perp = 0$.  

Moreover, we can distinguish two types of local operators: bulk operators $\mathcal{O}(\sigma,x_\perp)$, which are defined in $\mathcal{M}_d$ but away from the defect, and defect operators $\hat{\mathcal{O}}(\sigma)$, which lie on $\Sigma_p$. In a DCFT, a privileged role is played by primary operators. In particular, primary bulk operators are primaries with respect to the full conformal group $SO(d+1,1)$ while defect primary operators are defined through the subgroup $SO(p+1,1)$. Note that the rotation group $SO(d-p)$ will behave as a global symmetry group on those.

Evidently, the reduced symmetry is less constraining, leading to a richer structure of correlators. For instance, while the one-point functions in a CFT vanish (unless it is the identity operator), the bulk operators' one-point functions in a DCFT can be non-trivial, and constrained by the symmetry to take the form
\begin{equation}
\left<\mathcal{O}(\sigma, x_\perp) \right>=\frac{a_{\mathcal{O}}}{|x_\perp|^{\Delta_{\mathcal{O}}}}\,.
\end{equation}
Note that the one-point function of a scalar operator depends only on the distance to the defect. \\
Two-point functions of the defect operators have the same form, as in a normal CFT,
\begin{equation}
    \left< \hat{\mathcal{O}}_1(\sigma_1) \hat{\mathcal{O}}_2(\sigma_2) \right> = \frac{\delta_{12}}{|\sigma_1-\sigma_2|^{2 \Delta_{\hat{\mathcal{O}}}}},
\end{equation}
while the bulk-defect two-point function has the somewhat more complicated structure
\begin{equation}
   \left< \mathcal{O}_1(\sigma, y)\hat{\mathcal{O}}_2(0)\right> = \frac{b_{\mathcal{O}_1\hat{\mathcal{O}}_2}|x_\perp|^{\hat{\Delta}_2-\Delta_1}}{(x_\perp^2+\sigma^2)^{\hat{\Delta}_2}}.
\end{equation}

The quantities introduced above, namely the defect operator dimensions, bulk one-point function coefficients, and bulk-defect two-point function coefficients represent types of the CFT data associated with defects which are absent in usual CFTs.

In contrast to CFTs, two-point functions of bulk operators in the presence of a defect are fixed by the conformal symmetry only up to functions of an invariant cross-ratio, and so are similar in spirit to four-point functions in normal CFTs \cite{Billo:2016cpy,Herzog:2020bqw}. In particular, they can be expanded in terms of defect conformal blocks, and used in the Conformal Bootstrap Program \cite{Billo:2016cpy,Bissi:2018mcq,Behan:2020nsf,Behan:2021tcn,Gimenez-Grau:2022czc,Bianchi:2022sbz,Bianchi:2023gkk}.

In the characterisation of CFTs, other important quantities are the correlation functions of the stress tensor and the Weyl anomalies. Consider a QFT on an arbitrary background $\mathcal{M}_d$ with Euclidean metric $g_{\mu\nu}$. Given the effective action (or free energy) $F = - \log Z$, where $Z$ is the QFT's partition function, in our convention the stress tensor one-point function is given by
\begin{equation}
\left< T^{\mu\nu} \right> = - \frac{2}{\sqrt{g}} \frac{\delta F}{\delta g_{\mu\nu}}\,.
\end{equation}
While in a CFT in flat space the one-point function of the stress tensor vanishes, in the presence of a flat defect located at $x_\perp=0$ we have~\cite{Kapustin:2005py,Billo:2016cpy}
\begin{equation}
\label{eq:stress-tensor-one-pt-fn}
\langle T^{ab} \rangle = - a_T \frac{(q-1)\delta^{ab}}{|x_\perp|^d}\, ,\qquad \langle T^{ij} \rangle =a_T \frac{(d-q+1)\delta^{ij} - d \frac{x^i_\perp x_\perp^j}{|x_\perp|^2}}{|x_\perp|^d}\,.
\end{equation}
Notice that the symmetries completely determine the form of $T^{\mu\nu}$'s one-point function up to the single number $a_T$.

In a CFT, when the trace $T^\mu_\mu \ne 0$, we say that there is a Weyl anomaly, which produces a term that scales logarithmically with respect to a UV cut-off $\varepsilon$ in the limit $\varepsilon \rightarrow 0$. In particular, one finds
\begin{align}
\label{eq:W_log_A}
 F|_{\log \epsilon} =    \int d^d x \,  \sqrt{g} \left< T^{\mu}{}_\mu \right>  \,.
\end{align}

Such a Weyl anomaly exists only in the presence of a non-flat metric and it is a linear combination of scalars of the same dimension of the stress tensor, which is $d$, built out of the metric $g_{\mu\nu}$. By dimensional considerations, no Weyl anomaly is possible when $d$ is odd. When $d$ is even, WZ consistency implies that the anomaly takes the schematic form  \cite{Deser:1993yx}
\begin{equation}\label{eq:general-d-Weyl-anomaly}
T^\mu{}_\mu  = \frac{1}{(4\pi)^{\frac{d}{2}}}\Big( (-)^{\frac{d}{2}-1} a^{(d)}E_{d} + \sum_n c^{(d)}_n I_n\Big)\,,
\end{equation}
where we omitted any scheme-dependent contributions. $E_d$ is the $d$-dimensional Euler density, whose integral is a topological invariant, and the $I_n$ are a set of conformal invariants built from contractions of Weyl tensors and their derivatives.

If a defect occurs in the CFT, then there are Weyl anomalies localized on its manifold $\Sigma_p$. Schematically we may write 
\begin{align}
T^\mu{}_\mu = T^\mu{}_\mu\big|_{\mathcal{M}_d} + \delta^{(q)}(x_\perp) T^\mu{}_\mu\big|_{\Sigma_p}.
\end{align}
where $\left. T^{\mu\nu}\right|_{\mathcal{M}_d}$ and $\left. T^{\mu\nu}\right |_{\Sigma_p}$ denote the contributions to the stress tensor from the ambient CFT and the boundary/defect, respectively. Importantly, $T^\mu{}_\mu\big|_{\Sigma_p}$ is built out of structures that involve the metric $g_{\mu\nu}$, the embedding functions $x^\mu(\sigma^a)$, and their derivatives. The defect/boundary Weyl anomaly has been determined in four cases: $p=1$ in $d=2$~\cite{Polchinski:1998rq}, $p=2$ in $d \geq 3$~\cite{Berenstein:1998ij,Graham:1999pm,Henningson:1999xi,Gustavsson:2003hn,Asnin:2008ak,Schwimmer:2008yh,Cvitan:2015ana,Jensen:2018rxu}, $p=3$ in $d=4$~\cite{Herzog:2015ioa,Herzog:2017kkj}, and $p=4$ in $d \ge 5$~\cite{FarajiAstaneh:2021foi, Chalabi:2021jud}. \\
%
%
As an example, let us consider the Weyl anomaly of a $p=2$ defect in a $d \geq 3$ CFT is~\cite{Berenstein:1998ij,Graham:1999pm,Henningson:1999xi,Gustavsson:2003hn,Asnin:2008ak,Schwimmer:2008yh,Cvitan:2015ana,Jensen:2018rxu}
\begin{equation}
\label{eq:2d-defect-Weyl-anomaly}
T^\mu{}_\mu|_{\Sigma_2} = \frac{1}{24\pi}( b \bar{R} + d_{1} \mathring{\Pi}^2 + d_{2} W^{ab}{}_{ab})\,,
\end{equation}
where $\bar{R}$ is the intrinsic scalar curvature of the defect manifold, $\mathring{\Pi}_{ab}$ the traceless part of the second fundamental form $\Pi_{ab}$, and $W^{\mu\nu}_{\rho\sigma}$ the bulk Weyl tensor pulled back to $\Sigma_p$. The first term is the defect central charge analogous to the Weyl anomaly of a standalone $2d$ CFT proportional to the central charge $c$, while the remaining two are specific to defects (and boundaries). The term $d_{2}W^{ab}{}_{ab}$ does not exist for $q=1$ because the ambient Weyl tensor $W$ vanishes identically when $d= 3$.  Note that even if $d$ is odd, and the ambient CFT has no Weyl anomaly, the defect Weyl anomaly in \eqref{eq:2d-defect-Weyl-anomaly} still exists.

A privileged role is played by 
 the displacement operator defined from the broken Ward identities for translations normal to the defect,
\begin{align}\label{eq:displacement-operator-def}
\partial_\mu T^{\mu i} = \delta^{(d-p)}(x_\perp) \hat D^i\,.
\end{align}
We emphasize that generically the defect/boundary does not have its own intrinsically defined, conserved stress tensor. However, the components of the ambient stress tensor along $\Sigma_p$ are conserved everywhere on $\mathcal{M}_d$, including $\Sigma_p$, i.e. $\partial_\mu T^{\mu a} = 0$.
An interesting correlator, which we will compute in the following section of this work, is the two-point function of the displacement operator which reads
\begin{equation}
\label{eq:p2D2}
\langle \hat D^i (\sigma)\hat D^j(0)\rangle = \frac{C_{\hat D}}{|\sigma|^{2p+2}}\delta^{ij}\,.
\end{equation}

Similarly to CFT counterparts, the defect central charges appear in particular correlation functions containing the displacement operator and the bulk stress tensor. For example for $p=2$, the central charge $d_{1}$ controls the $\hat D^i$'s two-point function \cite{Lewkowycz:2014jia,Bianchi:2015liz}
\begin{equation}
\label{eq:p2D2}
C_{\hat D} = \frac{4}{3\pi^2} d_{1} \,.
\end{equation}
This relation also implies that $d_{1} \geq 0$, which follows from reflection positivity.
An analogous relation applies also for $p=4$ \cite{Chalabi:2021jud}. Moreover, when $q>1$, the central charge $d_{2}$ controls $T^{\mu\nu}$'s one-point function, which is allowed to be non-zero by the conformal symmetry preserved by the defect.  For $p=2$, $a_T$ is determined by $d_{2}$ \cite{Lewkowycz:2014jia,Bianchi:2015liz,Jensen:2018rxu}
\begin{equation}
\label{eq:2d1pt}
a_T =- \frac{1}{6\pi(d-1){\text{vol}}(S^{d-3})} \,d_{2}\,,
\end{equation}
where $\text{vol}(S^d)$ is the volume of a unit $d$-sphere. Again, an analogous result exists when $p=4$ \cite{Lewkowycz:2014jia,Chalabi:2021jud}.  If we Wick-rotate to Lorentzian signature, then applying the ANEC to eq.~\eqref{eq:stress-tensor-one-pt-fn} implies $d_{2}\leq 0$~\cite{Jensen:2018rxu}. Also, even though the $b$ central charge is usually extracted from the sphere free energy, in the boundary case it can be found in the two-point function of the stress tensor \cite{Prochazka:2018bpb}.

In a $d=4$ SCFT, a $p=2$ defect that preserves at least $\mathcal{N}=(2,0)$ two-dimensional SUSY has $d_1=d_2$, which is conjectured to extend to $p=2$ defects in $d>4$ SCFTs as well~\cite{Bianchi:2019sxz}. Furthermore, for such defects in a SCFT with $d\geq 3$, $b$ is fixed by a 't Hooft anomaly~\cite{Wang:2020xkc}, hence the IR defect R-symmetry can be identified by extremizing a trial $b$, similar to $a$-maximization~\cite{Intriligator:2003jj} or $c$-extremization~\cite{Benini:2012cz,Benini:2013cda}.

Another property of defect central charges is their connection to Entanglement Entropy. In particular, a linear combination of $b$ and $d_{2}$ determines the defect contribution to the universal part of the EE for a spherical region $A$ of radius $L$, centered on the defect, with the UV cutoff $\epsilon$~\cite{Kobayashi:2018lil,Jensen:2018rxu},
\begin{equation}
\label{eq:2dee}
S_{A,\Sigma_2} = \frac{1}{3}\left( b + \frac{d-3}{d-1}d_{2} \right)\log \left(\frac{L}{\e}\right)\,.
\end{equation}
There is an analogous relation for $p=4$ \cite{Chalabi:2021jud}.

One of the most important aspects of CFTs is the possibility of deforming the theory by adding relevant perturbations. This in general triggers an RG flow which may lead to another fixed point in the IR limit of the theory. Such a flow is claimed to be irreversible as degrees of freedom are integrated out along it. Monotonicity theorems provide a rigorous formulation of this intuition, contributing to the classification of CFTs. The quantities of interest are the sphere-free energy \cite{Klebanov:2011gs} and Weyl anomalies when $d$ is even \cite{Zamolodchikov:1986gt,CARDY1988749,Komargodski:2011vj}\footnote{We mention that monotonicity quantities can be constructed also out of the entanglement entropy and the relative entropy \cite{Casini:2012ei,Casini:2015woa,Casini:2016udt,Casini:2018cxg,Casini:2018nym,Casini:2022bsu,Casini:2023kyj}}. In a defect CFT, besides the common bulk deformation, there is also the possibility of turning on defect-localized relevant perturbations. In such cases, there are results proving monotonicity theorems for $p=1$ boundaries \cite{Friedan:2003yc} and line defects \cite{Cuomo:2021rkm}\footnote{The monotonicity theorem for line defects has also been discussed in explicit examples such as generalized free fields \cite{Nagar:2024mjz} and Wilson loops in ABJM theories \cite{Castiglioni:2022yes,Castiglioni:2023uus}.}, surface defects \cite{Jensen:2015swa}, $p=4$ defects \cite{Wang:2021mdq}, and a conjecture that the defect contribution to the sphere free energy is a monotonic quantity \cite{Kobayashi:2018lil}. Focusing again on $p=2$ defects, the defect central charge $b$, being the one associated with the Euler characteristic of $\Sigma_2$, shares important characteristics with the central charge $c$ of a two-dimensional conformal field theory (CFT). For example, under a defect renormalization group (RG) flow, $b$ satisfies an analogue of the $c$-theorem: $b_{\text{\tiny UV}} \geq b_{\text{\tiny IR}}$\cite{Jensen:2015swa,Casini:2018nym, Shachar:2022fqk}. Additionally, the Wess-Zumino (WZ) consistency conditions ensure that $b$ is independent of marginal couplings on the defect. However, unlike the central charge $c$, $b$ is not necessarily positive semi-definite, even in reflection-positive theories. As an example, a free, massless scalar with Dirichlet boundary conditions in $d=3$ has $b<0$~\cite{Nozaki:2012qd,Jensen:2015swa,Fursaev:2016inw}.
\vspace{10pt}

After this general introduction, let us come to more explicit constructions and examples. An important class of linear defects, specific to the gauge theories, are Wilson lines (and 't Hooft lines in the case of $4d$ gauge theories); while these operators will not be considered in this paper, see \cite{Cuomo:2022xgw, Gabai:2022vri, Gabai:2022mya, Aharony:2023amq} for recent developments. Another type of defects, that will be considered in quite some detail later on, are the \textit{monodromy} defects: they are associated with a symmetry group element $g$, and defined by turning on $g$-transformation along a cycle, linked with certain codimension-two submanifold (the defect submanifold). For now, let us discuss yet another way of introducing a defect to a CFT. One can pick up a CFT operator and add it to the action, but integrating only over a defect submanifold $\Sigma_p$,
\begin{eqnarray}
S_{\text{CFT}}+h\int_{\Sigma_p}d^p\sigma\sqrt{\gamma}\hat{\mathcal{O}}.
\end{eqnarray}
If this deformation is relevant, it will trigger a defect RG flow (the bulk will still stay at criticality), and in some cases may flow to a DCFT in the IR. We will sometimes refer to such defects as ``pinned" operator defects.

Recently, line-pinned operator defects were studied in the $O(N)$ model\footnote{See \cite{Allais, ParisenToldin:2016szc} for the previous numerical studies and \cite{Hanke, Allais:2014fqa, Pannell:2024hbu} for theoretical considerations.} \cite{Cuomo:2021kfm}. There, the only available operator to trigger a line RG flow happens to be $\phi_a$, and the defect thus breaks the global symmetry down to $O(N-1)$. $\epsilon$-expansion and large $N$ analysis, were performed, showing good agreement with each other. In particular, the existence of a (stable) IR fixed point was demonstrated, and anomalous dimensions of a few lowest defect primaries were computed. The authors also evaluated the $g$-function, with the result being consistent with the $g$-theorem.

Perhaps motivated by these results, a series of works studied planar defects, both in the theory of $N$ free scalar fields and in the $O(N)$ model \cite{Giombi:2023dqs, Raviv-Moshe:2023yvq, Trepanier:2023tvb}. The deformation
\begin{equation}
    g\int_{\Sigma_2}\sqrt{\gamma} \hat\phi^I \hat\phi_I
\end{equation}
is relevant for the bulk dimension $d<4$, and triggers an $O(N)$-symmetric defect RG flow (with obvious generalizations to $O(N)$-breaking cases). This deformation admits an exact solution in the case of the free bulk, and can be attacked with the aid of $\epsilon$-expansion and large $N$ techniques in the case of interacting bulk \cite{Giombi:2023dqs, Raviv-Moshe:2023yvq, Trepanier:2023tvb}; both methods uncover non-trivial IR fixed points. The authors then studied different observables, associated with these fixed points, including the $b$-coefficient \eqref{eq:2d-defect-Weyl-anomaly}, and explored the IR phases in the vicinity of the fixed points. Also, line defects have been studied in fermionic theories such as the Gross-Neveau universality class \cite{Giombi:2022vnz,Pannell:2023pwz,Barrat:2023ivo}.

In the present work, we focus on the defects in free scalar field CFT, which are not Gaussian: in other words, the resulting DCFT is not free, but the interactions are localized on the defect.\footnote{Models, possessing both bulk and defect/boundary interactions have recently been discussed in \cite{Harribey:2023xyv}.} A straightforward way of engineering such defects is to couple a bulk theory (taken here to be free) to a lower-dimensional matter, with the schematic form of the coupling being
\begin{equation}
    g\int_{\Sigma_p}\sqrt{\gamma} \hat\phi \hat{\mathcal{O}},
\end{equation}
where $\hat{\mathcal{O}}$ is a CFT$_p$ operator. When this deformation is relevant, it triggers a defect RG flow, and in certain cases, leads to fixed points that correspond to non-trivial IR DCFTs. This method was exploited previously to construct conformal boundaries (BCFTs) in \cite{Herzog:2017xha, Behan:2021tcn}

The fact that the bulk theory is free puts rather strong constraints on the spectrum of possible conformal defects. These constraints basically follow from the classical equation of motion, that the bulk field must satisfy, and their consequences were scrutinised in \cite{Lauria:2020emq}. In particular, it was shown that (a) non-trivial monodromy defects exist only if $d\geq4$ and (b) non-trivial defects without monodromy twist exist only if $q=3$ and $d\geq5$. As will be discussed in the Outline, and in more detail in the bulk of the paper, our findings are perfectly consistent with these no-go theorems.

\subsection*{Content of the paper}

In Section \ref{Section2} we start from a quite general discussion of $p-$dimensional defects in the free scalar CFT$_d$. Many results of this section have already been extensively discussed in the literature. However, some of the results to the best of our knowledge have not appeared in the literature so far. For instance, our results suggest that there is an unstable quadratic surface defect for $d=5$. We also provide the exact expressions for the $\phi^2$ one-point function, $T_{\perp\perp}$ one-point function and bull-to-bulk two-point function of $\phi$ with itself, valid for arbitrary $p$ and $d$. 

In Section \ref{Section3} we set up some conformal perturbation theory results, happening to be instrumental in the rest of the paper. In particular, we review the beta functions and free energy computations and compute the one- and two-point functions of a few lowest-dimension operators.

In Section \ref{Section4} we couple the higher-dimensional scalar field to $2d$ matter, a family of unitary Minimal Models $\mathcal{M}(p,p+1)$. There, one can develop the $1/p$ expansion, originally exploited to study RG flows between the Minimal Models \cite{Zamolodchikov:1987ti}, and used recently to construct conformal boundaries for the free scalar CFT$_3$ \cite{Behan:2021tcn}. This construction allowed us to find interacting surface defects in three and five dimensions.

In Section \ref{Section5} we proceed with the same approach, but couple the bulk to the $2d$ compact boson $\sigma$ via the coupling
\begin{equation}
    h\int\,d^2z\,\hat{\phi}\cos\alpha\sigma.
\end{equation}
Even though this interaction can be made weakly relevant by choosing $\alpha$ appropriately, one does not find any weakly coupled fixed point of the defect RG flow, with the situation being similar to Kosterlitz-Thouless transition. On the other hand, using the boson-fermion duality at the point $\alpha^2=2\pi$, the $2d$ matter can be equivalently described as a $2d$ fermion coupled to the bulk via a Yukawa-type coupling
\begin{equation}
    h\int\,d^2z\,\hat{\phi}\bar{\psi}\psi.
\end{equation}
This sort of system was again used to construct a BCFT for the free scalar CFT$_3$ \cite{Herzog:2017xha, Behan:2021tcn}, and was studied with the help of $\epsilon-$expansion at $d=4-\epsilon$, $p=3-\epsilon$. Straightforward application of this approach leads to an interacting fixed point on the surface defect in three dimensions. We have also considered $\epsilon-$expansion at $d=4+\epsilon$, $p=1+\epsilon$, which (at least formally) allows for a UV fixed point. If trusted, it provides another example of an interacting surface defect in free scalar CFT$_5$.

Finally, in Section \ref{sec:monodromy} we turn to another class of defects, known as the monodromy defects. These are co-dimension two defects, that can be defined whenever the bulk theory has a global 0-form symmetry: one just attaches the topological operator for a chosen symmetry transformation to a co-dimension two surface. In this section, we will work with the free complex scalar field, and study the defects with a $U(1)$ monodromy around them. The free monodromy defects, a convenient point for setting off, were studied in \cite{Bianchi:2021snj}. Given the $U(1)$ transformation parameter $\alpha$, the spectrum of defect primaries is generated by the operators $\hat{O}_s^{+}$ ($s\in\mathbb{Z}-\alpha$) of dimensions $\Delta_{s}^+=\frac{d}{2}-1+|s|$, $\hat{O}_{-\alpha}^{-}$ of dimension $\Delta_{-\alpha}^-=\frac{d}{2}-1-\alpha$ and their hermitian conjugates; other primaries can be obtained as composite operators of these elementary ones. With a judicious choice of the parameter $\alpha$, operators of the form $(O_{-\alpha}^-O_{-\alpha}^{-\dagger})^n$ are found to be (slightly) relevant, and can be used to trigger an RG flow out of the free defect. We computed the beta functions for the corresponding coupling and found a perturbative fixed point. We then compute several observables, associated with these fixed points, including certain correlators and the free energy. Finally, we coupled the monodromy defect to a few CFT$_2$, finding interacting DCFT fixed points.

\section{Defects in bulk-free real scalar fields}\label{Section2}
In this section, we analyse a massless, free-real scalar field in flat $d-$dimensional space, with a \textit{conformal} (non-monodromy) defect located on a submanifold of dimension 
$p$ and codimension $q\ge 2$\footnote{Monodromy defects are discussed in Section \ref{sec:monodromy}.}. While a similar analysis was previously conducted in \cite{Nishioka:2021uef} using a conformal mapping to hyperbolic space, which allows for the application of the well-established AdS/CFT techniques, we will derive the same results here through a purely QFT approach in flat space.

We will consider a conformal defect located on a hyperplane at $x_\perp=0$, starting with a quadratic theory that has no interactions localized on the defect. However, the results presented in this section can be extended to more general free scalar fields in the bulk, including cases with defect interactions, as will be discussed later. Furthermore, the restriction to quadratic theories will be completely relaxed in the following sections.

To be precise we take the classical action to be
\begin{equation}
S = \int\,d^dx \, \frac{1}{2}(\partial\phi)^2 \, ,
\end{equation}
and solve the bulk equation of motion and canonically quantizing the resulting classical solutions. 
To expose the symmetry of the problem, we find it convenient to employ the following form of the flat metric
\beq
\label{eq:cyl_coor}
ds^2 = d \sigma^2 + d x_\perp^2 + x_\perp^2 d\Omega_{q-1}^2 \, .
\eeq
We remind the reader that the coordinates parallel to the defect will be denoted as $\sigma^i$ with $i=1,\dots,p$, while the orthogonal coordinates as $x_\perp^i$ with $i=p+1,\dots,d$. Thus, we denote $x_\perp = \sqrt{x_{\perp \, i} x_\perp^i}$. Notice also that the codimension of the defect is $q=d-p$.  
The equation of motion for the scalar field $\phi(x)$  can be written as
\beq
\label{eq:eqofmotion}
- \partial^2 \phi = -\left( \partial^2_{||}+\frac{1}{x_\perp^{q-1}}\partial_{x_\perp} (x_\perp^{q-1}\partial_{x_\perp}) + \frac{1}{x_\perp^2}\partial^2_{\Omega_{q-1}} \right)\phi  =0
\eeq
By working in Lorenztian signature ($\sigma =(\sigma_0,\vec{\sigma})$) and employing the ansatz 
\beq
\phi \sim e^{-i\omega \sigma_0}e^{i {\vec{k}}{\vec{\sigma}}} Y_{\{l\}}(\{{\vec{\theta}}\}) f_{l}(k_\perp x_\perp)
\eeq
we find the following general solution
\beq
\phi \sim e^{-i\omega \sigma_0}e^{i {\vec{k}}{\vec{\sigma}}} Y_{\{l\}}(\{{\vec{\theta}}\}) \left( c^{+}_{k, \{l\}} \frac{J_{l-1+q/2}(k_\perp  x_\perp)}{x_\perp^{q/2-1}} +  c^{-}_{k, \{l\}} \frac{J_{-(l-1+q/2)}(k_\perp x_\perp)}{x_\perp^{q/2-1}} \right) + c.c. \, ,
\eeq
where $\omega = \sqrt{k^2_\perp + {\vec{k}^2}}$ and being $J_\alpha(x)$ the Bessel functions of the first kind. This field will be quantized later on by imposing the standard equal-time commutation relation $[\phi(\vec{\sigma}_1,\sigma_0),\dot{\phi}(\vec{\sigma}_2,\sigma_0)] = i \delta(\vec{\sigma}_1-\vec{\sigma}_2)$.

A defect can occur at $x_\perp = 0$ if suitable conditions for the scalar field in the limit $x_\perp \rightarrow 0$ are imposed. Those conditions are related to the choice of which mode for each $l$ will occur in the field expansion above.
One possible choice is that the field is regular in the $x_\perp \rightarrow 0$ limit. This would select only the $``+"$ modes, discarding all the $``-"$ ones, the latter being  divergent in this limit. This regularity condition corresponds to a trivial defect. However, this is not the end of the story. Indeed, we can relax the requirement of regularity, allowing for some $``-"$ mode. At this point, the defect unitarity comes into the game posing a strong restriction on which $``-"$ is really allowed. For scalars, the dimension of the defect primaries needs to satisfy 
\begin{align}
\label{eq:uni_def_bound}
\hat \Delta &\ge \frac{p}{2}-1 \,,  \qquad p \ge 2 \,,\nonumber\\
\hat \Delta &\ge 0 \,,  \qquad\qquad p < 2 \,.
\end{align}
The only exception is the identity operator, whose defect conformal dimension is $\hat \Delta = 0$.
    For a quadratic theory, we can then proceed with an approach similar to the one used in \cite{Lauria:2020emq,Bianchi:2021snj}. The bulk field $\phi$ can be written as a Laurent expansion around $x_\perp = 0$. For each mode $\{l\}$, this consists of a defect primary and its descendants. To be concrete, we may introduce defect operators by stripping off the leading behaviour of the Bessel functions in the $x_\perp \longrightarrow 0$ limit such that the remaining part is a quantity of order $\mathcal{O}(x_\perp^0)$. We write
\beq
\phi \sim  c^{+}_{k, \{l\}} O^{+}_{\{l\},k}(x_\perp, \sigma)  x_\perp^{l} Y_{\{l\}}(\{\theta\}) +  c^{-}_{k, \{l\}}  O^{-}_{\{l\},k}(x_\perp, \sigma ) x_\perp^{2-l-q}Y_{\{l\}}(\{\theta\}) \, .
\eeq
The idea is that the bulk operators $O^{+}_{\{l\},k}(x_\perp, \sigma)$ and $O^{-}_{\{l\},k}(x_\perp, \sigma)$ contain, for each $\{l\}$, a defect primary, which is $\hat O^{\pm}_{\{l\},k}(\sigma)\equiv O^{\pm}_{\{l\},k}(0, \sigma)$, and all its descendants, which appear as the coefficients in the higher order terms in the $x_\perp$-expansion of the fields $O^{\pm}_{\{l\},k}(x_\perp, \sigma)$.
For a quadratic theory, it is then easy to extract the conformal dimensions of such operators by applying dimensional analysis. They read
\beq
\label{eq:delta_O_scalar}
\hat \Delta^{+}_{\{l\}} =d/2 -1 +l  \, , \qquad
\hat \Delta^{-}_{\{l\}} = d/2+1 -l-q \, .
\eeq
More precisely, the bulk field $\phi$ can be written as
\begin{equation}\label{eq:defect_OPE}
\begin{split}
\phi =\;& \sum_{\{l\}} c_{\phi \hat O_{\{l\}}^+} x_\perp^{l} Y_{\{l\}}(\{\theta\}) \mathcal{C}^+_{l}\left(x_\perp^2 \partial^2_{\sigma}\right) \hat O^+_{\{l\}}(\sigma) + c_{\phi \hat O_{\{l\}}^-} x_\perp^{2-l-q} Y_{\{l\}}(\{\theta\}) \mathcal{C}^-_{l}\left(x_\perp^2 \partial^2_{\sigma}\right) \hat O^-_{\{l\}}(\sigma) \, .
\end{split}
\end{equation}
The differential operators $\mathcal{C}^{\pm}_{l}(x_\perp^2 \partial_{\sigma}^2)$ resum the contribution of all the conformal descendants, and are fixed by conformal invariance to be
\begin{equation}\label{eq:diff_op}
\mathcal{C}^{+}_{l}\left(x_\perp^2 \partial^2_{\sigma}\right) \equiv \sum_{k=0}^{+\infty} \frac{(-4)^{-k}(x_\perp^2 \partial^2_{\sigma})^k}{k!(1 + \hat\Delta_{\{l\}}^+-\frac{p}{2})_k}\,, \qquad \mathcal{C}^{-}_{l}\left(x_\perp^2 \partial^2_{\sigma}\right) \equiv \sum_{k=0}^{+\infty} \frac{(-4)^{-k}(x_\perp^2 \partial^2_{\sigma})^k}{k!(1 + \hat\Delta_{\{l\}}^--\frac{p}{2})_k}\,,
\end{equation}
where $(a)_k \equiv a(a+1)\dots(a+k-1)$ if $k \ne 0$ and $(a)_0\equiv 1$ is the Pochhammer symbol.
\subsubsection*{Constraints from unitarity}
From the unitarity bound \eqref{eq:uni_def_bound} we obtain, as expected, that all the $``+"$ modes are always allowed. The situation is more intricate for the $``-"$ modes, where it is  useful to distinguish between the cases $p\ge 2$ and $p=1$. 
For $p\ge 2$, both the $l=0$ and $l=1$ modes may be allowed. Regarding $l=0$, this is acceptable provided $q< 4$, while for the $l=1$ the condition happens to be $q \le 2$. However, for $q=2$, it can be argued that the $l=1$ case does not lead to a non-trivial defect \cite{Lauria:2020emq}. Indeed, the unitarity bound is saturated and thus the field must necessarily be a free field satisfying the equation $\partial^2_\sigma \hat O=0$\footnote{More precisely, when we deform the theory from $q=2$ to $q=2-\epsilon$, multiplet recombination is taking place \cite{Rychkov:2015naa, Bashmakov:2016pcg}, and the short multiplet of $\hat{O}_1^-$ is recombining with the long multiplet of $\hat{O}_1^+$ to form the long multiplet of $\hat{O}_1^-$.} (this can be actually checked directly from the equation of motion \eqref{eq:eqofmotion}).
For $p=1$, the condition in \eqref{eq:uni_def_bound} imposes that $q<3$ for $l=0$, while it gives $q<1$ for $l=1$, which also means no unitary defect in this case. In particular, for $q\ge 4$ (or $q\ge 3$ in the line defect case) no $``-"$ modes are allowed.

This means we are left with the following possibilities: either only $``+"$ modes are present, which means trivial defect, or we may swap the $l=0$ (or $l=1$)  $``+"$ mode for the $l=0$ (or $l=1$) $``-"$ mode. In those last scenarios, we obtain a free real scalar field in the presence of a \textit{conformal} defect.

Having discussed which are the allowed non-trivial defects, we can proceed by canonically quantizing the theory. For the sake of simplicity, in the following we will only consider the $l=0$ $``-"$ mode, keeping the $``+"$ mode for $l=1$. With this assumption, we may write
\begin{equation}
\label{eq:scalar_field_modes}
\begin{split}
\phi =   \sum_{\{l \}\ne \{0\}}^{\infty} & \int dk_\perp \int d^{p-1} {\vec k} \,  \frac{\sqrt{k_\perp}}{(\sqrt{2\pi})^{p}\sqrt{2\omega}}  \Bigg[  a_{\{l\}}( k) Y_{\{l\}}(\{\theta\}) \frac{J_{l-1+q/2}\left(k_\perp\,x_\perp\right)}{x_\perp^{q/2-1}} e^{-i \omega t+i {\vec k}\cdot{\vec{\sigma}}}  \Bigg] \\
&+ \int dk_\perp \int d^{p-1} {\vec k} \,  \frac{\sqrt{k_\perp}}{(\sqrt{2\pi})^{p}\sqrt{2\omega}}  \Bigg[ \sqrt{1-\xi} a^{+}_{\{0\}}( k)  \frac{J_{q/2-1}\left(k_\perp\,x_\perp\right)}{x_\perp^{q/2-1}} \\
& \hspace{3.cm}+ \sqrt{\xi}\, a^{-}_{\{0\}}( k)  \frac{J_{1-q/2}\left(k_\perp\,x_\perp\right)}{x_\perp^{q/2-1} }  \Bigg]  Y_{\{0\}}(\{\theta\}) e^{-i \omega t+i {\vec k}\cdot{\vec{\sigma}}} + \text{h.c.}\, ,
\end{split}
\end{equation}
where the ladder operators satisfy the standard commutation relations.
The mode expansion in terms of creation and annihilation operators has a more clear meaning in the case of a Gaussian theory, where either $\xi=0$ for the trivial defect, or $\xi = 1$. Notice that in the case of a non-conformal defect, we would have a more complicated structure which involves a scale, and only a particular linear combination of the modes is kept.

%
\subsection{Bulk-bulk propagator and correlation functions}

The bulk-bulk propagator, i.e. the two-point function of $\phi$, can be easily computed by noticing that with $\xi=0$ we need to recover the propagator without defects. The remaining part, proportional to $\xi$, can be computed by following, for instance, appendix B of \cite{Bianchi:2021snj}. In Euclidean signature, we obtain
\beq
\label{eq:prop_scal}
\begin{split}
\left<\phi(x_1)\phi(x_2) \right> =  \frac{C_{\phi}}{|x_1-x_2|^{d-2}} - \frac{\xi}{4 \pi^{\frac{p}{2}+1}} \frac{ \Gamma \left(\frac{q}{2}\right)}{\pi ^{\frac{q}{2}-1}} \left(\frac{1}{x_{\perp 1} x_{\perp 2}}\right)^{\frac{d}{2}-1}&\Bigg[ \frac{\Gamma\left(\frac{d}{2}-1\right)}{\Gamma\left(\frac{q}{2}\right)} F_{{\frac{d}{2}-1}}(\eta) \\
&- 
\frac{\Gamma\left(\frac{p-q}{2}+1\right)}{\Gamma\left(2-\frac{q}{2}\right)} F_{{\frac{p-q}{2}+1}}(\eta)  \Bigg]
\end{split}
\eeq
where $F$ is the function defined in \eqref{defectblock}, we used the normalisation $Y_{\{0\}}(\{\theta\})=\sqrt{\pi ^{1-\frac{q}{2}} \Gamma \left(\frac{q}{2}\right)}$, and $C_{\phi}=\Gamma(\frac{d-2}{2})/(4\pi^{\frac{d}{2}})$. 

Furthermore, notice that for $q=2$, the two functions $F$ in the propagator \eqref{eq:prop_scal} cancel exactly producing the one for a trivial defect. This is expected since, in agreement with eq. \eqref{eq:delta_O_scalar}, the two defect operators $\hat{\mathcal{O}}^{+}_{\{l=0\}}$ and $\hat{\mathcal{O}}^{-}_{\{l=0\}}$ reduce to the same operator. This is, in particular, the case for a line defect in the $3d$ free scalar CFT, where the quadratic deformation is easily seen to be marginally irrelevant.

While the propagator has been computed for the quadratic theory, meaning either $\xi=0$ or $\xi=1$, this result is expected to be more general and extendable to cases involving defect interactions, as long as the bulk theory remains free. In such cases, $\xi$ can possibly be different from $0$ and $1$, as we will show in the following sections. The extended applicability of the propagator can be understood by following a different approach, i.e. by solving directly  for the Green function as done in \cite{Lauria:2020emq} for the $\mathbb{Z}_2$ monodromy defect. By following similar steps, we end up with \eqref{eq:prop_scal}, with the range of $\xi$ constrained by unitarity to be 
\begin{equation}
\label{eq:xi_range}
    \xi \in [0,1] \, .
\end{equation}
As a consistency check of this claim, in Appendix \ref{app:BB_prop_DC} we present the leading correction to the propagator in the specific case of the Dirichlet coupling (see Sec. \ref{Section3}) recovering the propagator \eqref{eq:prop_scal} with a specific value of $\xi$ compatible with the range \eqref{eq:xi_range}.

\subsubsection*{One-point functions}
From the propagator \eqref{eq:prop_scal}, we can compute the $\phi^2$ one-point function. By taking the coincidence limit and neglecting the usual UV divergence coming from the identity exchange in the $\phi-\phi$ OPE,  we obtain
\beq
\label{eq:phi_sqr}
\left< \phi^2 (x) \right> =  - \frac{\pi ^{\frac{1}{2}-\frac{d}{2}} 2^{-p-1} \Gamma \left(\frac{d}{2}-1\right) \Gamma \left(-\frac{d}{2}+p+1\right)}{\Gamma \left(\frac{p+1}{2}\right) \Gamma \left(\frac{1}{2}
   (-d+p+2)\right)}\xi \left(\frac{1}{x_\perp}\right)^{d-2} \, .
\eeq

Another relevant correlator is the stress-tensor one-point function. Since this correlator is fixed by conformal invariance up to a constant as shown in  \eqref{eq:stress-tensor-one-pt-fn}, we only need one of its components. In the coordinates \eqref{eq:cyl_coor}, we choose to compute the $|x_\perp||x_\perp|$ component of the stress tensor, which we call $T_{\perp\perp}$. Since this component can only depend on $|x_\perp|$, we merely need to compute the following expectation value
\begin{equation}
\label{eq:T_corr}
\left<T_{\perp\perp}(x_\perp) \right> = \left<\partial_\perp \phi \, \partial_\perp \phi \right> -\frac{1}{4(d-1)} \left[ (d-1)\partial^2_{\perp} + (q-1) \frac{1}{x_\perp} \partial_\perp \right]\left<\phi^2 \right> \, .
\end{equation}
Thus, we need to compute the one-point function of $\phi^2$ and of $(\partial_\perp \phi)^2$. 
The first one has already been computed in \eqref{eq:phi_sqr}, while the latter is found to be
%
%
%
\begin{equation}
\begin{split}
\left<\left( \partial_\perp \phi(x_\perp)\right)^2 \right> =
- \frac{\pi ^{-d/2} \Gamma \left(\frac{d}{2}+1\right) \Gamma \left(\frac{p}{2}+1\right) \Gamma \left(-\frac{d}{2}+p+1\right)}{2 (p+1) \Gamma (p) \Gamma \left(\frac{1}{2}
   (-d+p+2)\right)} \xi \, \frac{1}{x_\perp^{d}} \, .
\end{split}
\end{equation}
Plugging the above results together according to \eqref{eq:T_corr}, we obtain the one-point function of the stress tensor
\begin{equation}
\begin{split}
\left<T_{\perp\perp} \right> & = \frac{ (1-q) (2-q) \Gamma \left(\frac{d}{2}\right) \Gamma \left(\frac{p-q}{2}+1\right)}{\pi ^{\frac{d-1}{2}} 2^{p+3}(d-1) \Gamma \left(\frac{p+3}{2}\right)
   \Gamma \left(1-\frac{q}{2}\right)} \xi \, \frac{1}{x_\perp^d} \, .
\end{split}
\end{equation}
Notice that when the codimension is 1, namely $d=p+1$, then this one-point function is vanishing while the ones above are not. This is indeed expected from the conservation of the stress tensor.
From this result, we can extract the quantity $a_T$, which reads
\begin{equation}
a_T = \frac{ (2-q) \Gamma \left(\frac{d}{2}\right) \Gamma \left(\frac{p-q}{2}+1\right)}{\pi ^{\frac{d-1}{2}} 2^{p+3}(d-1) \Gamma \left(\frac{p+3}{2}\right)
   \Gamma \left(1-\frac{q}{2}\right)} \xi \, .
\end{equation}
Notice that the coefficient $h \ge 0$ in the unitarity range $q \leq 4$ for $p \ge 2$ and $q<p+2$ for $p<2$, with $\xi \in(0,1]$. 

As discussed in the introduction, when $p=2$ there is the Weyl anomaly \eqref{eq:2d-defect-Weyl-anomaly}. One of its coefficients, namely $d_2$, is related to $a_T$ according to \eqref{eq:2d1pt}. Thus, in this case we find
\begin{equation}
d_2 = -\frac{1}{8}(d-2)(d-4)^2 \xi \, .
\end{equation}
\subsubsection*{Displacement operator}
The displacement operator for the non-trivial defect can be constructed from the lightest defect primaries $O^{+}_{\{l=1\}}$ and $O^{-}_{\{l=0\}}$. Indeed, notice that both the conformal dimension ($\Delta_{\hat{D}}=p+1$) and transversal spin ($s=1$) match if we take the product of the two fields. However, for the purpose of finding its precise form, we find it more convenient to work in Poincaré coordinates rather than cylindrical ones.  Thus, we can define new defect primary as the following limit of $\phi$:
\beq
\hat{\tilde{O}}_{l=0}^{-} \equiv \text{lim}_{x_\perp\rightarrow 0}x_\perp^{q-2} \phi \, , \qquad \left(\hat{\tilde{O}}^{+ }_{l=1}\right)^i \equiv \text{lim}_{x_\perp\rightarrow 0} \partial^i\phi \, .
\eeq
The ansatz for the displacement operator is thus
\beq
\hat D^i \equiv \kappa_{\hat D}\, \hat{\tilde{O}}_{l=0}^{-}\left(\hat{\tilde{O}}^{+ }_{l=1}\right)^i\, ,
\eeq
where $\kappa_{\hat D}$ is a coefficient that needs to be fixed. To check this ansatz and fix $\kappa_{\hat D}$, we require the Ward identity 
\beq
\int d^p \sigma \, \left< \phi^2(x_j,0) \hat D_i(\sigma) \right> = \partial_i \left< \phi^2(x_j,0)\right> \, .
\eeq
To compute the above expression we need the correlators
\beq
\left<\hat{\tilde{O}}_{l=0}^{-}(\sigma)\phi(x_\perp,0) \right>= \xi \frac{   \Gamma \left(\frac{d-p}{2}\right)
   \Gamma \left(-\frac{d}{2}+p+1\right) }{4 \pi ^{\frac{d}{2}} \Gamma \left(2-q/2\right)} \frac{1}{x_\perp^{q-2}} \left(\frac{1}{x_\perp ^2+\sigma
   ^2}\right)^{p+1-\frac{d}{2}} \, ,
\eeq
and
\beq
\left<\hat{\tilde{O}}_{l=1}^{+, i}(\sigma)\phi(x_\perp,0) \right>= -\frac{\Gamma \left(\frac{d}{2}\right)}{2\pi ^{d/2}}    \frac{x^i}{\left(x_\perp
   ^2+\sigma ^2\right)^{d/2}} \, .
\eeq
Solving for $\kappa_{\hat D}$ gives 
\beq
\kappa_{\hat D} = -4 \pi ^{ q/2-1} \sin \left( \pi  q/2\right)
   \Gamma \left(2-q/2\right) \, .
\eeq
Then, the two-point function of the displacement operator is 
\beq
\begin{split}
\left<\hat D^i(\sigma) \hat D^j(0) \right>& =\kappa_{\hat D}^2 \left<\hat{\tilde{O}}_{l=0}^{-}(\sigma) \hat{\tilde{O}}_{l=0}^{-}(0) \right> \left<\hat{\tilde{O}}_{l=1}^{+, i}(\sigma) \hat{\tilde{O}}_{l=1}^{+, j}(0) \right> \\ 
&=\kappa_{\hat D}^2  \frac{\Gamma \left(\frac{d}{2}\right)}{2\pi ^{d/2}}     \frac{   \Gamma \left(\frac{d-p}{2}\right)
   \Gamma \left(-\frac{d}{2}+p+1\right) }{4 \pi ^{\frac{d}{2}} \Gamma \left(2-q/2\right)} \xi  \frac{\delta^{ij}}{|\sigma|^{2p+2}} \, . 
   \end{split}
\eeq
Thus, we find
\begin{equation}
C_{\hat D} = \frac{1}{\pi^{p+1}}    (2-q) \Gamma \left(\frac{d}{2}\right) \sin
   \left(\pi \frac{q}{2}  \right) \Gamma \left(\frac{p-q}{2}+1\right)\xi \, ,
\end{equation}
and for $p=2$, by using the relation to the anomaly coefficient \eqref{eq:p2D2}, we  get
\begin{equation}
d_1 =3\pi\frac{(d-2)(d-4)^2}{16}\xi \, .
\end{equation}
We refer the reader to appendix \ref{app:displacement} for an alternative discussion of the displacement operator.

\subsection{RG flows}

Now we discuss the possible RG flows in the free scalar theory with a defect. We focus on the quadratic deformations, and this means we have only two possible relevant deformations. The operator $\left(\hat O^{+}_{l=0}\right)^2$ is relevant if $p>d-2$ while $\left(\hat O^{-}_{l=0}\right)^2$ is relevant if $q>2$. Due to the unitarity constraint, we are left with the following possibilities:
\beq
\begin{cases} 
\left(\hat O^{+}_{l=0}\right)^2 \qquad \text{relevant if}\qquad q<2 \, ,\\
\left(\hat O^{-}_{l=0}\right)^2 \qquad \text{relevant if}\qquad 2 < q \le 4 \, .\\
\end{cases}
\eeq
Now we study those RG flows and show that the cases $\xi = 0$ and $\xi = 1$ are exactly connected by them. The relevant perturbation $(\hat{O}^{+})^2$ in the case of a surface defect has been studied in other works \cite{Giombi:2023dqs,Raviv-Moshe:2023yvq,Trepanier:2023tvb, Brax:2023goj}, where the perturbation is written as $\phi^2$ integrated on the $2$-dimensional submanifold. To the best of our knowledge, the other relevant deformation $(\hat{O}^{-})^2$ as well as the extension to a generic $p-$dimensional defect has not been considered previously in the literature. 

To simplify the computations below, we find it convenient to be slightly more general and only assume that the perturbation is quadratic and generated by a defect conformal primary $\hat O_{\hat \Delta}$ of conformal dimension $\hat \Delta$. More precisely we will consider 
\beq
 \label{eq:quadr_pert}
 \delta S_\text{pert} = h_c \int d^p\sigma \, \left(\hat O_{\hat \Delta}\right)^2 \, .
 \eeq
We can proceed by assuming the defect primary is normalized as 
\beq
\left< \hat O_{\hat \Delta}(\sigma)\hat O_{\hat \Delta}(0) \right> = \frac{1}{|\sigma|^{2\hat \Delta}}
\eeq
and with a bulk-defect two-point function given by
\beq
 \left<\phi(x) \hat  O_{\hat\Delta}(0) \right> =c_{\phi \hat O_{\hat \Delta}} \frac{1}{x_\perp^{\Delta_\phi - \hat\Delta}} \left(\frac{1}{x_\perp ^2+\sigma
   ^2}\right)^{\hat \Delta} \, .
\eeq
Notice that specialising to the defect primaries $\hat O^{+}_{\{l=0\}}$ and $\hat O^{-}_{\{l=0\}}$ we have
\beq
c_{\phi \hat O^{+}_{\{0\}}} = \sqrt{C_\phi} \, , \qquad c_{\phi \hat O^{-}_{\{0\}}} = \sqrt{ \frac{   \Gamma \left(\frac{q}{2}\right)
   \Gamma \left(\frac{d}{2}-q+1\right) }{4 \pi ^{\frac{d}{2}} \Gamma \left(2-q/2\right)}}\, .
\eeq
To proceed we consider the Fourier representations 
\begin{equation}
\label{eq:fourier_xp}
\frac{1}{(\sigma^2+x_\perp^2)^{\hat\Delta}} = \int \frac{d^p k}{(2\pi)^p} \, f(k \, x_\perp) k^{-p+2 \hat \Delta} e^{i k \cdot \sigma}, \qquad f(\zeta) \equiv \frac{2 \pi^{p/2}}{\Gamma( \hat\Delta )}(2 \zeta)^{\frac{p}{2}-\hat\Delta} K_{\frac{p}{2}-\hat\Delta}( \zeta ) \,,
\end{equation}
and
\begin{equation}
\label{eq:fourier_sigma}
\frac{1}{|\sigma|^{2\hat\Delta}} = f_0\, \int \frac{d^p k}{(2\pi)^p} \,  k^{-p+2 \hat \Delta} e^{i k \cdot \sigma}, \qquad f_0\equiv f(0) = \frac{2^{p-2\hat\Delta}\pi^\frac{p}{2}\Gamma\left( \frac{p}{2}-\hat \Delta \right) }{\Gamma ( \hat\Delta)}  \,.
\end{equation}
In this way, we can easily compute the two-point function of the field $\phi$ in the presence of a quadratic perturbation \eqref{eq:quadr_pert}. At this end, let us expand the correlator in powers of $h_c$
\begin{equation}
\left< \phi(x_1)\phi(x_2) \right>_{h_c}=\left< \phi(x_1)\phi(x_2) \right>_{0}+\sum_{k=1}^{\infty}\frac{(-1)^kh_c^{k}}{2^{k}(k)!}\left<\phi(x_1)\phi(x_2)\prod_{i=1}^{k}\,\int d^p\sigma_i \, \hat O_{\hat \Delta}^2(\sigma_i) \right>^{\text{\tiny conn}}_0.
\end{equation}
where in the correlators we only keep the connected terms.
Applying Wick contractions and passing to the momentum space, the $k^{\text{th}}$ term in the sum can be evaluated to
\begin{equation}
(-1)^kh_c^{k}c^2_{\phi \hat O_{\hat \Delta}}f_0^{k-1}\int\frac{d^pk}{(2\pi)^p}\frac{e^{ik(\sigma_1-\sigma_2)}}{k^{(p-2\hat\Delta)(k+1)}}f(kx_{\perp,1})f(kx_{\perp,2}),
\end{equation}
We can then resum the series, with the result
\begin{equation}
\label{eq:prop_h_scalar}
\begin{split}
\left< \phi(x_1)\phi(x_2) \right>_{h_c} = &\left< \phi(x_1)\phi(x_2) \right>_{0} - 4 \pi^p \frac{2^{p-2\hat\Delta}}{\Gamma^2 (\hat\Delta )} c^2_{\phi \hat O_{\hat \Delta}} \left(\frac{1}{x_{1 \perp}x_{2 \perp}}\right)^{\frac{q}{2}-1} \times \\
\times &\int \frac{d^p k}{(2\pi)^p} e^{i k \cdot \sigma} h_c \frac{k^{-p+2\hat \Delta}}{1+h_c  f_0 k^{-p+2\hat \Delta}} K_{\frac{p}{2}-\hat\Delta}(k \, x_{\perp,1} )K_{\frac{p}{2}-\hat\Delta}(k \, x_{\perp,2} )\,.
\end{split}
\end{equation}
We notice that the deformation becomes marginal when $p-2\hat\Delta=0$. In this case, we obtain a pole in the denominator of the above expression due to the constant $f_0$. Exploiting this fact, we can find a $\beta$-function for this deformation. What follows is a generalisation of the discussion in \cite{Giombi:2023dqs} to a generic defect dimension $p$. By defining $2\hat\Delta=p-\epsilon$ we can define the renormalized coupling as 
\beq
h_c = \mu^\epsilon\frac{h_r}{1-A \, h_r /\epsilon} \, ,
\eeq
where, with the choice of the minimal subtraction scheme, the constant $A$ is fixed to be $A=2  \pi^{p/2}/\Gamma(p/2)$. Thus, we obtain
\beq
\beta = -\epsilon h_r + \frac{2  \pi^{\frac{p}{2}}}{\Gamma\left(\frac{p}{2}\right)}\, h_r^2 \, ,
\eeq
with a fixed point at
\beq
h_r = \frac{\Gamma\left(\frac{p}{2}\right)}{2  \pi^{\frac{p}{2}}} \epsilon \, .
\eeq
For $2\hat\Delta-p<0$, the coupling $h_c$ is relevant and the IR conformal point is obtained in the limit $h_c \rightarrow +\infty$. The above two-point function then reduces to
\begin{equation}
\label{eq:phi-phi-conf}
\begin{split}
\left< \phi(x_1)\phi(x_2) \right>_{+\infty} = &\left< \phi(x_1)\phi(x_2) \right>_{0}-   \frac{4\pi^{\frac{p}{2}}}{\Gamma \left(\frac{p}{2}-\hat\Delta \right)\Gamma \left(\hat\Delta \right)} c^2_{\phi \hat O_{\hat \Delta}} \left(\frac{1}{x_{1 \perp}x_{2 \perp}}\right)^{\frac{d}{2}-1} \times \\
\times & \frac{\pi}{4 \pi^{\frac{p}{2}}\sin \pi \left( \frac{p}{2}-\hat\Delta \right) } \left[\frac{\Gamma \left(\hat\Delta\right)}{\Gamma \left(1-\frac{p}{2}+\hat\Delta\right)} \,    F_{\hat\Delta}(\eta)	 - \frac{\Gamma \left(p-\hat\Delta\right)}{\Gamma \left(1+\frac{p}{2}-\hat\Delta\right)} \,    F_{p-\hat\Delta}(\eta)  \right] \,.
\end{split}
\end{equation}
%
%
where we used the integral formula in \eqref{eq:KK_int} and the definition of the function $F$ reported in \eqref{defectblock}. We notice that this result matches the propagator \eqref{eq:prop_scal} if we consider either the relevant perturbation $(O^{+}_{\{l=0\}})^2$ or $(O^{-}_{\{l=0\}})^2$, starting from the trivial defect $\xi=0$ or the non-trivial one $\xi=1$ respectively.

It is interesting to come back to the full propagator and take the defect limit. In this case, we find
\begin{equation}
\begin{split}
\left< \hat O_{\hat \Delta}(\sigma_1)\hat O_{\hat \Delta}(\sigma_2) \right>_{h_c} = \left<  \hat O_{\hat \Delta}(\sigma_1)\hat O_{\hat \Delta}(\sigma_2) \right>_{0} -&  f_0^2 \,  \int \frac{d^p k}{(2\pi)^p} e^{i k \cdot \sigma}\, \frac{1}{k^{p-2\hat\Delta}}  \frac{ h_c \, k^{-p+2\hat\Delta}}{1+h_c f_0 k^{-p+2\hat\Delta}}\, .
\end{split}
\end{equation}
The IR fixed point is reached by sending $|\sigma|\rightarrow +\infty$, or alternatively sending $h_c \rightarrow +\infty$. By expanding the above equation, we get
\begin{equation}
\begin{split}
\left< \hat O_{\hat \Delta}(\sigma_1)\hat O_{\hat \Delta}(\sigma_2) \right>_{h_c} = \left<  \hat O_{\hat \Delta}(\sigma_1)\hat O_{\hat \Delta}(\sigma_2) \right>_{0} -&  f_0^2 \,  \int \frac{d^p k}{(2\pi)^p} e^{i k \cdot \sigma}\, \left[\frac{k^{2\hat\Delta-p}}{f_0} -\frac{1}{f_0^2 \, h_c} + \frac{k^{p-2\hat\Delta}}{f_0^3\, h_c^2} + \dots \right]\, .
\end{split}
\end{equation}
Now, assuming $2\hat\Delta-p\neq0$, it is straightforward to see that the first term in the expansion cancels exactly the unperturbed correlator $<\cdot>_0$, the second term in the expansion is a contact term we neglect, while the third term gives the leading IR behaviour\footnote{We note that the resulting two-point function has positive normalization for $0<d-p<2$.}
\begin{equation}
\label{eq:OO_defect_scal}
\begin{split}
\left< \hat O_{\hat \Delta}(\sigma_1)\hat O_{\hat \Delta}(\sigma_2) \right>_{h_c} & = -\frac{ 2^{p-2 \hat\Delta } \Gamma (p-\hat\Delta )}{\pi ^{\frac{p}{2}}\Gamma \left(\Delta
   -\frac{p}{2}\right)}  \frac{1}{f_0\, h^2_c}\frac{1}{|\sigma|^{2p-2\hat\Delta}}+ \mathcal{O}\left(\frac{1}{h_c^3} \right)  \\
& = \frac{\Gamma (\hat\Delta ) (p-2 \hat\Delta ) \sin
   \left( \pi  \left(\frac{p}{2}- \hat\Delta \right)\right) \Gamma (p-\hat\Delta )}{2 \pi ^{p+1}} \frac{1}{ h^2_c}\frac{1}{|\sigma|^{2p-2\hat\Delta}}+ \mathcal{O}\left(\frac{1}{h_c^3} \right)\, .
\end{split}
\end{equation}
From this correlator we deduce the conformal dimension of the operator $\hat \phi$ at the IR fixed point, which happens to be
\begin{equation}
\hat \Delta_{\text{\tiny IR}} = p-\hat\Delta \, .
\end{equation}
This is in agreement with the papers \cite{Giombi:2023dqs, Trepanier:2023tvb, Raviv-Moshe:2023yvq} when $p=2$ and $\hat\Delta = d/2-1$.

\subsection{Defect free energy}
We can compute the defect free energy variation due to the relevant perturbation \eqref{eq:quadr_pert}.
 Following \cite{Klebanov:2011gs}, we perform the Hubbard-Stratonovich transformation which usually is done with a quartic coupling. In this case, we want to reduce the power of the boundary perturbation to a linear coupling.
 We obtain
 \beq
 \begin{split}
 Z[h] & = \int \mathcal{D} \phi \exp \left\{ - S_0  - h_c \int d^{p} \sigma \, \hat O^2_{\hat\Delta} \right\}\\
 & =  \frac{1}{\mathcal{Z}}  \int \mathcal{D} \hat \lambda \,\mathcal{D} \phi  \, \exp \left\{ - S_0  - \frac{1}{4 \, h_c} \int d^{p} \sigma \, \hat \lambda^2  - i \int d^{p} \sigma \, \hat \lambda \, \hat{O}^2_{\hat\Delta} \right\} 
 \end{split}
 \eeq
 where we defined 
 \beq
 \mathcal{Z} \equiv \int \mathcal{D} \hat \lambda \, e^{ - \frac{1}{4 \, h} \int d^{p} \sigma \, \hat \lambda^2}
 \eeq
 where the $\hat \lambda$-integration is parallel to the real axis.
 Now we can integrate over the field $O_{\hat\Delta}$ to get
 \beq
 \begin{split}
 Z[h]  &  =\frac{Z[0]}{\mathcal{Z}}   \int \mathcal{D} \hat \lambda  \, \exp \left\{   - \frac{1}{2} \int d^{p} \sigma_1 \, d^{p} \sigma_2  \, \hat \lambda \left(\sigma_1 \right) \left[ \hat G\left(\sigma_1- \sigma_2\right) + \frac{\delta^{p}(\sigma_1-\sigma_2)}{2\, h_c} \right] \hat \lambda \left(\sigma_2\right)  \right\},
 \end{split}
 \eeq
 where in the last step we integrated over the field $\hat O_{\hat\Delta}$, and $\hat G$ is the defect-defect propagator. By integrating over the auxiliary field $\hat \lambda$ we finally obtain
 \beq
 \label{eq:Delta_G_Id}
  Z[h]  = Z[0] \frac{1}{\sqrt{\text{Det}\left(\, \hat G(\cdot)+\frac{1}{2\, h_c}\right)}} .
 \eeq
Thus, the variation in the free energy is given by
 \beq
 \Delta F \equiv  F_{\text{IR}}-F_{\text{UV}}=-\log\left|\frac{Z[h]}{Z[0]}\right|=  \frac{1}{2} \Tr \log\left( \hat G(\cdot)+ \frac{1}{2\, h_c} \right)
 \eeq
 In general the defect-defect propagator can be written as
 \beq
 \hat G(\sigma_1,\sigma_1) = \frac{A}{s^{2\hat\Delta}(\sigma_1,\sigma_2)}
 \eeq
being $2\hat \Delta$ the dimension of the perturbing operator, $s(\cdot)$ is the invariant distance on the $p$-sphere of radius $R$, and $A$ is an unessential normalization.
In \cite{Gubser:2002vv}, it is reported the spectral decomposition of the propagator, which reads
\beq
\frac{1}{s^{2\hat\Delta}(\sigma_1,\sigma_2)} = \sum_{\{l\}} g_l \, Y_{\{l\}}(\sigma_1) Y_{\{l\}} (\sigma_2)
\eeq 
where the eigenvalues are\footnote{We note that the spherical harmonics are taken to be normalized as $\int\,d^px\sqrt{g}\,Y_{l,m}Y^*_{l',m'}=\delta_{l,l'}\delta_{m,m'}$, and so have dimension $p/2$. This gives a factor of $R^p$ in the expression for $g_l$.}
\beq
g_l = R^{p-2\hat \Delta}\, \pi^{\frac{p}{2}} 2^{p-\hat \Delta} \frac{\Gamma\left( \frac{p}{2} - \hat \Delta \right)}{\Gamma\left( \hat \Delta \right)} \frac{\Gamma\left( l+\hat\Delta \right)}{\Gamma\left( p + l - \hat \Delta \right)}, \qquad  l \ge 0,
\eeq
with multiplicity
\beq
m_l = \frac{(2 l+p-1)(l+p-2)!}{(p-1)! \, l!} .
\eeq
In the IR limit $h_c\rightarrow+\infty$ the identity contribution in \eqref{eq:Delta_G_Id} can be neglected, and we get
\beq
 \Delta F = \frac{1}{2} \sum_{l=0}^{+\infty} m_l   \log\left(\frac{2 A h_c}{R^{2\hat \Delta-p}}g_l\right)  .
\eeq
The usual approach to treat such a sum is to exploit the identity
\beq
 \Delta F = -\frac{1}{2} \frac{d}{ds} \left[ \sum_{l=0}^{+\infty} m_l   \left(\frac{2 A h_c}{R^{2\,\hat\Delta-p}}g_l \right)^{-s} \right]_{s=0} = -\frac{1}{2} \frac{d}{ds} \left[ \sum_{l=0}^{+\infty} m_l \,   g_l^{-s} \right]_{s=0} + \left. \Delta F\right|_{\text{anom}} \log (\mu R ) \,,
\eeq
where
\beq
\begin{split}
\left. \Delta F\right|_{\text{anom}} & = \left. \frac{p -2\hat \Delta}{2} \sum_{l=0}^{+\infty} m_l    \left(\frac{\Gamma\left( l+\hat \Delta \right)}{\Gamma\left( p + l - \hat \Delta \right)}\right)^{-s} \right|_{s=0} \, .
\end{split}
\eeq
If we focus on the case $p$ even then $\left. \Delta W\right|_{\text{anom}} \ne 0$, and there is a conformal anomaly producing a logarithmic dependence on $R$. 

While it is still a difficult problem to evaluate this sum, we can extract the logarithmic divergence, relevant for the case of even fixed $p$. The ways to do it are extensively covered in the literature, and below we choose to follow Appendix C of \cite{Brust:2016gjy}. There, the authors notice that the only contribution to the anomaly (namely, in the limit $s\rightarrow 0$) is given by the pieces in the sum which do not converge. Thus, the idea is to first separate the $l = 0$ mode and then, expanding the ``integrand" for large $\ell$, keep the first few contributing terms, and evaluate the sum with the aid of zeta-function regularisation. This procedure can be implemented on a case-by-case basis \textit{e.g.} with the help of a computer. We do this for the case $p=2$ to  illustrate the method, while we directly give the result for the other cases of interest.

\paragraph{$p=2$ case}
We need to regularise the sum
\beq
\begin{split}
\left. \Delta F\right|_{\text{anom}} & =\left(1-\hat\Delta \right) \sum_{l=0}^{+\infty} (1+2 l) l^{2(1-\hat\Delta)s}\left(\frac{l^{2(\hat\Delta-1)} \Gamma \left(l+2 -\hat\Delta\right)}{\Gamma
   \left(\hat\Delta+l\right)} \right)^{s}  \\
& =  \left(1-\hat\Delta \right)\left[ \left(\frac{\Gamma \left(2-\hat\Delta\right)}{\Gamma
   \left(\hat\Delta\right)}\right)^s+ \sum_{l=1}^{+\infty} (1+2 l) l^{2(1-\hat\Delta)s}\left(\frac{l^{2(\hat\Delta-1)} \Gamma \left(l+2-\hat\Delta\right)}{\Gamma
   \left(\hat\Delta+l\right)} \right)^{s}\right] .
\end{split}
\eeq
We expand the function depending on $s$ and $\ell$ at large $\ell$, namely
\beq
\left(\frac{l^{2(\hat\Delta-1)} \Gamma \left(l+2-\hat\Delta\right)}{\Gamma
   \left(\hat\Delta+l\right)} \right)^{s} = 1-\frac{\hat\Delta -1  }{l} s+ \frac{(\hat\Delta -1)  (2 (\hat\Delta -2) \hat\Delta +3 (\hat\Delta -1) s+3)s}{6
   l^2}+\mathcal{O}\left(l^{-3}\right)
\eeq
As one can see the sum can be done, and the $s\rightarrow 0$ limit gives 
\beq\label{eq:anom_pinnp2}
\left. \Delta F\right|^{p=2}_{\text{anom}}  = - \frac{1}{3} (\hat\Delta-1)^3 \, .
\eeq
While for $p=4$ and $p=6$ we get
\begin{eqnarray}
&& \left. \Delta F\right|^{p=4}_{\text{anom}}= -\frac{1}{180} (\hat\Delta -2)^3 (3 (\hat\Delta -4) \hat\Delta +7) \, ,\label{eq:anom_pinnp4} \\
&& \left. \Delta F\right|^{p=6}_{\text{anom}}= -\frac{(\hat\Delta -3)^3 (3 (\hat\Delta -6) \hat\Delta  ((\hat\Delta -6) \hat\Delta
   +11)+82)}{7560} \, \label{eq:anom_pinnp6}.
\end{eqnarray}
We note that these expressions are odd under the change $\hat\Delta \rightarrow p-\hat\Delta$. This is consistent with the fact that the $+$ and $-$ deformation are constructed out of defect operators with dimension such that $\hat \Delta^{+} = p-\hat \Delta^{(-)}$. In particular, as discussed above, when in the defect spectrum the operator $\hat O_{l=0}^{(-)}$ is present the non-trivial defect theory is in the UV.   

In terms of the conformal anomalies we obtain for $p=2$
\begin{equation}
 \Delta b = (\hat \Delta -1)^3 = \begin{cases} 
-\frac{(4-d)^3}{8} \qquad \text{if}\qquad 2<d<4 \, ,\\
\frac{(4-d)^3}{8} \qquad \text{ if}\qquad 4 < d \le 6 \, .\\
\end{cases}
\end{equation}
Note that when $d=3$ we obtain $b=-1/8$ which is twice the value of the boundary case with Dirichlet boundary conditions \cite{Nozaki:2012qd,Jensen:2015swa}. This is in perfect agreement with the observation done in \cite{Raviv-Moshe:2023yvq} that the relevant defect deformation $\phi^2$ in $d=3$ and $p=2$ produces an interface with two decoupled BCFT with Dirichlet boundary condition.\footnote{The fact that the IR fixed point corresponds to the two copies of the half-space with Dirichlet boundary conditions can be seen from the exact propagator \eqref{eq:prop_scal}, which for $d=3$, $p=2$ and $\xi=1$ reduces to
\begin{equation}
    \left<\phi(x_1)\phi(x_2)\right>\,=\,\frac{C_{\phi}}{((x_1-x_2)^2+(\sigma_1-\sigma_2)^2)^{1/2}}\,-\,\frac{C_{\phi}}{((x_1+x_2)^2+(\sigma_1-\sigma_2)^2)^{1/2}}.
\end{equation}} 
In the case of $p=4$, the anomaly of a defect theory on a sphere is given by \cite{Chalabi:2021jud}
\begin{equation}
\left. \left<T^\mu_\mu\right>\right|_{\text{\tiny sphere}} =-    \frac{a}{16 \pi^2} E_{\Sigma_4}
\end{equation}
being $E_{\Sigma_4}$ the Euler density in $4-$dimensions. The anomaly coefficient $a$ is then
\begin{equation}
 \Delta a =  \begin{cases} 
-\frac{(d-6)^3 (3 (d-12) d+88)}{23040} \qquad \text{if}\qquad 4<d<6 \, ,\\
\frac{(d-6)^3 (3 (d-12) d+88)}{23040} \qquad \text{ if}\qquad 6 < d \le 8 \, .\\
\end{cases}   
\end{equation}
Also in this case, when $d=5$ we obtain $a= -17/1440$ twice the central charge of a free scalar with Dirichlet boundary conditions, in perfect analogy with the $d=3$ case discussed above.

The computation, with the due modifications, can be extended to not even $p$, where there is no Weyl anomaly for spherical defects. We do not delve into this case and we refer to the work \cite{Nishioka:2021uef} where the defect-free energy for the free scalar is extensively analysed. 

%
%
%


\section{Some Conformal Perturbation Theory} \label{Section3}

In the following, we are going to consider the free scalar field, coupled to some low-dimensional matter, living on the defect, and Conformal Perturbation Theory provides a convenient tool for analysing this setup. Let us start by reviewing the beta function computation.  While our main focus is on two-dimensional defects, it does not require much effort to consider a more general $p$-dimensional defect here.

We consider an unperturbed CFT$_p$, and deform it by a few slightly relevant operators, which also include coupling to the bulk:

\begin{equation}
S=S_{\text{CFT}_p}+\int\,d^pz\sum_{k}\,g_{0,k}\,\mathcal{O}_k.
\end{equation}

We will now compute the beta functions for the corresponding renormalized couplings $g_k$, using conformal perturbation theory. In order to study the generated RG flow, we will (following \cite{Komargodski:2016auf}) treat the new interactions as a perturbation, which is turned on in a finite volume around the origin, assumed to be a round ball $B_p$ of radius $R$: this creates a state

\begin{equation}
    |0\rangle_{g_{0,k}, B_p}\,=\,e^{-\int_{B_p}d^pz \,\sum_k\,g_{0,k}\mathcal{O}_k}|0\rangle.
\end{equation}

We then would like to compute the overlap of this state with the states $\langle\mathcal{O}_{i}|$, obtained by inserting the operators $\mathcal{O}_{i}$ at infinity,
\begin{equation}
    \mathcal{O}(\infty)_{i}\,=\,\lim\limits_{x\rightarrow\infty}x^{2\Delta_{i}}\mathcal{O}_{i}(x)  \, . 
\end{equation}
This can be done by expanding in powers of $g_{0,k}$, namely

\begin{eqnarray}\label{ConfPertTh1}
\langle\mathcal{O}_{i}|0\rangle_{h,g,B_p}&=&\langle\mathcal{O}_{i}(\infty)\rangle-\sum_k\,g_{0,k}\int_{B_p} d^pz\,\langle \mathcal{O}_{i}(\infty)\mathcal{O}_k(z) \rangle+\nonumber\\
   & +&\frac{1}{2}\sum_{k,l}g_{0,k}g_{0,l}\int_{B_p} d^pz\int_{B_p} d^pw\langle \mathcal{O}_{i}(\infty)\mathcal{O}_k(z)\mathcal{O}_l(w) \rangle+\mathcal{O}(g^3) + \dots \, .
\end{eqnarray}
We have restricted ourselves to the next-to-leading contribution, which is enough for the one-loop beta function computation.
In computing the beta functions, we are interested in the short-distance divergence structure of the expressions at hand, and so we substitute $z\rightarrow0$ and $\int_{B_p} dz\rightarrow V_p R^2$ ($V_p=\pi^{\frac{p}{2}}/\Gamma(\frac{p}{2}+1)$), obtaining
\begin{eqnarray}\label{ConfPertTh2}
    \frac{1}{V_p R^2}\langle\mathcal{O}_{i}|0\rangle_{g_{0,k},D}\sim\,-N_i g_{0,i}+\frac{1}{2}\sum_{k,l}g_{0,k}g_{0,l}\int_D d^p w\,\langle \mathcal{O}_i(\infty)\mathcal{O}_k(0)\mathcal{O}_l(w) \rangle+\mathcal{O}(g^3).
\end{eqnarray}
Above, $\sim$ means that both sides have the same UV divergence structure. Moreover, we have used that the one-point functions vanish, $\langle \mathcal{O}_{i}(\infty) \rangle=0$, and the operators are assumed to be orthogonal, such that the two-point functions are given by $\langle \mathcal{O}_{i}(\infty)\mathcal{O}_{j}(0) \rangle=N_{i}\delta_{i,j}$ (our operators are not necessarily normalized to one). Note that in what follows, we will denote the normalization of possibly composite operators such as $ \phi \mathcal{O}_1$ by $N_i$, while when we mean the normalization of just a single operator such as $\mathcal{O}_1$ we will use $C_{{\mathcal{O}}_1}$. In particular, $N_1= C_\phi C_{{\mathcal{O}}_1}$. Similarly, the structure constants will be denoted $C_{ijk}$ when we mean the one corresponding to the products of operators and $C_{{\mathcal{O}}_i{\mathcal{O}}_j{\mathcal{O}}_k}$ for single ones.

Let now the operator dimensions be such that $\delta_{k} \, \epsilon=p-\Delta_{k}$ are positive and close to zero, $0<\epsilon\ll1$. With $\epsilon$ being the UV regulators, the integrals are now finite and evaluate to
\begin{eqnarray}
  \frac{1}{V_p R^2}\langle\mathcal{O}_{i}|0\rangle_{g_{0,k},D}&\sim -N_i\,g_{0,i}+\frac{S_{p-1}}{2\epsilon}\sum_{k,l} g_{0,k}g_{0,l}C_{ikl}\frac{R^{\epsilon(\delta_k+\delta_l-\delta_i)}}{\delta_k+\delta_l-\delta_i}+\mathcal{O}(g^3),
\end{eqnarray}
where $C_{ikl}$ are the 3-point function coefficients and $S_{p-1}=2\pi^{\frac{p}{2}}/\Gamma(\frac{p}{2})$.
We then introduce the renormalized couplings $g_{k}$ via
\begin{equation}\label{RenormCoupling}
    g_{0,i}=\mu^{\epsilon\delta_i}\left(g_i+\sum_{k,l}\frac{S_{p-1} C_{ikl}\, g_{0,k}g_{0,l}}{2N_i\epsilon(\delta_k+\delta_l-\delta_i)}+\mathcal{O}(g^3)\right),
\end{equation}
which are designed to absorb the poles in $\epsilon$\footnote{We assume that $\delta_i$ are sufficiently generic, in particular, $\delta_i+\delta_j-\delta_k\neq0$ for any $i$, $j$, $k$.}. Bare couplings and UV regulators do not depend on $\mu$, and imposing this condition in the form $\mu\frac{dg_{0,i}}{d\mu}=0$, we obtain the beta functions
\beq\label{BetaFuncGeneral_d}
\beta_{g_i}=-\epsilon \, \delta_i g_i+\frac{\pi ^{p/2}}{\Gamma \left(\frac{p}{2}\right)}\frac{1}{N_i}\sum_{k,l}C_{ikl}\, g_{k}g_{l}.
\eeq
When $p=2$, this reads as
\begin{equation}\label{BetaFuncGeneral}
    \beta_{g_i}=-\epsilon \, \delta_{i}g_i+\frac{\pi}{N_i}\sum_{k,l}C_{ikl}\, g_{k}g_{l}\,.
\end{equation}

We now specialize to the case in which the bulk free scalar field is coupled to a lower dimensional conformal field theory through two slightly relevant perturbations. Our interactions will be either of the form
\begin{equation}
\delta S_{\text{\tiny D}} = \int d^p x \, g_{0,1}\, \hat \phi \, \hat{\mathcal{O}}_1 +\int d^p x \, g_{0,2}\, \hat{\mathcal{O}}_2 \,,
\end{equation}
referred as the Dirichlet coupling, or in a codimension$-1$
\begin{equation}
\delta S_{\text{\tiny N}} =
\int d^p x \, g_{0,1}\,\partial_{\perp}\hat \phi \hat{\mathcal{O}}_1 + \int d^p x \, g_{0,2}\,\hat{\mathcal{O}}_2  \, ,
\end{equation}
referred as the Neumann coupling, where $\hat{\mathcal{O}}_1$, $\hat{\mathcal{O}}_2$ are operators in the putative CFT$_p$. 
In this specific case the $\beta$-functions have zeros given by 
\begin{eqnarray}
&& g_{1 *} = 0 \, \qquad g_{2 *} = 0 \, ,\\
&& g_{1 *} = 0 \, , \qquad g_{2 *} = \frac{ C_{\hat{\mathcal{O}}_2}  \Gamma
   \left(\frac{p}{2}\right)}{  \pi ^{p/2}C_{\hat{\mathcal{O}}_2 \hat{\mathcal{O}}_2 \hat{\mathcal{O}}_2}} \delta_2 \, \epsilon \, , \\
&&g_{1 *} = \pm \frac{\sqrt{\delta_1 C_{\hat{\mathcal{O}}_1}}  \Gamma
   \left(\frac{p}{2}\right) \sqrt{2 C_{\hat{\mathcal{O}}_1 \hat{\mathcal{O}}_1 \hat{\mathcal{O}}_2} \delta_2
   C_2-C_{\hat{\mathcal{O}}_2 \hat{\mathcal{O}}_2 \hat{\mathcal{O}}_2} \delta_1 C_{\hat{\mathcal{O}}_1}}}{2 \pi ^{p/2} C_{\phi/\partial\phi}^{1/2}C_{\hat{\mathcal{O}}_1 \hat{\mathcal{O}}_1 \hat{\mathcal{O}}_2}^{3/2}}\, \epsilon , \quad g_{2 *} = \frac{ C_{\hat{\mathcal{O}}_1}  \Gamma
   \left(\frac{p}{2}\right)\delta_1 \, \epsilon}{ 2 \pi ^{p/2}C_{\hat{\mathcal{O}}_1 \hat{\mathcal{O}}_1 \hat{\mathcal{O}}_2}}  \, .
   \label{eq:fix_point_ntriv}
\end{eqnarray}
Here $C_{\hat{\mathcal{O}}_i}$ is the normalization of $\hat{\mathcal{O}}_i$. Notice that the second solution corresponds to a CFT$_p$ decoupled from the bulk, while the last solutions are genuine non-trivial defect fixed points. These fixed points are real when the condition
\begin{equation}
    \frac{2 C_{\hat{\mathcal{O}}_1 \hat{\mathcal{O}}_1 \hat{\mathcal{O}}_2} \delta_2C_{\hat{\mathcal{O}}_2 }}{C_{\hat{\mathcal{O}}_2 \hat{\mathcal{O}}_2 \hat{\mathcal{O}}_2} \delta_1 C_{\hat{\mathcal{O}}_1 }}>1
\end{equation}
is satisfied.

\subsection{Correlators and Anomalous Dimensions}

In what follows, we continue with the set-up described above, namely, we assume two deformation operators, and either Dirichlet, or Newmann type of coupling to the bulk, and study different types of correlators. The easiest ones are the following: defect-defect propagators like $\left<\hat{\mathcal{O}}_i(\sigma)\hat{\mathcal{O}}_i(0) \right>$ and bulk expectation values like $\left<\phi^2(x_\perp)\right>$.  
Let us start with the one-point function of $\phi^2$. For the Dirichlet coupling and at the first non-trivial order we get
\begin{equation}
\left< \phi^2(x_\perp) \right>_{\text{\tiny D}} = \frac{g_1^2}{2} \int d^p
\sigma_1 d^p\sigma_2 \left< \phi^2(x_\perp) \hat{\mathcal{O}}_1(\sigma_1)\hat\phi(\sigma_1)\hat{\mathcal{O}}_1(\sigma_2)\hat\phi(\sigma_2)  \right>_{0,c} +\hat{\mathcal{O}}(g_1^2 g_2)\, ,
\end{equation}
where the subscript $(0,c)$ means connected correlator of the free-theory. By using Wick theorem we get
\begin{eqnarray}
\left< \phi^2(x_\perp) \right>_{\text{\tiny D}} && = g_1^2 \int d^p\sigma_1 d^p\sigma_2 \left< \phi(x_\perp)\hat\phi(\sigma_1)\right>_{0} \left< \phi(x_\perp)\hat\phi(\sigma_2)\right>_{0}\left<  \hat{\mathcal{O}}_1(\sigma_1)\hat{\mathcal{O}}_1(\sigma_2)  \right>_{0} +\hat{\mathcal{O}}(g_1^2 g_2) \nonumber\\
&& = g_1^2 \int d^p\sigma_1 d^p\sigma_2 \frac{C_{\phi}}{|x_\perp^2+\sigma_1^2|^{\frac{d-2}{2}}} \frac{C_{\phi}}{|x_\perp^2+\sigma_2^2|^{\frac{d-2}{2}}} \frac{C_{\hat{\mathcal{O}}_1}}{|\sigma_1-\sigma_2|^{2\hat\Delta_{1}}} +\hat{\mathcal{O}}(g_1^2 g_2)
\end{eqnarray}
The integral can be performed obtaining

\begin{eqnarray}
\label{eq:one-point_Dirich}
\left< \phi^2(x_\perp) \right>_{\text{\tiny D}} &&=g_1^2\, C_{\hat{\mathcal{O}}_1}\frac{ \Gamma \left(\frac{p}{2}-\hat\Delta_1 \right) \Gamma^2
   \left(\frac{d-p}{2}-1+\hat\Delta_1 \right) \Gamma (d-p+\hat\Delta_1 -2)}{16 \pi ^{d-p}
   \Gamma \left(\frac{p}{2}\right) \Gamma (d-p+2 \hat\Delta_1 -2)}  \frac{1}{|x_\perp|^{2d+2\hat\Delta_1-2p-4}} \nonumber\\
   &&=g_{1*}^2C_{\hat{\mathcal{O}}_1}\frac{\Gamma(\frac{d-p-2}{2})\Gamma(\frac{p}{2})\Gamma(\frac{d-2}{2})}{16\pi^{d-p}\Gamma(p)}\frac{1}{|x_{\perp}|^{d-2}}+\hat{\mathcal{O}}(\epsilon^2)\,.
\end{eqnarray}
On the last line, we took the IR limit and extracted the leading in $\epsilon$ contribution. 
When the defect codimension is one, we can also consider the Neumann-type coupling, with the result 
\begin{eqnarray}
\left< \phi^2(x_\perp) \right>_{\text{\tiny N}} &&=g_1^2 C_{\hat{\mathcal{O}}_1}\frac{ \Gamma \left(\frac{p}{2}-\hat\Delta_1 \right) \Gamma^2
   \left(\frac{d-p}{2}+\hat\Delta_1 \right) \Gamma (d-p+\hat\Delta_1 ) }{4 \pi ^{d-p} \Gamma \left(\frac{p}{2}\right) \Gamma (d-p+2 \hat\Delta_1
   )} \frac{1}{|x_\perp|^{2d+2\hat\Delta_1 -2p-2}}\nonumber\\
   &&=g_{1*}^2C_{\hat{\mathcal{O}}_1}\frac{\Gamma(\frac{d-p}{2})\Gamma(\frac{p}{2})\Gamma(\frac{d}{2})}{4\pi^{d-p}\Gamma(p)}\frac{1}{|x_{\perp}|^{d-2}}+\hat{\mathcal{O}}(\epsilon^2)\,.
\end{eqnarray}

In the following we are going to consider only the Dirichlet case, being the Neumann one completely analogous.
We consider the defect-defect propagators. Let us start with\footnote{Here we assume for simplicity that $C_{\hat{\mathcal{O}}_1\hat{\mathcal{O}}_1\hat{\mathcal{O}}_1}=0$.}
\begin{eqnarray}
&&\left< \hat{\mathcal{O}}_1(\sigma)\hat{\mathcal{O}}_1(0)\right>_{\text{\tiny D}} = \left< \hat{\mathcal{O}}_1(\sigma)\hat{\mathcal{O}}_1 (0)\right>_{0} - g_{0,2} \int d^p \sigma_1 \left< \hat{\mathcal{O}}_1(\sigma)\hat{\mathcal{O}}_1 (0) \hat{\mathcal{O}}_2(\sigma_1)  \right>_{0,c} + \hat{\mathcal{O}}(g_2^2) \nonumber \\
&& = \frac{C_{\hat{\mathcal{O}}_1}}{|\sigma|^{2\hat\Delta_{1}}} -g_{0,2} \pi^{p/2} \, C_{\hat{\mathcal{O}}_1\hat{\mathcal{O}}_1\hat{\mathcal{O}}_2} \frac{\Gamma^2\left( \frac{p}{2}-\frac{\hat\Delta_2}{2} \right) \Gamma\left( \hat\Delta_2-\frac{p}{2} \right)}{\Gamma^2\left( \frac{\hat\Delta_{2}}{2} \right) \Gamma\left( p-\hat\Delta_{2} \right)}\frac{1}{|\sigma|^{2\hat\Delta_{1}+\hat\Delta_2-p}}+ \hat{\mathcal{O}}(g_2^2)\nonumber\\
&& =\frac{C_{\hat{\mathcal{O}}_1}}{|\sigma|^{2\hat\Delta_1^{(0)}}}\left(1+2\delta_1 \epsilon \log|\sigma| \right)-\frac{4\pi^{p/2} g_2C_{\hat{\mathcal{O}}_1\hat{\mathcal{O}}_1\hat{\mathcal{O}}_2}}{\Gamma\left(\frac{p}{2}\right)|\sigma|^{2\Delta_1^{(0)}}}\left(\frac{1}{\delta_2 \epsilon} +\log\mu+\frac{\delta_2+2\delta_1}{\delta_2}\log|\sigma| \right)\nonumber\\
&& +\hat{\mathcal{O}}(\epsilon^2)\, .
\end{eqnarray}
In the last step we have switched to the renormalized coupling {$g_2=g_{0,2}\mu^{-\delta_2 \epsilon}$} and expanded in powers of $\delta_2\epsilon$ , assuming that $\delta_1\sim \delta_2$. We have also rewritten the dimension of the first operator as $\hat\Delta_1=\hat\Delta_1^{(0)}-\delta_1 \epsilon$, and $\hat\Delta_1^{(0)}$ is the critical dimension: it depends on the bulk dimension, as well as on the type of coupling (Dirichlet or Neumann)  (recall also $\hat\Delta_2=p-\delta_2 \epsilon$). There is a pole in $\delta_2\epsilon$, and in order to deal with it we introduce the renormalized operator
\begin{equation}
\label{eq:Z_wafunc}
    \hat{\mathcal{O}}_1=Z_1\hat{\mathcal{O}}_1^{\text{ren}},\qquad Z_1=1-\frac{2\pi^{p/2} g_2C_{\hat{\mathcal{O}}_1\hat{\mathcal{O}}_1\hat{\mathcal{O}}_2}}{\Gamma\left(\frac{p}{2}\right) C_{\hat{\mathcal{O}}_1}\delta_2 \, \epsilon}.
\end{equation}
The renormalized correlator is thus found to be

\begin{eqnarray}
    \left< \hat{\mathcal{O}}_1(\sigma)^{\text{ren}}\hat{\mathcal{O}}_1(0)^{\text{ren}}\right>_{\text{\tiny D}}=\frac{C_{\hat{\mathcal{O}}_1}}{|\sigma|^{{2}\hat\Delta_1^{(0)}}}\left(1+2\delta_1\epsilon\log|\sigma| \right)-\frac{4\pi^{p/2} g_{2}C_{\hat{\mathcal{O}}_1\hat{\mathcal{O}}_1\hat{\mathcal{O}}_2}}{\Gamma\left(\frac{p}{2}\right) |\sigma|^{{2}\hat\Delta_1^{(0)}}}\log(\mu|\sigma|)\nonumber\\
\end{eqnarray}

We can extract the corrected operator dimension from the expression above:
\begin{equation}
    \hat\Delta_{1}^{\text{ren}}=\hat\Delta_1+\hat\gamma_1 \,  g_{2}, \qquad \hat\gamma_1=\frac{2\pi^{p/2} C_{\hat{\mathcal{O}}_1\hat{\mathcal{O}}_1\hat{\mathcal{O}}_2}}{\Gamma\left(\frac{p}{2}\right) C_{\hat{\mathcal{O}}_1}}+\mathcal{O}(g_2g_1^2).
\end{equation}
The Eqs. \eqref{BetaFuncGeneral}, and the condition $\beta_1=0$ in particular, determines the value of the coupling at the interacting fixed point to be

\begin{equation}
    g_{2*}=\frac{\Gamma(p/2) N_1\delta_1}{2\pi^{p/2} C_{\hat{\mathcal{O}}_1\hat{\mathcal{O}}_1\hat{\mathcal{O}}_2}} \epsilon \, .
\end{equation}
Observing that $N_{1}=C_{\phi}C_{\hat{\mathcal{O}}_1}$ and $C_{112}=C_{\hat{\mathcal{O}}_1\hat{\mathcal{O}}_1\hat{\mathcal{O}}_2}C_{\phi}$, we find for the dimension at the fixed point:
\begin{equation}\label{HalfIntegerDim}
\hat \Delta_{1}^{\text{ren}}=\hat \Delta_1+\delta_1 \epsilon=\hat \Delta_1^{(0)}\,.
\end{equation}
This result is expected, since the dimension of $\hat{\mathcal{O}}_1$ is determined by the equation of motion for the bulk scalar, which in the case of Dirichlet-type coupling reads as
\begin{equation}
    \Box\phi=g_{1*}\hat{\mathcal{O}}_1\,\,\delta^{d-p}(y)\,,
\end{equation}
and for the Neumann-type coupling in three dimensions it is
\begin{equation}
    \Box\phi=g_{1*}\hat{\mathcal{O}}_1\,\,\partial_{\perp}\delta^{d-p}(y)\,.
\end{equation}
The result of Eq. \eqref{HalfIntegerDim} is thus supposed to be exact, and will not receive corrections at higher order.

Next, we consider the bulk-defect correlators, with the simplest non-trivial choice being  $< \hat{\mathcal{O}}_1\phi >$. For the Dirichlet-type coupling the leading and next-to-leading orders in perturbation theory give us
\begin{eqnarray}
  \left< \hat{\mathcal{O}}_1(0)\phi(x_\perp) \right>_{\text{\tiny D}}  &&= - g_{0,1} \int d^p\sigma_1 \left<\hat{\mathcal{O}}_1(0)\hat{\mathcal{O}}_1(\sigma_1)\hat  \phi(\sigma_1)\phi(x_\perp)  \right> \nonumber \\
  && + g_{0,1} g_{0,2} \int d^p\sigma_1 d^p\sigma_2 \left<\hat{\mathcal{O}}_1(0)\hat{\mathcal{O}}_1(\sigma_1)\hat \phi(\sigma_1)\hat{\mathcal{O}}_2(\sigma_2)\phi(x_\perp)  \right> \nonumber \\
  && = - g_{0,1} \frac{\pi^{p/2} C_{1} C_{\phi}    \Gamma (p/2-\hat\Delta_{1}) \Gamma \left(\hat\Delta_1 +\hat\Delta_{\phi} - p/2\right)}{\Gamma\left(\Delta_{\phi}\right)\Gamma\left(\frac{p}{2}\right)}
    \frac{1}{x_\perp^{2\hat\Delta_{1}+2\Delta_{\phi}-p}}  \nonumber \\
   &&\hspace{-3.cm}+ g_{0,1} g_{0,2} C_{\hat{\mathcal{O}}_1\hat{\mathcal{O}}_1\hat{\mathcal{O}}_2}C_{\phi} \frac{\pi ^p \Gamma^2 \left(\frac{p-\hat\Delta_2}{2}\right) \Gamma
   \left(\hat\Delta_2-\frac{p}{2}\right) \Gamma \left(p-\hat\Delta_1-\frac{\Delta_2}{2}\right) \Gamma \left(-p+\hat\Delta_1+\frac{\hat\Delta_2}{2}+\Delta_{\phi} \right)}{\Gamma
   \left(\frac{\hat\Delta_2}{2}\right)^2 \Gamma (\Delta _{\phi} ) \Gamma
   \left(\frac{p}{2}\right) \Gamma (p-\hat\Delta_2)} \frac{1}{x_\perp^{2\hat\Delta_{1}+\hat\Delta_{2}+2\Delta_{\phi}-2p}}\nonumber \\
\end{eqnarray}
The second term has poles in $\delta_2 \epsilon$, and, as before, we can switch to renormalized operators and couplings. In particular, now we will also need the first subleading term of $g_{0,1}$. From \eqref{RenormCoupling} we have
\begin{equation}
\label{eq:ren_g2}
g_{0,1} = \mu ^{\delta_1 \epsilon } \left(g_1 + \frac{2 \pi ^{p/2} C_{\hat{\mathcal{O}}_1\hat{\mathcal{O}}_1\hat{\mathcal{O}}_2} 
    }{C_{\hat{\mathcal{O}}_1} \delta_2  
   \Gamma \left(\frac{p}{2}\right)} \frac{g_1 g_2}{\epsilon}\right) \, .
\end{equation}
By employing the same wave-function renormalization as in \eqref{eq:Z_wafunc} and the value of $g_2$ at the fixed point as in \eqref{eq:fix_point_ntriv} we finally obtain

\begin{equation}
\begin{split}
\left< \hat{\mathcal{O}}^{\text{\tiny ren}}_{1}(0)\phi(x_\perp) \right>_{\text{\tiny D}}  = -  \frac{g_{1 *}\pi^{p/2} \Gamma \left(\frac{d-p-2}{2}\right) C_{\hat{\mathcal{O}}_1} C_\phi   }{\Gamma (\frac{d-2}{2})} \frac{1}{x_\perp^{\hat\Delta^{(0)}_1 +\Delta_\phi}}\Big[1-\delta_1 \epsilon  \log (\mu \, x_\perp )\Big]\, .
\end{split}
\end{equation}
In particular, this leads to the same anomalous dimension, as obtained above. The Neumann-type coupling case is obtained by applying $\partial_{\perp}$ to the above result.

Now we consider the two-point function of $\hat{\mathcal{O}}_2$. Following the same computation as above we obtain
\begin{eqnarray}
    \left< \hat{\mathcal{O}}_2(\sigma)^{\text{ren}}\hat{\mathcal{O}}_2(0)^{\text{ren}}\right>_{\text{\tiny D}}=\frac{C_{\hat{\mathcal{O}}_2}}{|\sigma|^{{2}\Delta_2^{(0)}}}\left(1+2\delta_2\log|\sigma| \right)-\frac{4\pi^{p/2} g_{2 *}C_{\hat{\mathcal{O}}_2\hat{\mathcal{O}}_2\hat{\mathcal{O}}_2}}{\Gamma\left(\frac{p}{2}\right) |\sigma|^{{2}\hat\Delta_2^{(0)}}} \log(\mu|\sigma|)\nonumber\\
\end{eqnarray}
Thus we get the anomalous dimension
\begin{equation}
    \Delta_{2}^{\text{ren}}=\hat\Delta_2+\hat\gamma_2 g_{2 *}, \qquad \hat\gamma_2=\frac{2\pi^{p/2} C_{\hat{\mathcal{O}}_2\hat{\mathcal{O}}_2\hat{\mathcal{O}}_2}}{C_{\hat{\mathcal{O}}_2}\Gamma(p/2)}+\mathcal{O}(g_2g_1^2).
\end{equation}
with
\begin{equation}
\hat\gamma_2 g_{2 *} = \frac{ C_{\hat{\mathcal{O}}_1} C_{\hat{\mathcal{O}}_2\hat{\mathcal{O}}_2\hat{\mathcal{O}}_2}  }{C_{\hat{\mathcal{O}}_2}C_{\hat{\mathcal{O}}_1\hat{\mathcal{O}}_1\hat{\mathcal{O}}_2}} \delta_1\,  \epsilon \, .
\end{equation}

Finally, we can compute the two-point function of the defect operator $\hat \phi$. We have at the first non-trivial order
\begin{equation}
\begin{split}
\left< \hat \phi(\sigma)\hat \phi(0) \right>_{\text{\tiny D}} & = \left< \hat \phi(\sigma)\hat \phi(0) \right>_0 +\frac{1}{2}g_{0,1}^2 \int d^p\sigma_1 d^p\sigma_2 \, \left<\hat \phi(\sigma) \hat{\mathcal{O}}_1(\sigma_1)\hat \phi(\sigma_1)\hat{\mathcal{O}}_1(\sigma_2)\hat \phi(\sigma_2) \hat \phi(0)\right> \\ & =\frac{C_\phi}{|\sigma|^{2\Delta_\phi}} +  g_{0,1}^2 C_\phi^2 C_{\hat{\mathcal{O}_1}}\int d^p\sigma_1 d^p\sigma_2 \, \frac{1}{|\sigma-\sigma_1|^{2\Delta_\phi}}\frac{1}{|\sigma_2|^{2\Delta_\phi}}\frac{1}{|\sigma_1-\sigma_2|^{2\Delta_{\hat{\mathcal{O}}}}} \\
&\hspace{-2cm} = \frac{C_\phi}{|\sigma|^{2\Delta_\phi}} +  g_{0,1}^2 C_\phi^2 C_{\hat{\mathcal{O}_1}} \frac{\pi ^p \Gamma \left(\frac{p}{2}-\Delta_{\hat{\mathcal{O}_1}} \right) \Gamma^2
   \left(\frac{p}{2}-\Delta_\phi\right)  \Gamma (\Delta_{\hat{\mathcal{O}_1}} +2 \Delta_\phi -p )}{\Gamma (\Delta_{\hat{\mathcal{O}_1}} )
   \Gamma^2 (\Delta_\phi ) \Gamma \left(\frac{3 p}{2}-\Delta_{\hat{\mathcal{O}_1}} -2 \Delta_\phi
   \right)} \frac{1}{|\sigma|^{2\Delta_{\hat{\mathcal{O}_1}}
   +4 \Delta_\phi -2p}}
\end{split}
\end{equation}
In this case we do not find any divergence for small $\epsilon$ (assuming $q\equiv d-p\neq0\text{mod}2$), and we get
\begin{equation}
\left< \hat \phi(\sigma)\hat \phi(0) \right>_{\text{\tiny D}}  =\frac{C_\phi}{|\sigma|^{2 \Delta_\phi
   }} \left( 1 + g_{1*}^2\,  C_{\mathcal{O}_1} C_\phi \frac{  (d-p)\pi^{p+1}}{2\sin\frac{(d-p)\pi}{2}\Gamma (\frac{d}{2}-1) \Gamma
   (1+p-\frac{d}{2} )} \right)+O\left(\epsilon ^1\right)  \, .
\end{equation}
As a final check, we can consider the two-point function of the two defect operators $\hat\phi$ and $\hat{\mathcal{O}}_1$, i.e. $<\hat\phi (\sigma) \hat{\mathcal{O}}_1(0) >_{\text{\tiny D}}$. Since in general the two operators have different conformal dimensions, this two-point function needs to be vanishing. And indeed, by employing the renormalized coupling \eqref{eq:ren_g2} with the wave-function renormalization \eqref{eq:Z_wafunc} we find that this is the case at the order $\epsilon^2$. 

We conclude this section by mentioning that in  Appendix \ref{app:BB_prop_DC} we computed the leading correction to the propagator in the case of the Dirichlet coupling showing explicitly that the form in eq. \eqref{eq:prop_scal} remains valid in the presence of defect interactions.
%

\subsubsection*{Displacement Operator}

Now we study the displacement operator of the theory. First of all, we need to find its operatorial form. Since the bulk is free, we can simply use the classical conservation of the stress-energy tensor and the equation of motion.

The two are given by
\begin{equation}
\partial^\mu T_{\mu \nu} = \partial_\nu\phi \, \partial^2 \phi \, .
\end{equation}
and
\begin{equation}
-\partial^2 \phi  + g_1 \, \hat{\mathcal{O}}_1  \, \delta^{d-p}(x_\perp)=0 \, .
\end{equation}
for the Dirichlet-type coupling and 

\begin{equation}
-\partial^2 \phi - g_1 \, \hat{\mathcal{O}}_1  \, \partial_{\perp}\delta(x_\perp)=0 \, .
\end{equation}
for the Neumann-type coupling.
Thus, we find
\begin{equation}
 \begin{cases}
 \hat D^i = g_1 \, \hat{\mathcal{O}}_1 \partial^i \hat{\phi} \,,\qquad & \text{Dirichlet c.}\\
\hat D = g_1 \, \hat{\mathcal{O}}_1 \partial^2_{\perp}\, \hat{\phi} \,,\qquad & \text{Neumann c.}.
\end{cases}
\end{equation}
Now we proceed to the two-point function, starting from the Dirichlet case. We get
\begin{equation}
\begin{split}
\left<\hat D^i(\sigma_1) \hat D^j(\sigma_2)\right>_{\text{\tiny D}} = g_1^2 \, \left<\hat{\mathcal{O}}_{\sigma_1} \partial^i \hat\phi_{\sigma_1} \hat{\mathcal{O}}_{\sigma_2} \partial^j \hat\phi_{\sigma_2}\right>= g_1^2 \frac{ 2\Delta_{\phi}\,C_{\phi}\, C_{\hat{\mathcal{O}}}\delta^{ij}}{|\sigma_1-\sigma_2|^{2p+2}} \,+\mathcal{O}(g_i^3),
\end{split}
\end{equation}
where in the final expression we, working at leading order, tuned the operator dimensions to marginality. Analogously, for the Neumann case we get 

\begin{equation}
\begin{split}
\left<\hat D(\sigma_1) \hat D(\sigma_2)\right>_{\text{\tiny N}} = g_1^2 \, \left<\hat{\mathcal{O}}_{\sigma_1} \partial^2_{\perp} \hat\phi_{\sigma_1} \hat{\mathcal{O}}_{\sigma_2} \partial^2_{\perp} \hat \phi_{\sigma_2}\right>=g_1^2 \frac{ 12\Delta_{\phi}(\Delta_{\phi}+1)\,C_{\phi}\, C_{\hat{\mathcal{O}}}}{|\sigma_1-\sigma_2|^{2p+2}} \,+\mathcal{O}(g_i^3)\, .
\end{split}
\end{equation}
From these expressions we can read off the coefficients $C_{\hat D}$ at leading order in the coupling constants.

\subsection{Defect free energy}

The defect sphere free energy in perturbation theory has been discussed in \cite{Kobayashi:2018lil} (see also a previous work about bulk sphere-free energy \cite{Klebanov:2011gs}) where a slightly relevant perturbation deforms the theory. Here, we repeat their argument, keeping the number of deformations generic. Again, we conformally map the bulk theory on the $d$-sphere $S_d$, and put the defect on an equatorial $p$-sphere $S_p$.  We follow the work \cite{Klebanov:2011gs}, which can be easily adapted to the case of defect RG-flows. We start by assuming a defect perturbation of the form
\begin{equation}
    S=S_{\text{DCFT}}+\int\,d^p \,\sigma \sum_{k}\,g_{k}\,\hat{\mathcal{O}}_k \,,
\end{equation}
with $\beta$-functions given in eq. \ref{BetaFuncGeneral_d}. Conformal perturbation theory provides the expansion for $\delta F$ in terms of integrated $n$-point correlators:

\begin{equation}\label{FreeEnergyExpansion}
    \delta F =-\log \left|\frac{Z(g_{i,0})}{Z(0)}\right|=\sum_{n=1}^{\infty}\sum_{i_1,...,i_n}\frac{(-1)^n g_{i_1}...g_{i_n}}{n!}\int d^p\sigma_1\sqrt{G}...\int d^p\sigma_N\sqrt{G}\left< {\hat{\mathcal{O}}}_{i_1}(\sigma_1)...{\hat{\mathcal{O}}}_{i_n}(\sigma_n) \right>.
\end{equation}
To obtain the leading order result, it is enough to retain the two-point and three-point functions contributions (the one-point correlators vanish). The required integrals were evaluated in \cite{Cardy:1988cwa} with the result 
\begin{eqnarray}\label{SphereIntegrals}
    &I_2&=\int d^p x\sqrt{G}\int d^p y\sqrt{G}\left<{\hat{\mathcal{O}}}(x){\hat{\mathcal{O}}}(y) \right>=\frac{(2a)^{2\delta \epsilon}\pi^{p+1/2}\Gamma(-p/2+\delta \epsilon)}{2^{p-1}\Gamma(\frac{p+1}{2})\Gamma(\delta \epsilon)},\\
    &I_3&=\int d^p x\sqrt{G}\int d^p y\sqrt{G}\int d^p z\sqrt{G}\left<{\hat{\mathcal{O}}}_1(x){\hat{\mathcal{O}}}_2(y){\hat{\mathcal{O}}}_3(z) \right>\nonumber\\
    &=&C_{123}\frac{\pi^{\frac{3p}{2}}(2a)^{(\delta_1+\delta_2+\delta_3)\epsilon}}{\Gamma(p)} \, \frac{\Gamma(\frac{-\delta_1+\delta_2+\delta_3}{2}\epsilon)\Gamma(\frac{\delta_1-\delta_2+\delta_3}{2}\epsilon)\Gamma(\frac{\delta_1+\delta_2-\delta_3}{2}\epsilon)\Gamma(\frac{\delta_1+\delta_2+\delta_3}{2}\epsilon-\frac{p}{2})}{\Gamma(\delta_1 \epsilon)\Gamma(\delta_2 \epsilon)\Gamma(\delta_3 \epsilon)}.
\end{eqnarray}
Substituting these results into eq. \ref{FreeEnergyExpansion} and switching to the renormalized couplings \eqref{RenormCoupling}, we obtain 
\beq
\label{eq:free_energ_gen}
\begin{split}
\delta F &=   -\frac{2\pi ^{p+1}}{\sin  \left(\frac{\pi  p}{2}\right)\Gamma(p+1) }  \left(\frac{ 1
   }{2}\sum_i \, \delta_i N_{i} \epsilon (g^i)^2 \,-\frac{ \pi ^{p/2}}{3\,  \Gamma
   \left(\frac{p}{2}\right)}\sum_{i j k} C_{ijk}\, g^i g^j g^k+\mathcal{O}(g_i^4)\right) = \\
    &=  -\frac{\pi ^{p+1}}{\sin  \left(\frac{\pi  p}{2}\right) } \frac{1}{3\, \Gamma\left( p+1 \right)} \sum_i N_{i} \, \delta_i \, \epsilon  \left( g_*^i \right)^2  \,+\mathcal{O}(g_i^4) ,
\end{split}
\eeq
where in the second step we have substituted the fixed point values of the couplings.
As discussed in \cite{Kobayashi:2018lil}, we also mention that when the defect theory is unitary, the quantity defined as $\left. \tilde F\right|_{\text{\tiny def}} =\left. \sin(\pi \, p/2) F \right|_{\text{\tiny def}}$ does not increase under a defect RG flow. Being the quantity $\left. F\right|_{\text{\tiny def}}$ mentioned above the defect contribution to the sphere-free energy.  

It is important to notice that if $p$ is even, then there is a pole due to a zero in the sine. This is related to the Euler-type anomaly which occurs only for even-dimensional defects (or boundaries).
In this case, the anomaly coefficient is just the coefficient of the pole.\footnote{More precisely, we take $p=2m-x$ with $m\in \mathbb{Z}$, and identify the residue in from of $1/x$.} Thus, we have
\beq\label{aAnomaly}
\left. \Delta F\right|_{\text{anom}} =(-1)^{\frac{p}{2}+1} \,    \frac{2\pi ^{p}}{3\, \Gamma\left( p+1 \right)} \sum_i N_{i} \, \delta_i \, \epsilon  \left( g_*^i \right)^2 \,, \qquad p \; \text{even} \,  .
\eeq

We can apply those expressions to the pinned defect for the scalar field studied at the beginning of this work. For $p$ even we find
\beq
\left. \Delta F\right|_{\text{anom}} = (-1)^{\frac{p}{2}+1} \frac{\Gamma\left( \frac{p}{2} \right)}{12 \, \Gamma(p+1)} \epsilon^3 \,, \qquad p \; \text{even} \, ,
\eeq
in agreement with equations \eqref{eq:anom_pinnp2}, \eqref{eq:anom_pinnp4}, and \eqref{eq:anom_pinnp6} to the leading order in $\epsilon$.

In the case $p=2$ and $p=4$, we can link the expression \eqref{aAnomaly} to the variation of the defect central charges $b$ and $a$. For $b$, by performing the spherical integral of the anomaly in eq. \eqref{eq:2d-defect-Weyl-anomaly} and comparing with eq. \eqref{eq:W_log_A} we find that $\left.  F\right|_{\text{anom}} =  -b/3$. Thus, we get
\begin{equation}
\Delta b =  -\pi^2 \sum_i N_{i} \, \delta_i \, \epsilon  \left( g_*^i \right)^2 \,, \qquad p=2 \, .
\end{equation}
In the $p=4$ case, the relevant anomaly corresponds to the Euler density in four dimensions \cite{Chalabi:2021jud,FarajiAstaneh:2021foi}, where $\left.  F\right|_{\text{anom}} = 4 \,a$, resulting in   
\begin{equation}
\Delta a=  -\frac{\pi^4}{9} \sum_i N_{i} \, \delta_i \, \epsilon  \left( g_*^i \right)^2 \,, \qquad p=4 \, .
\end{equation}

\section{Coupling to the Minimal Models}
\label{Section4}
Having developed general machinery, we now specialize to $2-$dimensional defects. Our first choice of 2d matter will be the series of unitary diagonal minimal models $\mathcal{M}(p, p+1)$. This family of models was originally used in \cite{Zamolodchikov:1987ti} to construct perturbative RG flows in 2d, and also recently in \cite{Behan:2021tcn} for constructing conformal boundary conditions for a three-dimensional free scalar field CFT. Our setup will be close to this last instance, with a difference that we are interested in defects, rather than boundary conditions.

Coupling the bulk scalar to a 2d minimal model, we can provide a perturbative construction of surface defects for certain bulk dimensions.

Given the model $\mathcal{M}(p, p+1)$, the spectrum of primary operators is found to be
\begin{equation}
    h = \bar{h} = \frac{((p+1) m - p n)^2 - 1}{4 p (p + 1)},
\end{equation}
where $m=0,...,p-1$ and $n=0,...,p$, and the integers $(n,m)$ label the conformal families. The idea is to consider the limit $p\rightarrow\infty$ with $m,n$ kept fixed \cite{Zamolodchikov:1987ti}. Then operator dimensions behave like $\Delta=h+\bar{h}=\frac{(m-n)^2}{2}+\mathcal{O}(1/p)$. The case of interest for us is $n=m+1$, which can be used to deform Dirichlet boundary conditions in three dimensions and Neumann boundary conditions in five dimensions. Let us describe these two cases in order.

\subsection{Deforming Dirichlet b.c. in 3d}
We couple the 3d bulk and  the 2d surface defect via the interaction of the form
\begin{equation}
  \int d^2\sigma\,\hat\Phi_{(m,m+1)}\partial_\perp \hat\phi.
\end{equation}
This operator is slightly relevant, with the dimension $\Delta=2-\epsilon$ with $\epsilon=\frac{2m+1}{2(p+1)}-\frac{m^2}{2p(p+1)}\ll1$, and so generates an RG flow, which may possibly be treated perturbatively.

The $\hat\Phi_{(m,m+1)}\times \hat\Phi_{(m,m+1)}$ OPE contains, among others, a family of operators of the form $\hat\Phi_{(2k+1, 2k+3)}$ with $0<k<m$. These operators also become marginal in the limit $p\rightarrow\infty$, and need to be included in the Lagrangian for the sake of renormalizability, together with nearly marginal operators appearing this time in their OPE. This proliferation of couplings makes the analysis rather complicated unless we choose the special case of $m=1$. In this case the only extra operator we need to add is $\hat\Phi_{(1,3)}$\footnote{The simplification is provided by the fact that the operators $\hat\Phi_{(1,m)}$ form a closed subalgebra under the fusion}, with the dimension $\Delta_{(1,3)}=2-\frac{4}{p+1}$. Our deformation thus takes the form\footnote{One may be worried by the operator $\phi^4$ is marginal in the UV, and should also be considered in the RG analysis. Note though that it is not allowed by the bulk shift symmetry, and thus can be ignored.}
\begin{equation}
 \delta S =   \int d^2x\,g_1 \hat\Phi_{(1,2)}\partial_y\hat\phi\,+\,g_2\hat\Phi_{(1,3)}.
\end{equation}
We can now apply eqs. \eqref{BetaFuncGeneral}, and the one-loop beta functions are easily found to be \cite{Behan:2021tcn}
\begin{eqnarray}
   && \beta_{g_1}=-\frac{3}{2p}g_1+2\pi g_1g_2C_{(1,2)(1,2)}^{(1,3)}+\mathcal{O}(1/p^2, g_i^3),\\
  &&  \beta_{g_2}=-\frac{4}{p}g_2+\pi g_1^2 C_{(1,2)(1,2)}^{(1,3)}C_{\partial_y\phi}+\pi g_2^2 C_{(1,3)(1,3)}^{(1,3)}+\mathcal{O}(1/p^2, g_i^3),\\
\end{eqnarray}
where the OPE coefficients are \cite{Dotsenko:1985hi}
\begin{equation*}
    C_{(1,2)(1,2)}^{(1,3)}=-\frac{\sqrt{3}}{2}+\mathcal{O}(1/p^2),\qquad  C_{(1,3)(1,3)}^{(1,3)}=-\frac{4}{\sqrt{3}}+\mathcal{O}(1/p),
\end{equation*}
and $C_{\partial_y\hat \phi}=2\Delta_{\phi}C_{\phi}$ is the normalization of the $\left<\partial_y\hat\phi\partial_y\hat\phi\right>$ 2-point function; the operators $\hat \Phi_{(1,2)}$, $\hat \Phi_{(1,3)}$ are assumed to be normalized to one. 
Apart from the trivial UV fixed point with $g_1=g_2=0$, we find the following fixed points:
\begin{itemize}\label{3dMinBeta}
    \item $g_1=\pm\frac{1}{\pi p}\sqrt{\frac{2}{C_{\partial_y\hat\phi}}}+\mathcal{O}(1/p),\quad g_2=-\frac{\sqrt{3}}{2\pi p}+\mathcal{O}(1/p^2)$: a couple of interacting fixed points, with the defect being coupled to the bulk.
    \item $g_1=0+\mathcal{O}(1/p^2),\quad g_2=-\frac{\sqrt{3}}{\pi p}+\mathcal{O}(1/p^2)$: a fixed point with defect being decoupled from the bulk, and the defect dynamics is described by a pure 2d CFT. This 2d CFT is the Minimal Model $\mathcal{M}(p-1,p)$ \cite{Zamolodchikov:1987ti}.
\end{itemize}

Having the beta functions, we can easily extract the dimensions of the corresponding operators, which are found to be
\begin{equation}
    \Delta_{\pm}=2\pm\frac{\sqrt{6}}{p}.
\end{equation}
We can now apply the Eq. \eqref{aAnomaly} to compute the change of the $a$-coefficient along the RG flow. The result for the interaction fixed point is $\Delta b_{\text{\tiny Int}}=-\frac{6}{p^3}+\mathcal{O}(1/p^4)$, and for the decoupled fixed point it is $\Delta b_{\text{\tiny Decoup}}=-\frac{12}{p^3}+\mathcal{O}(1/p^4)$. Taking into account that for a decoupled defect $b=c$, where $c$ is the 2d central charge, this last result is consistent with the difference of central charges of two successive Minimal Models: $c_{(p-1,p)}-c_{(p,p+1)}=-\frac{12}{p^3}+\mathcal{O}(1/p^4)$. We note that the $b$-coefficient at the interacting fixed point is greater than at the decoupled IR fixed point, consistently with the expectation that the interacting fixed point is unstable.

    \begin{table}[]
        \centering
        \begin{tabular}{|c|c|c|c|c|c|}
        \hline
            \text{dim} & $\Delta b_{\text{\tiny Int}}$ & $a_{\phi^2}$ & $a_T$ & $b_{\mathcal{O}_1\phi}$ & $C_{\hat D }$ \\
        \hline
        \hline
            $d=3$ & $-\frac{6}{p^3}$ & $\frac{1}{\pi p^2}$ & $-$ & $\pm\sqrt{\frac{2}{\pi}}\frac{1}{p}$ & $\frac{18}{\pi^2 p^2}$ \\
        \hline
        $d=5$ & $-\frac{6}{p^3}$ & $\frac{1}{2\pi^2 p^2 }$ & $\frac{1}{32\pi^2 p^2}$ & $\pm\frac{1}{\pi p}$ & $\frac{6}{\pi^2 p^2}$ \\
        \hline
        \end{tabular}
        \caption{Some DCFT data for a surface defect in three dimensions, obtained by coupling to a $\mathcal{M}(p, p+1)$ Minimal Model at the leading order in $1/p$ expansion.}
        \label{}
    \end{table}

\subsection{Deforming Neumann b.c. in 5d}
A completely analogous discussion applies to 2d defects of the 5d free scalar, with the only difference being that the Minimal Model is coupled to the operator $\phi$, rather than $\partial_y\phi$. So, the relevant deformation under consideration is
\begin{equation}
    \int d^2x\,g_1  \hat\Phi_{(1,2)}\hat\phi\,+\,g_2\hat\Phi_{(1,3)}.
\end{equation}
The beta functions are the same as in \eqref{3dMinBeta} with the substitution $C_{\partial_y\hat\phi}\rightarrow C_{\phi}$, and the fixed points are:
\begin{itemize}\label{3dMinBeta}
    \item $g_1=\pm\frac{1}{\pi p}\sqrt{\frac{2}{C_{\phi}}}+\mathcal{O}(1/p),\quad g_2=-\frac{\sqrt{3}}{2\pi p}+\mathcal{O}(1/p^2)$: a couple of interacting fixed points, with the defect being coupled to the bulk.
    \item $g_1=0+\mathcal{O}(1/p),\quad g_2=-\frac{\sqrt{3}}{\pi p}+\mathcal{O}(1/p)$: a fixed point with defect being decoupled from the bulk, and the defect dynamics is again described by $\mathcal{M}(p-1,p)$.
\end{itemize}
The change in the $b$-coefficient is exactly the same, as in the previous case: $b_{\text{\tiny Int}}=-\frac{2}{p^3}+\mathcal{O}(1/p^4)$.

\section{Coupling to the 2d Scalar}\label{Section5}

Another example of the coupling between a 2d defect and a higher-dimensional scalar field is provided by the coupling to a two-dimensional scalar field.\footnote{We assume the scalar to be compact, possibly with the radius going to infinity.} We will consider different bulk dimensions and sometimes different ways to couple the defect to the bulk, which can be summarised as follows:
\begin{eqnarray}
    S_{d=3,d}&=&\int\,d^3x\,\frac{1}{2}(\partial\phi)^2+\int\,d^2z\,\left[\partial\sigma\bar{\partial}\sigma\,+\,h\hat\phi\,\cos(\alpha\sigma)\,+\,g_1\hat\phi^2+g_2\hat\phi^4\right],\label{3dDirichlet}\\
    S_{d=3,n}&=&\int\,d^3x\,\frac{1}{2}(\partial\phi)^2+\int\,d^2z\,\left[\partial\sigma\bar{\partial}\sigma\,+\,h\partial_{\perp}\hat\phi\,\cos(\alpha\sigma)\,+\,g\hat\phi^2+g_2\hat\phi^4\right],\label{3dNeumann}\\
    S_{d=4}&=&\int\,d^4x\,\frac{1}{2}(\partial\phi)^2+\int\,d^2z\,\left[\partial\sigma\bar{\partial}\sigma\,+\,h\hat\phi\,\cos(\alpha\sigma)\,+\,g\hat\phi^2\right],\\
    S_{d=5}&=&\int\,d^5x\,\frac{1}{2}(\partial\phi)^2+\int\,d^2z\,\left[\partial\sigma\bar{\partial}\sigma\,+\,h\hat\phi\,\cos(\alpha\sigma)\right].
\end{eqnarray}
We remind the reader the conformal dimension of the vertex operator $\cos \alpha$ is $\Delta_\alpha = \alpha^2/(4\pi)$. Thus, depending on the value of $\alpha$, some other operators may become relevant, and so should be added to the Lagrangian. These models resemble the extremely well-studied sine-Gordon model, and the occasional comparisons between the two will be useful.

\subsection{Semiclassical analysis}
In the regime of $\alpha\ll1$ we can analyse the system semiclassically. Let us start with the four-dimensional case. Indeed, by rescaling the fields as $\sigma\rightarrow\sigma/\alpha$, $\phi\rightarrow\phi/\alpha$, as well as redefining the couplings as $h\rightarrow h/\alpha^2$, $g\rightarrow g$, we find that $\alpha^2$ factor out of the action, and serves as an effective Planck constant. This limit is somewhat subtle due to the proliferation of relevant operators: $\hat\phi\cos k\alpha\sigma$ for $k$ odd and $\cos l\alpha\sigma$ for $l$ even are compatible with the symmetries of the problem, and become relevant for more and more values of $k,l$, while $\alpha$ is going to zero. For this reason, we will consider a general case of interaction,
\begin{equation}
    \hat\phi\sum_{l=1,\,l\,\text{odd}}^{l_{\text{max}}}h_l\cos l\sigma\,+\,\sum_{k=1,\,k\,\text{even}}^{k_{\text{max}}}g_k\cos k\sigma\equiv\hat\phi\, V_1(\sigma)+V_2(\sigma),
\end{equation}
for some $k_{\text{}mac}$ and $l_{\text{max}}$. Classical equations of motion take the form
\begin{eqnarray}
    \Box_4\phi&-&(V_1(\sigma)+2g\phi)\delta^{(2)}(x_{\perp})=0,\\
    \Box_2\sigma&-&\, \phi V_1'(\sigma)- V_2'(\sigma)=0.
\end{eqnarray}
The precise vacuum structure depends on the details of the potentials $V_1$, $V_2$. A solution describing a conformal defect at long distance would correspond to $\phi_{\text{vac}}=0$, and so the vacuum value of $\sigma$ should satisfy $V_1(\sigma_{\text{vac}})=V_2'(\sigma_{\text{vac}})=0$. 
Moreover, for the solution to be (marginally) stable we need to require that the Jacobian is non-negative:
\begin{eqnarray}
2g V_2''(\sigma_{\text{vac}})-(V_1'(\sigma_{\text{vac}}))^2\geq0
\end{eqnarray}
Finally, if we insist on an interacting theory, we do not want $\sigma$ to be gapped, and so $V_2''(\sigma_{\text{vac}})=0$. All these conditions cannot be satisfied simultaneously, unless $V_1$ and $V_2$ have a rather special form, or in other words unless the coefficients $g_l$ and $g_k$ are chosen in a special way; if done so, this choice in any case would not be RG-invariant. We thus generically expect to have gapped $\sigma$, similarly to the case of sine-Gordon model.

If we try to apply the quasiclassical considerations beyond the range of small $\alpha$, we find that the picture holds even in the range $\sqrt{\pi/2}<\alpha<\sqrt{\pi}$, where $V_2=h_2\cos2\alpha\sigma$: in the vacua $\phi=0$, $\sigma=\pi k/(2\alpha)$ ($k\in\mathbb{Z}$) $\sigma$ gets mass and decouples in the IR. This can happen even in the absence of tree-level potential $V_2$, due to an effective Colemann-Weinberg potential generated for $\sigma$, but we are not presenting the relevant computation here. A similar picture of the gapped $\sigma$ field holds for other dimensions.

\subsection{Perturbative RG flow}
Next, we consider the cases when the interaction is close to criticality, and the defect RG flow can be studied perturbatively. The discussion splits depending on the dimension, so we consider different cases in order.

The first case we consider is $\hat \phi\cos\alpha\sigma$ interaction in $4-\epsilon$. When $d=4$ and $\alpha=\sqrt{4\pi}$, the interaction term is classically marginal. We can thus consider $\alpha^2=4\pi(1-\zeta)$ with $\zeta\ll1$, such that the operator is slightly relevant, and use Eq. \eqref{BetaFuncGeneral} to compute perturbatively the beta functions. Apart from the couplings $g$ and $h$, we also need to take the marginal operator $\partial\sigma\bar{\partial}\sigma$ into account: let us denote the corresponding couplings $t$. We obtain the following system of beta functions:\footnote{We use the  3-pnt function coefficient $ \left<\mathcal{V}_{\alpha}(\infty)\mathcal{V}_{-\alpha}(1)\partial\sigma\bar{\partial}\sigma(0) \right>=-\frac{\alpha^2}{(4\pi)^2}.$.}

\begin{eqnarray}
    \beta_h &=& -\left(\zeta+\frac{\epsilon}{2}\right) h + \frac{2\pi C_{\phi\phi\phi^2}gh}{C_{\phi}} - \frac{\alpha^2 ht}{4\pi},\\
    \beta_g &=&-\epsilon g + \frac{g^2}{\pi}+\frac{\pi C_{\phi\phi\phi^2}h^2}{2C_{\phi}^2},\\
    \beta_t &=&= -\alpha^2 \pi C_{\phi} h^2.
\end{eqnarray}

Evidently, due to the presence of a marginal operator the system does not have any perturbative interacting fixed point. The situation is similar to the sine-Gordon model at $\alpha=8\pi$, where there is no non-trivial perturbative fixed point found, again due to the presence of a marginal operator.

More precisely, we find two lines of fixed points. In both lines $h_*=0$ and $t_*$ is not fixed, and $g_*$ 
is either $0$, or $\pi \epsilon$: this is the conformal manifold of the free scalar, decoupled from the bulk, tensored with the decoupled defect (either trivial or not). Linearizing around the first family of fixed points, we find that it is always unstable (assuming $\epsilon>0$), essentially due to the relevant coupling $g$. The neighbourhood of the second line is organized in a more complicated manner: there is a critical value $t_*^{cr}=\frac{4\pi}{\alpha^2}\left( 2\pi^2\epsilon C_{\phi\phi\phi^2}/C_{\phi}-\zeta-\epsilon/2\right)$\footnote{This critical value is determined by linearization of beta functions around the line of fixed points $g=g_*=\pi\epsilon$, $h=h_*=0$, $t=t_*$. At $t_*=t_*^{cr}$ the coefficient in $\beta_h$ in front of the term linear in $h$ changes the sign, thus signifying the stability-instability transition.}, such that for $t_*<t_*^{cr}$ the line is stable under sufficiently small perturbations, while for $t_*>t_*^{cr}$ it becomes unstable. This is the analogue of Kosterlitz-Thouless transition, and we would be tempted to conclude that the IR physics is described either by a defect with a decoupled free scalar, or we fall into the massive phase. 

One can also consider the model with $\phi\, cos(\alpha\sigma)$ interaction around $d=5$, where $\alpha^2=2\pi(1-\zeta)$. There the operator $\phi^2$ becomes irrelevant, and does not play a role in the RG flow. On the other hand, one has to take into account the marginal operator $cos(2\alpha\sigma)$: let us call the corresponding coupling $h'$. The beta functions are readily computed:
\begin{eqnarray}
    \beta_h &=& -\left(\zeta+\frac{\epsilon}{2}\right) h + \pi h h' - \frac{\alpha^2 ht}{4\pi},\\
    \beta_{h'} &=&-2\zeta h'+ \pi C_{\phi} h^2-\frac{\alpha^2 h't}{\pi},\\
    \beta_t &=& -\alpha^2 \pi C_{\phi} h^2-4\alpha^2 \pi  h'^2.
\end{eqnarray}
Again, due to the marginal operator one does not find any perturbative fixed points. Also, as in the previous case, there is a line of fixed points with $h_*=h'_*=0$ and $t_*$ unconstrained. This line is stable for $t_*<-4\pi(\zeta+\epsilon/2)/\alpha^2$, and we again find a Kosterlitz-Thouless-like transition.

A very similar picture is found for the $\partial_{\perp}\phi\,\cos(\alpha\sigma)$ coupling around $d=3$ with $\alpha^2=2\pi(1-\zeta)$. Here $\phi^2$ is a relevant operator, and we tune the corresponding coupling to zero. There is an extra marginal operator, $\phi^4$, but it is not generated at one loop. The beta functions are as above, with the substitution $C_{\phi}\rightarrow C_{\partial\phi}$.

Finally, we may have the interaction $\phi \,\cos(\alpha\sigma)$ with $\alpha^2=6\pi(1-\zeta)$ in $d=3$ dimensions. Similarly to the previous case, $\phi^2$ is a relevant operator, that can be tuned to zero, and $\phi^4$ is marginal and generated at one loop. On the other hand, $\cos 2\alpha\sigma$ is irrelevant, and so is not generated; we are thus left just with two couplings $h$ and $t$, and the RG flow is again of Thouless-Kosterlitz type.

\subsection{Fermionization and $\epsilon-$expansion}

For the special value $\alpha^2=4\pi$, the free scalar in two dimensions is equivalent to a single Dirac fermion, with the following operator matching (up to c-number factors):
\begin{equation}
    \cos(\sqrt{4\pi}\sigma)\,\longleftrightarrow\,\bar{\psi}\psi.
\end{equation}
In this language our model reads
\begin{equation}
    S=\int\,d^dx\,\frac{1}{2}\left(\partial\phi\right)^2+\int\,d^2z\,\left[i\bar{\psi}\slashed{\partial}\psi\,+\,h\hat{\phi}\,\bar{\psi}\psi\,+\,g\hat\phi^2\right]
\end{equation}

As we will see momentarily, this model does not have an interacting fixed point for $d=4$, in agreement with the general results about 2d conformal defects in 4d free scalar theory. Our interest is thus in considering the cases of $d=3$ and $d=5$. In both cases we would like to exploit $\epsilon-$expansion, albeit around different dimensions of the bulk and the defect; the story thus splits into two cases.

\subsubsection{$\epsilon$-expansion around $d=4$, $p=3$}
We study the fixed points of the model
\begin{equation}
    S=\int\,d^{d-1}xdy\,\frac{1}{2}(\partial\phi)^2+\int\,d^{d-1}x\,\left[i\bar{\psi}\slashed{\partial}\psi\,+\,h\hat{\phi}\,\bar{\psi}\psi\right]
\end{equation}
with $d=4-\epsilon$. We restrict ourselves to the one-loop analysis, so the operator $\phi^3$, even though nearly marginal, is not generated at the defect, while $\phi^2$ is now relevant, and we tune it to zero. We now compute the renormalization factors, needed for the beta function computation. We will need the fermion propagator
\begin{equation}
    \Delta_{\psi}^{3d}(p)=\frac{i\slashed{p}}{p^2},
\end{equation}
as well as the $4d$ scalar propagator, restricted to the $3d$ defect, and transformed into the momentum space:
\begin{equation}
    \Delta_{\phi}^{3d}(p)=\frac{1}{2|p|}.
\end{equation}
We can now calculate the wave function and coupling renormalizations, and extract the beta function.

The relevant computation was first performed in \cite{Herzog:2017xha}, with the only difference being that the computation is done for the boundary, while we are interested in the defect (interface). This modifies the scalar propagator in momentum space, restricted to the boundary/defect as
\begin{equation}
    \frac{-i}{|p|}\,\rightarrow\frac{-i}{2|p|}.
\end{equation}
For completeness, in the following we repeat their computation taking into account this modification.

The one-loop correction to the scalar propagator is
\begin{eqnarray}
    i\Pi_{\phi}(q)&=&(-1)(-ih)^2\int\frac{d^pp}{(2\pi)^p}\,\frac{\text{tr}[i\slashed{p}\,i(\slashed{p}+\slashed{q})]}{p^2(p+q)^2}\nonumber\\
    &=&ig^2\frac{2^{3-2p}\pi^{\frac{3-p}{2}}}{\sin\left(\frac{\pi p}{2}\right)\Gamma\left(\frac{p-1}{2}\right)}q^{p-2}.
\end{eqnarray}
This result is finite for $p=3$ and reduces to
\begin{equation}
    \Pi_{\phi}(q)=-\frac{h^2}{8}|q|.
\end{equation}
The correction to the fermion propagator takes the form
\begin{eqnarray}
    i\Pi_{\psi}(q)&=&(-ih)^2\int\frac{d^pp}{(2\pi)^p}\frac{i\slashed{p}(-i)}{2p^2|p-q|}\nonumber\\
    &=&-ih^2\frac{\slashed{q}}{q^{3-p}}\frac{2^{1-2p}\pi^{-p/2}\Gamma(\frac{3-p}{2})\Gamma(p-1)}{\Gamma(p-\frac{1}{2})}.
\end{eqnarray}
When the defect dimension is close to three, $p=3-\epsilon$, we find the following leading divergence:
\begin{equation}
    \Pi_{\psi}(q)=-\slashed{q}h^2\left[\frac{1}{12\pi^2\epsilon}+\text{finite} \right].
\end{equation}
Finally, we need the vertex one-loop correction:
\begin{eqnarray}
    -ih\Gamma(q_1,q_2)=(-ih)^3\int\frac{d^pp}{(2\pi)^d}\frac{i(\slashed{p}+\slashed{q_1})\,i(\slashed{p}+\slashed{q_2})(-i)}{2(p+q_1)^2(p+q_2)^2|p|}.
\end{eqnarray}
We will not compute the full result, but restrict ourselves with finding the leading divergence:
\begin{equation}
    h\Gamma(q_1,q_2)=-h^3\left[\frac{1}{4\pi^2\epsilon}+\text{finite}\right].
\end{equation}
We can now write down the renormalization factors:
\begin{eqnarray}
    Z_{\psi}&=&1+h^2\left(-\frac{1}{12\pi^2\epsilon}+\text{finite}\right),\\
    Z_{\phi}&=&1+h^2\left(\text{finite}\right),\\
    Z_g&=&1+h^2\left(\frac{1}{4\pi^2\epsilon}+\text{finite}\right).
\end{eqnarray}
The bare coupling is expressed as $h_0=h\mu^{\epsilon/2}Z_{\phi}^{-1/2}Z_{\psi}^{-1}Z_g$, and requiring its $\mu$-independence, we get the beta function
\begin{equation}
    \beta=-\frac{\epsilon}{2}h+\frac{1}{3\pi^2}h^3.
\end{equation}
For positive $\epsilon$ this leads to the non-trivial IR fixed point with
\begin{equation}
    h_*^2=\frac{3\pi^2}{2}\epsilon.
\end{equation}
At this point we can compute the correlator $\phi^2(x)$. It reads
\begin{equation}
\begin{split}
\left< \phi^2(x) \right> & =  \frac{h^2}{2} \int d^{d-1}\sigma_1 d^{d-1}\sigma_2 \, \left<\phi(\sigma_1) \bar \psi(\sigma_1)\psi(\sigma_1)\phi(\sigma_2) \bar \psi(\sigma_2)\psi(\sigma_2) \phi^2(x) \right>_0 +\mathcal{O}(h^4)  \\
& = h^2 C_\phi^2 C_\psi^2\, 2^{\lfloor\frac{d-1}{2}\rfloor}  \int d^{d-1}\sigma_1 d^{d-1}\sigma_2 \, \frac{1}{|x_\perp^2 + \sigma_1^2|^{d/2-1}} \frac{1}{|x_\perp^2 + \sigma_2^2|^{d/2-1}} \frac{1}{|\sigma_1-\sigma_2|^{2d-4}}+\mathcal{O}(h^4) \\
& = - h^2 C_\phi^2 C_\psi^2 2^{\lfloor\frac{d-1}{2}\rfloor} \frac{\pi ^{d-\frac{1}{2}}  
   \Gamma \left(d-\frac{5}{2}\right)\sec \left(\frac{\pi  d}{2}\right)}{(d-3)^3 \Gamma^2 (d-3)}\frac{1}{x_\perp^{2d-6}} \, ,
\end{split}
\end{equation}
where we used the following propagator of a Dirac fermion living in a $p-$dimensional space
\begin{equation}
\left<\psi(x_1)\bar\psi(x_2) \right> =C_\psi \frac{i \, \gamma \cdot (x_1-x_2)}{|x_1-x_2|^{p}} \, , \qquad C_\psi = \frac{\Gamma\left(\frac{p}{2}\right)}{2 \pi^{\frac{p}{2}}} \, .
\end{equation}
At the leading order in $\epsilon$ we find
\begin{equation}
\begin{split}
\left< \phi^2(x) \right> & = -\frac{  h^2}{256\, \pi^2 x_\perp^2} =- \frac{3  }{512 }  \frac{1}{x_\perp^2} \epsilon \, .
\end{split}
\end{equation}

While for the displacement operator, by following the discussion given in Sec. \ref{Section3} we obtain
\begin{equation}
\begin{split}
\left<\hat D(\sigma_1) \hat D(\sigma_2)\right> & = h^2 \, \left<(\bar\psi \psi)_{\sigma_1} \partial_\perp \phi_{\sigma_1} (\bar\psi \psi)_{\sigma_2} \partial_\perp \phi_{\sigma_2}\right>=  \frac{ 2\Delta_{\phi}\,\,h^2\,C_{\phi}\, C^2_{\psi}2^{\lfloor\frac{d-1}{2}\rfloor}}{|\sigma|^{2d}} \,+\mathcal{O}(h^3) \\
& = \frac{3 \, \epsilon }{32 \pi ^2 \sigma ^8} \, .
\end{split}
\end{equation}

Having found the fixed point, we can evaluate the generalized $F$-function $\tilde{F}=-\sin\frac{\pi p}{2}\log Z$ \cite{Giombi:2014xxa, Fei:2015oha}, or more precisely its change along the flow, with the result

\begin{equation}
  \Delta\tilde{F}=-\frac{\pi^2\epsilon^2}{384} \, .
\end{equation}

\begin{table}[]
        \centering
        \begin{tabular}{|c|c|c|c|c|c|}
        \hline
            \text{dim} & $\Delta\tilde{F}$ & $a_{\phi^2}$ & $a_T$ & $b_{\mathcal{O}_1\phi}$ & $C_{\hat D}$ \\
        \hline
        \hline
            $d=4-\epsilon$, $p=3-\epsilon$ & $-\frac{\pi^2\epsilon^2}{384}$
            & $-\frac{3\epsilon}{512}$ & $0$ & $-\frac{1}{32\pi}\sqrt{\frac{3\epsilon}{2}}$ & $\frac{3\epsilon}{32\pi^2}$ \\
        \hline
        $d=4+\epsilon$, $p=1+\epsilon$ & $-\frac{\pi^2\epsilon^2}{24}$ & $\frac{\epsilon}{32}$ & $-\frac{\epsilon^2}{192}$ & $-\frac{\sqrt{2\epsilon}}{16}$ & $\frac{\epsilon}{4}$  \\
        \hline
        \end{tabular}
        \caption{Some DCFT data for a free scalar in $4-\epsilon$ dimensions, coupled to a $3--\epsilon$-dimensional fermion, and for a free scalar in $4+\epsilon$ dimensions, coupled to a fermion in $1+\epsilon$ dimensions, at the leading order in $\epsilon$ expansion.}
        \label{}
    \end{table}


\subsubsection{$\epsilon$-expansion around $d=4$, $p=1$}

We now develop $\epsilon$-expansion around four-dimensional bulk and one-dimensional defect. In order to match the fermionic degrees of freedom, we consider two complex fermions, living on the one-dimensional defect, and symmetrically coupled to the bulk scalar. Gamma matrices in $1d$ are one-dimensional, and the fermion propagator in momentum space is
\begin{equation}
    \Delta_{\psi}(p)=\frac{ip}{p^2}=\frac{i}{p}.
\end{equation}
Transforming the $4d$ scalar propagator to $1d$ momentum space, we obtain
\begin{equation}
    \Delta_{\phi}^{1d}=-\frac{|p|}{4\pi}.
\end{equation}
The correction to the scalar propagator is 
\begin{eqnarray}
    i\Pi_{\phi}(q)=2(-1)(-ih)^2\int\frac{d^pp}{(2\pi)^p}\frac{ip\,i(p+q)}{p^2(p+q)^2}\nonumber\\
    =-2ih^2\,q^{p-2}\frac{\Gamma^2(p/2)\Gamma(1-p/2)}{\Gamma(p-1)},
\end{eqnarray}
and vanishes for $p=1$. The one-loop contribution to the fermion propagator is
\begin{eqnarray}
    i\Pi_{\psi}(q)=(-ih)^2\int\frac{d^p p}{(2\pi)^p}\frac{ip}{p^2}\left(-(-i)\frac{|p-q|}{4\pi}\right)\nonumber\\
    =-ih^2\frac{q}{q^{1-p}}\frac{\Gamma(\frac{1-p}{2})\Gamma(p)}{2^{2p+2}\pi^{p/2+1}\Gamma(p+1/2)}
\end{eqnarray}
For $p=1+\epsilon$ we obtain
\begin{equation}
    \Pi_{\psi}(q)=\frac{h^2q}{4\pi^2\epsilon}+\text{finite}.
\end{equation}
Finally, the one-loop correction to the vertex reads as follows:
\begin{eqnarray}
    -ih\Gamma(q_1,q_2)=(-ih)^3\int\frac{d^pp}{(2\pi)^p}\frac{i(p+q_1)i(p+q_2)}{(p+q_1)^2(p+q_2)^2}\left(-(-i)\frac{|p|}{4\pi} \right).
\end{eqnarray}
From this, the leading divergence can be extracted:
\begin{equation}
    \Gamma(q_1,q_2)=\frac{h^2}{4\pi^2\epsilon}+\text{finite}.
\end{equation}
Collecting the results above, we obtain the following renormalization factors:
\begin{eqnarray}
Z_{\phi}&=&1,\\
Z_{\psi}&=&1+h^2\left(\frac{1}{4\pi^2\epsilon}+\text{finite}\right),\\
Z_h&=&1+h^2\left(-\frac{1}{4\pi^2\epsilon}+\text{finite}\right).
\end{eqnarray}
This allows us to compute the beta function
\begin{equation}
    \beta=\frac{\epsilon}{2}h-\frac{h^3}{4\pi^2}+\mathcal{O}(h^4).
\end{equation}
Aside from the trivial IR fixed point with $h_*=0$, we also find a UV fixed point with
\begin{equation}
    h_*^2=2\pi^2\epsilon.
\end{equation}
In this case, the one-point function of $\left<\phi^2 \right>$ reads 
\begin{equation}
\begin{split}
\left< \phi^2(x) \right> & = \frac{h^2_*}{64 \pi^2 \, x_{\perp}^2}= \frac{1} {32 \, x_{\perp}^2}\epsilon.
\end{split}
\end{equation}
The two-point function of the displacement operator gives instead
\begin{equation}
\begin{split}
\left<\hat D^i(\sigma_1) \hat D^j(\sigma_2)\right> &  = \frac{ 2\Delta_{\phi}\,\,h^2\,C_{\phi}\, C^2_{\psi} p}{|\sigma|^{2p+2}} \,+\mathcal{O}(h^3) = \frac{\epsilon}{4 |\sigma|^4}\, .
\end{split}
\end{equation}

\subsection{Phase diagrams}

\begin{figure}

\begin{center}
    \begin{tikzpicture}
        \draw[->, color=black] (-5, 0) -- (5, 0) node[anchor=north, color=black]{$\alpha$};
        \node[left] at (0.5, -0.5) {$\sqrt{4\pi}$};
        \shade[line width=2pt, top color=green!65!blue,opacity=0.4] 
        (0,0) to [out=90, in=-90]  (0, 2)
        to [out=0,in=180] (4.8, 2)
        to [out = -90, in =90] (4.8, -0)
        to [out=190, in =0]  (0, 0);
        \shade[line width=2pt, top color=red!65!yellow,opacity=0.4] 
        (-4.8,0) to [out=90, in=-90]  (-4.8, 2)
        to [out=0,in=180] (0, 2)
        to [out = -90, in =90] (0, 0)
        to [out=190, in =0]  (-4.8, 0);
        \draw[line width=1pt, color = black, dashed] (0,0) -- (0,2);
        \node[left] at (0.5, -1) {\textbf{a)}};
        \node[left] at (0.5, 2.4) {\text{\scriptsize critical.}};
        \node[left] at (0.6, 2.2) {\text{\scriptsize fermion.}};
    
        \draw[->, color=black] (-5, -4) -- (5, -4) node[anchor=north, color=black]{$\alpha$};
        \node[left] at (2.5, -4.5) {$\sqrt{6\pi}$};
        \node[left] at (0.5, -4.5) {$\sqrt{4\pi}$};
        \shade[line width=2pt, top color=green!65!blue,opacity=0.4] 
        (2,-4) to [out=90, in=-90]  (2, -2)
        to [out=0,in=180] (4.8, -2)
        to [out = -90, in =90] (4.8, -4)
        to [out=190, in =0]  (2, -4);
        \shade[line width=2pt, top color=red!65!yellow,opacity=0.4] 
        (-4.8,-4) to [out=90, in=-90]  (-4.8, -2)
        to [out=0,in=180] (0, -2)
        to [out = -90, in =90] (0, -4)
        to [out=190, in =0]  (-4.8, -4);
        \draw[line width=1pt, color = black, dashed] (0,-4) -- (0,-2);
        \draw[line width=1pt, color = black, dashed] (2,-4) -- (2,-2);
        \shade[line width=2pt, top color=red,opacity=0.4] 
        (0,-4) to [out=90, in=-90]  (0, -2)
        to [out=0,in=180] (2, -2)
        to [out = -90, in =90] (2, -4)
        to [out=190, in =0]  (0, -4);
        \node[left] at (0.5, -5) {\textbf{b)}};

        \node[left] at (2.6, -1.8) {\text{\scriptsize critical.}};
        \node[left] at (0.6, -1.8) {\text{\scriptsize fermion.}};
        
        \draw[->, color=black] (-5, -8) -- (5, -8) node[anchor=north, color=black]{$\alpha$};
        \node[left] at (-1.5, -8.5) {$\sqrt{2\pi}$};
        \shade[line width=2pt, top color=green!65!blue,opacity=0.4] 
        (-2,-8) to [out=90, in=-90]  (-2, -6)
        to [out=0,in=180] (4.8, -6)
        to [out = -90, in =90] (4.8, -8)
        to [out=190, in =0]  (-2, -8);
        \shade[line width=2pt, top color=red!65!yellow,opacity=0.4] 
        (-4.8,-8) to [out=90, in=-90]  (-4.8, -6)
        to [out=0,in=180] (-2, -6)
        to [out = -90, in =90] (-2, -8)
        to [out=190, in =0]  (-4.8, -8);
        \draw[line width=1pt, color = black, dashed] (-2,-8) -- (-2,-6);
        \node[left] at (0.5, -9) {\textbf{c)}};

        \node[left] at (-1.5, -5.6) {\text{\scriptsize critical.}};
        \node[left] at (-1.4, -5.8) {\text{\scriptsize fermion.}};

        \draw[->, color=black] (-5, -12) -- (5, -12) node[anchor=north, color=black]{$\alpha$};
        \node[left] at (-1.5, -12.5) {$\sqrt{2\pi}$};
        \node[left] at (0.5, -12.5) {$\sqrt{4\pi}$};
        \shade[line width=2pt, top color=green!65!blue,opacity=0.4] 
        (0,-12) to [out=90, in=-90]  (0, -10)
        to [out=0,in=180] (4.8, -10)
        to [out = -90, in =90] (4.8, -12)
        to [out=190, in =0]  (0, -12);
        \shade[line width=2pt, top color=red!65!yellow,opacity=0.4] 
        (-4.8,-12) to [out=90, in=-90]  (-4.8, -10)
        to [out=0,in=180] (-2, -10)
        to [out = -90, in =90] (-2, -12)
        to [out=190, in =0]  (-4.8, -12);
        
        \shade[line width=2pt, top color=blue!65!red,opacity=0.4] 
        (-2,-12) to [out=90, in=-90]  (-2, -10)
        to [out=0,in=180] (0, -10)
        to [out = -90, in =90] (0, -12)
        to [out=190, in =0]  (-2, -12);

        \draw[line width=1pt, color = black, dashed] (-2,-12) -- (-2,-10);
        \draw[line width=1pt, color = black, dashed] (0,-12) -- (0,-10);
        \node[left] at (0.5, -13) {\textbf{d)}};

        \node[left] at (-1.4, -9.8) {\text{\scriptsize critical.}};
        \node[left] at (0.6, -9.8) {\text{\scriptsize fermion.}};

    \end{tikzpicture}
\end{center}

\caption{Phase diagrams of the $2d$ scalar field coupled to the bulk scalar as a function of the parameter $\alpha$. a) Four-dimensional bulk scalar with $\hat\phi\cos\alpha\sigma$ coupling. b) Three-dimensional bulk scalar with $\hat\phi\cos\alpha\sigma$ coupling. c) Three-dimensional bulk scalar with $\partial_{\perp}\hat\phi\cos\alpha\sigma$ coupling. d) Five-dimensional scalar with $\hat\phi\cos\alpha\sigma$ coupling.
	}
	\label{Phases}

\end{figure}
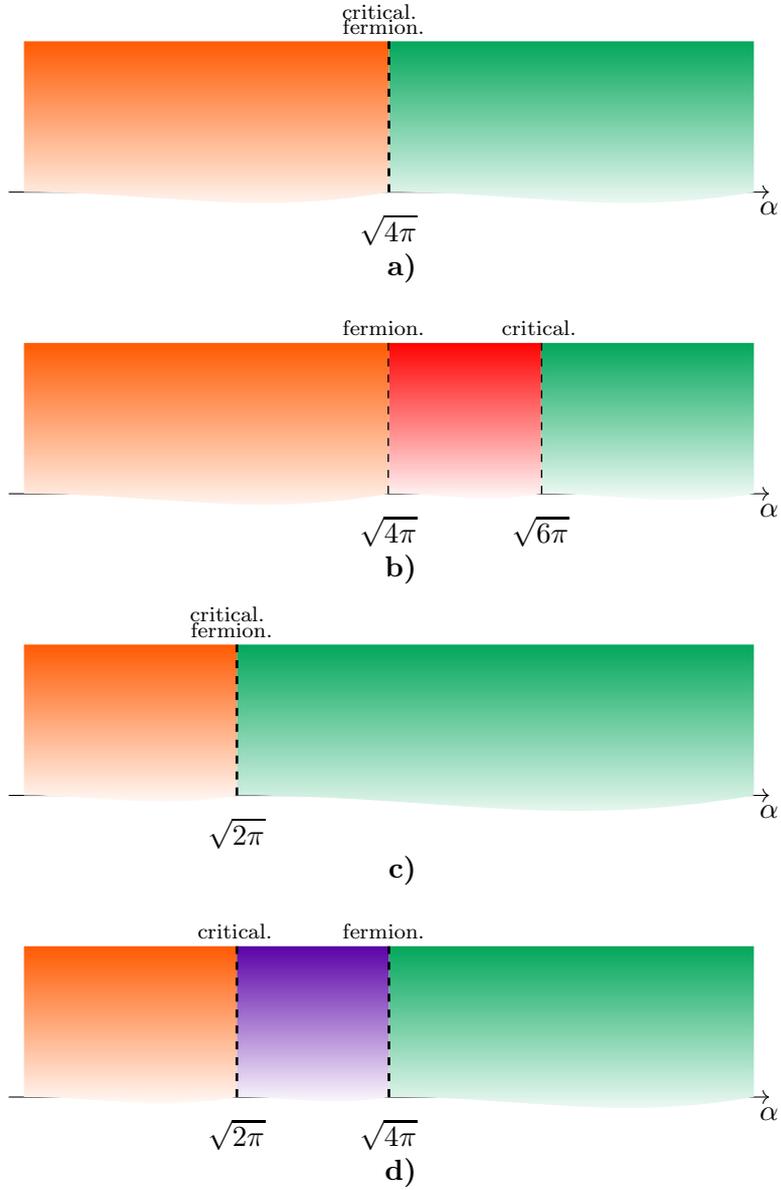

Let us now summarise our findings. When $\alpha\ll1$, the semiclassical considerations suggest that either the defect degrees of freedom are gapped, and the bulk scalar is subject to Dirichlet boundary conditions in the IR, or there is no stable vacua at all (namely, when $d=5$). We would like to emphasise again that while varying the value of $\alpha$, we also change the model quite significantly since the number of relevant operators also varies. Thus, we really switch from a model to a model, rather than changing a parameter in a single chosen model. Having said that, we would like to conjecture that this picture holds for a finite range of the values of $\alpha$: on Fig. \eqref{Phases} this phase is denoted by orange colour. On the other hand, when alpha is large enough, the interaction operator becomes irrelevant, and the long-distance dynamics of the defect is the one of the trivial bulk defect with the decoupled 2d sector: on Fig. \eqref{Phases} this is denoted by green colour.

The two above-mentioned phases are separated by the region, where the interaction term is nearly marginal. The RG flow analysis does not reveal any perturbative interacting fixed points. All these pieces of information are put together into a picture very similar to the one known from the analysis of sine-Gordon model.

A new aspect of the problem gets revealed when some special values of $\alpha$ are considered. More precisely, for $\alpha=\sqrt{4\pi}$ we can use an equivalent description in terms of $2d$ Dirac fermions, coupled to the bulk scalar field via the Yukawa interaction localised on the defect. This Yukawa interaction is strongly relevant for the $3d$ bulk and strongly irrelevant for the $5d$ bulk, with the interaction in both cases being $\hat\phi\cos\alpha\sigma$. Nevertheless, we can refer to the $\epsilon$-expansion, starting from the dimension where the interaction is nearly marginal, and look for fixed points of the beta functions. Indeed, we found an IR fixed point for the $3d$ bulk and a UV fixed point for the $5d$ bulk. It is tempting to conjecture that the fixed point exists not just for a single point, but for some range of $\alpha$, when the interaction gets closer to marginality. In Fig. \eqref{Phases} the conjectured range, where the IR fixed point exists, is denoted by red, and the conjectured interval, where the UV fixed point exists, is denoted by purple.

Finally, we note that due to the absence of conformal defects for a free four-dimensional scalar, we do not have any fixed point in this case (Fig. \eqref{Phases}, a)).

\section{Monodromy defects for bulk-free complex scalars}
\label{sec:monodromy}

\subsection{Free-scalar monodromy defect}

In this section, we study the $U(1)_f$ monodromy defect in the theory of a bulk-free, complex scalar field, $\phi(x)$, in $d$ dimensions. Monodromy defects can be thought of as surface operators on which co-dimension 1 topological domain walls that implement flavour
symmetry rotations can end. In the present case, it can be engineered by turning on a background gauge field for the $U(1)_f$ global symmetry which is closed but not exact. We take the Euclidean action of the theory to be
\begin{equation}
\label{eq:scalar_action}
S=\int d^d x \, \sqrt{g} \left[    D^\mu\phi (D_\mu \phi)^\dagger + \frac{d-2}{4(d-1)} \mathcal{R} \left| \phi\right|^2 \right],
\end{equation}
where the coupling to the scalar curvature $\mathcal{R}$ is needed to have a conformal and Weyl invariant action. We also define the gauge covariant derivative $D_\mu \equiv \nabla_\mu -i eA_\mu$.  

Varying $S$ with respect to the gauge field $A_\mu$ gives the conserved $U(1)_f$-current
\begin{equation}
\label{eq:scalar_current}
J_\mu = \frac{1}{\sqrt{g}}\frac{\delta S}{\delta A^\mu} = - i \left( \phi \nabla_\mu \phi^\dagger -\nabla_\mu \phi \phi^\dagger  + 2 i e A_\mu \left|\phi \right|^2\right),
\end{equation} 
while the variation with respect to the metric $g_{\mu\nu}$ produces the stress energy-momentum tensor
\begin{equation}
\begin{split}
T_{\mu \nu}   
&= \frac{2}{\sqrt{g}} \frac{\delta I_{\mathrm{scalar}}}{\delta{g^{\mu\nu}}}  \\ 
&= D_\mu \phi  (D_\nu\phi)^\dagger + (D_\mu\phi)^\dagger D_\nu  \phi  -\frac{d-2}{2(d-1)} \left[ \nabla_\mu \nabla_\nu  +\frac{ g_{\mu\nu}}{d-2} \nabla^2 - \mathcal{R}_{\mu\nu}  \right]\left|\phi \right|^2,
\end{split}
\end{equation}
where $\mathcal{R}_{\mu\nu}$ is the Ricci tensor for the background geometry.

A monodromy defect can be introduced by taking the connection $A_\mu$
\begin{equation}
A = \alpha \, d\theta \, ,
\end{equation}
where we restrict $\alpha\in (0,1)$. In addition, for simplicity we set the charge $e=1$. 
In \cite{Bianchi:2021snj}, the authors provided a detailed derivation of the mode expansion of the bulk field $\phi$ and its propagator. By adopting complex coordinates transverse to the defect, i.e. $z = x_\perp e^{i \theta}$ and $\bar z = x_\perp e^{-i \theta}$, the mode expansion of the field can be written as a sum over the modes 
\begin{equation}
\label{eq:oper_O}
\phi_{m\pm\a} \equiv \int d k_\perp \int d^{d-3} \vec k \left[ f ( k)  a_{\mp m}( k)  + f^* ( k)  b_{\mp m}^\dagger( k)      \right]  \frac{J_{m\pm\alpha}(k_\perp x_\perp)}{x_\perp^{m\pm\alpha}}\,,
\end{equation}
where $J_\nu (\zeta)$ is the Bessel function of the first kind, and 
\begin{equation}
f(k) = \frac{\sqrt{k_\perp}}{(\sqrt{2\pi})^{d-2}\sqrt{2\omega}}e^{-i \omega t+i {\vec k\cdot \vec \sigma}}\,,
\end{equation}
with $\omega^2=k_{\perp}^2+\vec k^2$. 

The behaviour of the field $\phi$ in the defect limit $x_\perp \rightarrow 0$ fixes the ``boundary condition". Similarly to the real scalar case discussed in Sec. \ref{Section2}, 
unitarity imposes strong constraints on the allowed modes. In particular, only the modes with a divergence milder than 
$\left(x_\perp^{-1}\right)$ as $x_\perp\to 0$ are allowed. This means that the modes $\phi_{m+\a}$ are restricted to $m\geq -1$, while the modes $\phi_{m-\a}$  to $m\geq 0$. Note that the only modes that are divergent in the defect limit are $\phi_{-\alpha}$ corresponding to the quantum number $m=0$, and $\phi_{-1+\alpha}$ for $m=-1$. For the sake of simplicity, in this work we only consider the singular mode $\phi_{-\alpha}$ and we impose regularity on the $|m|=1$ mode.
With this last assumption, we obtain
\begin{equation}
\label{eq:mod_exp_salar_defect}
\phi = \sqrt{\xi} \, \phi_{-\a} \,  z^{-\alpha} + \sqrt{1-\xi}\, \phi_{\a} \, \bar z^{\alpha} +  \sum_{m=1}^{\infty} \Big[ \phi_{m-\alpha}  z^{m-\alpha} + \phi_{m+\alpha} \bar z^{m+\alpha} \Big] \,,
\end{equation}
where $\xi=0$ or $\xi=1$.
Moreover, in \cite{Bianchi:2021snj} the authors showed that the bulk field can be written in terms of some defect operators $\hat O^+_s$ of dimensions $\hat\Delta_s^{+} = d/2-1+|s|$ and $\hat O^{-}_{-\alpha}$ of dimension $\hat\Delta^{-}_\alpha = d/2-1-\alpha$. In particular we may write 
\begin{equation}\label{eq:defect_OPE}
\begin{split}
\phi =\;& \sum_{s\in \mathbb{Z}-\a} c_{\phi \hat O_s^+} x_\perp^{|s|} e^{is\theta} \mathcal{C}^+_{s}\left(x_\perp^2 \partial^2_{\sigma}\right) \hat O^+_{s}(\sigma) +c_{\phi \hat O_{-\a}^-}\frac{e^{-i\a\theta}}{x_\perp^{\a}} \mathcal{C}^-_{s}\left(x_\perp^2 \partial^2_{\sigma}\right) \hat O^-_{-\a}(\sigma)\, ,
\end{split}
\end{equation}
where the differential operators $\mathcal{C}^{\pm}_{s}(x_\perp^2 \partial_{\sigma}^2)$ resum the contribution of all the conformal descendants, and are fixed by conformal invariance to be
\begin{equation}\label{eq:diff_op}
\mathcal{C}^{\pm}_{s}\left(x_\perp^2 \partial^2_{\sigma}\right) \equiv \sum_{k=0}^{+\infty} \frac{(-4)^{-k}(x_\perp^2 \partial^2_{\sigma})^k}{k!(1\pm|s|)_k}\,,
\end{equation}
where $(a)_k \equiv a(a+1)\dots(a+k-1)$ if $k \ne 0$ and $(a)_0\equiv 1$ is the Pochhammer symbol. In addition, the defect-bulk OPE coefficients read
\begin{equation}\label{bulkdefcoeff}
c_{\phi \hat O_s^+}= \sqrt{\frac{\Gamma\left( \frac{d}{2}-1+|s|  \right)}{4\pi^{d/2}\Gamma\left(1+ |s| \right)}} \qquad \text{for   } s\neq-\a \, ,
\end{equation}
with the special cases
\begin{subequations}
\begin{align}
 c_{\phi \hat O_{-\alpha}^+}&= \sqrt{(1-\xi)\frac{\Gamma\left( \frac{d}{2}-1+\a  \right)}{4\pi^{d/2}\Gamma\left(1+ \a \right)}}\,, & c_{\phi \hat O_{-\alpha}^-}&= \sqrt{\xi\frac{\Gamma\left( \frac{d}{2}-1-\a  \right)}{4\pi^{d/2}\Gamma\left(1- \a \right)}}\,. \label{bulkdefcoeff1}
\end{align}
\end{subequations}

The above-mentioned defect operators are defect primaries of conformal dimension $\hat  \Delta^\pm_s = \frac{d}{2}-1 + m \pm \a$ and orthogonal spin $s =  \a - m$ for ``$-$" and $s = m+ \a$ for ``$+$" ($s\in \mathbb{Z}-\alpha$). Notice their normalization is chosen such that their two-point function is
%
%
%
\begin{align}\label{eq:defdefprop}
\left<\hat O_{\nu}^{\pm}(\sigma)\hat O_{\nu'}^{\dagger \pm}(0)\right> =\frac{\delta_{\nu, \nu'}}{|\sigma^a|^{d-2\pm2|\nu|}} \, .
\end{align}
As anticipated above, the allowed defect primaries operators with conformal dimension $\hat{\Delta}^{\pm}_s$ can be found by employing the unitarity bound \eqref{eq:uni_def_bound} \cite{Lauria:2020emq,Bianchi:2021snj}. While $\hat{\Delta}^{+}_s$ are always allowed, the ``$-$" ones  violate the unitarity bound for a co-dimension 2 defect in $d > 4 $ unless $|s|<1$. For $d\leq 4$, unitarity requires $|s|<\frac{d-2}{2}$. In $d<4$, the unitarity bounds above \eqref{eq:defect_OPE} impose further conditions on the range of $\a$ at $\xi=1$. As an example consider $d=3$, for which the unitarity bound reads $|s|<\frac{1}{2}$, which means  $\a\in \left[0,\frac{1}{2}\right)$ if $\xi=1$.

In the following we report some of the correlators computed in \cite{Bianchi:2021snj,Giombi:2021uae}.

\subsubsection*{Bulk-bulk propagator}
The bulk-bulk propagator has also been computed and the result is
\begin{align}
\label{eq:two-point function_gaug_transf}
\left< \phi(x)\phi^\dagger(0,x_\perp') \right> = \left(\frac{1}{x_\perp x_\perp'}\right)^{\frac{d}{2}-1} \Bigg(\sum_{s\in \mathbb{Z}-\a} c_s^+ \,    F_{\hat{\Delta}^+, s}(\eta,\theta)		+  c_{-\a}^- \,    F_{\hat{\Delta}^-, -\a}(\eta,\theta)\Bigg)\,,
\end{align}
where the coefficients $c_s^\pm = |c_{\phi \hat O_s^\pm}|^2$, and
\begin{equation}\label{defectblock}
F_{{\hat \Delta}, s}(\eta,\theta)=\left(\frac{\eta}{2} \right)^{\hat{\Delta}_s} \phantom{}_2 F_1\left( \frac{\hat{\Delta}_s}{2},\frac{\hat \Delta_s+1}{2}; \hat{\Delta}_s+2-\frac{d}{2}; \eta^2  \right) e^{i s\theta}\,,
\end{equation}
being ${}_2 F_1 (a,b;c;z)$ the ordinary hypergeometric function, and $\eta$ the cross ratio
\begin{equation}\label{eq:crossratio-def}
\eta\equiv \frac{2 x_\perp x_\perp'}{x_\perp^2+x_\perp'^2 +  |\sigma^a|^2} \,.
\end{equation}
Notice that the coefficients of the expansion \eqref{eq:two-point function_gaug_transf} are the bulk to defect couplings.
We have set $\sigma'{}^a=0$ by translational invariance along the defect, and $\theta'=0$ for rotational invariance around the defect.

Finally, we report another important correlator, namely the bulk-defect two-point function between $\phi$ and $\hat O_\nu$. With the normalisation \eqref{eq:defdefprop}, it reads
\beq
\left< \phi(x_1) \hat O^\dagger_\nu(\sigma_2)  \right> =\sqrt{\frac{\Gamma\left( \frac{d}{2}-1+\nu \right)}{4 \pi^{d/2} \Gamma(1+\nu)}} \frac{1}{x_\perp^{-\nu}} \left[ \frac{1}{x_\perp^2 + (\sigma_1 - \sigma_2)^2} \right]^{d/2-1+\nu} .
\eeq
\subsubsection*{One-point function of $\bold{|\phi|^2}$}
From the coincident limit of the bulk-bulk propagator \eqref{eq:two-point function_gaug_transf} we obtain the one-point function of $|\phi|^2$, which reads
\begin{equation}\label{oneptphi2full}
 \left< |\phi(x)|^2 \right>= \frac{ \Gamma (\frac{d}{2}-\alpha ) \Gamma
	(\frac{d}{2}+\alpha -1 ) \sin (\pi  \alpha )}{2^{d-1} \pi^{\frac{d+1}{2}}\Gamma \left(\frac{d-1}{2}\right)} \frac{1}{x_\perp^{d-2}} \left(-\frac{1}{d-2}+\frac{\xi}{\frac{d}{2}-\a-1}\right).
\end{equation}
\subsubsection*{One-point function of the stress tensor}
The one-point function of the stress tensor has been found to be \cite{Dowker:1987mn}
\begin{align}
\label{eq:T_mon}
\langle T_{\perp\perp}(x)\rangle &=-\frac{ \Gamma \left(\frac{d}{2}-\alpha \right) \Gamma \left(\frac{d}{2}+\alpha -1\right)\sin (\pi  \alpha )  \left(\frac{\alpha  (1-\alpha )}{d}+\frac{\alpha ^2 \xi }{ \frac{d}{2}-\alpha-1}\right)}{2^{d-1}\pi^{\frac{d+1}{2}}\Gamma \left(\frac{d+1}{2}\right)}\frac{1}{x_\perp^d}\,,
\end{align}
which is vanishing for $\alpha=0$ and $\alpha=1$ when $d>4$, as expected.  Comparing to \eqref{eq:stress-tensor-one-pt-fn}, we find that $h$ is expressed as
\begin{equation}
\label{eq:h_scalar}
a_T =   \frac{ \Gamma \left(\frac{d}{2}-\alpha \right) \Gamma \left(\frac{d}{2}+\alpha -1\right)\sin (\pi  \alpha )  \left(\frac{\alpha  (1-\alpha )}{d}+\frac{\alpha ^2 \xi }{\frac{d}{2}-\alpha-1}\right)}{2^{d-1}\pi^{\frac{d+1}{2}}\Gamma \left(\frac{d+1}{2}\right)}\,.
\end{equation}
\subsubsection*{One-point function of the current and sphere-free energy}
One interesting correlator is the one-point function of the U(1)$_f$ current, which is non-zero due to the presence of the monodromy. One finds \cite{Bianchi:2021snj,Giombi:2021uae}
\begin{equation}
\label{eq:J_mon}
\langle J_\theta \rangle=  \frac{C_J(\alpha)}{x_\perp^{d-2}}\,,
\end{equation}
with 
\begin{equation}
C_J(\alpha) = \frac{ \Gamma \left(\frac{d}{2}-\alpha \right) \Gamma \left(\frac{d}{2}+\alpha -1\right) \sin (\pi  \alpha )\left(1-2 \alpha+\frac{2 \alpha  (d-1) \xi }{ \frac{d}{2}-\alpha-1}\right)}{2^d\pi^{\frac{d+1}{2}}\Gamma \left(\frac{d+1}{2}\right) } \, .
\end{equation}
From this correlator it is possible to find the derivative with respect to $\alpha$ of the free-energy, namely \cite{Bianchi:2021snj,Giombi:2021uae}
\begin{equation}
\label{eq:F_deriv}
\frac{d}{d\alpha} F = -  C_J(\alpha) \frac{2 \pi ^{\frac{d}{2}+1}}{\Gamma \left(\frac{d}{2}\right) \sin \left( \frac{\pi}{2} d \right)} \,.
\end{equation}
For $d=4$, when the defect is $p=2$, we can then extract the central charge $b$ to be \cite{Bianchi:2021snj}
\begin{equation}
\label{eq:b-free-scalar}
b =  \frac{ (1-\alpha)^2 \alpha ^2 + 4 \xi \alpha^3}{2} \,,
\end{equation}
while for $d=6$ (and $p=4$) we get the $a$ defect central charge 
\begin{equation}
a= -\frac{\alpha ^2}{720}  \Big[\alpha ^4-3 \alpha ^3 +5 \alpha -3 -2 \alpha \left(5-3 \alpha
   ^2\right) \,  \xi\Big] \, .
\end{equation}
In \cite{Bianchi:2021snj}, it has been observed that the two possible boundary conditions in the free theory $\xi =0,1$ are connected by a quadratic RG flow from $\xi=1$ to $\xi =0$ triggered by the relevant operator $(\hat O^{-}_{-\alpha}\hat O^{- \dagger}_{-\alpha})$. We can indeed notice that, in the range of validity $\alpha \in(0,1)$, the defect central charges $b$ and $a$ take on smaller values in the IR compared to the UV.

\subsubsection*{Displacement operator}
The displacement operator is another important defect primary. For the free scalar monodromy defect, the displacement has been found in \cite{Bianchi:2021snj,Giombi:2021uae}, and it can be written as  
\beq
\label{eq:disp_mon}
\hat D_z = \hat O^{-}_{-\alpha} \hat O^{+ \dagger}_{1+\alpha} \, , \qquad \hat D_{\bar z} = \hat O^{- \dagger}_{-\alpha} \hat O^{+}_{1+\alpha} \, .
\eeq
Their non-trivial two-point function reads\footnote{The $1/2$ factor in the definition of $C_{\hat D}$ appears from the transformation to the complex coordinates $z,\bar z$. }
\begin{equation}
\label{eq:CD-h_relation}
\left< \hat D_z(\sigma)\hat D_{\bar z}(0)\right> = \frac{C_{\hat D}}{2|\sigma|^{2(d-1)}} \, , \qquad C_{\hat D} = 2^{d} d \frac{\Gamma\left( \frac{d+1}{2}\right)}{\pi^{\frac{d-3}{2}}} a_T \, .
\end{equation}
Note that the coefficient of the two-point function of the displacement operator and the one corresponding to the one-point function of the stress tensor are proportional in the free theory case. This relation holds also for theories with $\mathcal{N}=2$ supersymmetry where the defect is half-BPS \cite{Bianchi:2019sxz}. In the following, we will show that the relation is generically broken by defect interactions.

We conclude this discussion pointing out that the validity of the propagator \eqref{eq:two-point function_gaug_transf} goes beyond the purely quadratic theory as it only requires a free bulk. This means that even in the presence of defect localised interactions, if the theory is at a defect fixed point, the same expression \eqref{eq:two-point function_gaug_transf} applies where the only modification is the value of $\xi$ which now will be a real number in the range $\xi \in [0,1] $ to be fixed by an independent computation. This also means that the expression of correlators quadratic in $\phi$ such as $|\phi|^2$ in \eqref{oneptphi2full}, $<T_{\mu\nu}>$ in \eqref{eq:T_mon}, and $<J_\mu>$ in \eqref{eq:J_mon} are still valid in the presence of defect interactions.  

\subsection{Defect Interaction in $\epsilon$-expansion}
In the section above we reviewed the quadratic monodromy defect. We discussed how there is a possible choice when the defect spectrum contains the defect primary operator  $\hat O_{-\alpha}^-$ of dimension $\Delta^{-}_{-\alpha} = d/2-1-\alpha$. This means that this theory can be deformed by adding suitable defect-relevant perturbations. The easiest case is the quadratic deformation discussed above in sec. \ref{}, which leads to a flow toward the other boundary condition where only the operator $\hat O_{-\alpha}^+$ is present. A more general possibility is going beyond the free theory and considering non-quadratic defect deformations.
In particular, in this section we consider the perturbation
\begin{equation}
S_{\lambda_n} = \lambda_n \int d^{d-2}\sigma \, \left(\hat O^-_{-\alpha}(\sigma) \hat O^{\dagger -}_{-\alpha}(\sigma) \right)^n .
\end{equation}
By dimensional considerations, we have that the coupling constant $\lambda_n$ is relevant provided
\begin{equation}
\alpha > \bar \alpha, \qquad  \bar \alpha \equiv \frac{(n-1)(d-2)}{2n},
\end{equation}
and classically marginal for $\alpha =\bar \alpha$.
Since $\alpha$ is a continuous parameter, we can obtain a slightly relevant perturbation by defining $\alpha = \bar \alpha + \epsilon$ with $0<\epsilon \ll 1$. In this way we can proceed with a perturbative approach as described in Sec. \ref{Section3}. 

Notice that in general, for a fixed value of $\alpha$ and $d$ there is more than one relevant perturbation. To study the most general flow is thus quite involved. For this reason, we will only study the slightly relevant perturbation tuning all the other ``strongly" relevant ones to zero.

\subsubsection{Defect $\beta-$function}

The one-loop defect $\beta-$function can be obtained as in the previous sections by applying the general rule \eqref{BetaFuncGeneral}. In order to use that, we need both the two and three-point functions of the operator $ \left(\hat O^-_{-\alpha}(\sigma) \hat O^{\dagger -}_{-\alpha}(\sigma) \right)^n$. In the following we provide such information. 
For the two-point function we just have
\beq
\left<  \left(\hat O^-_{-\alpha}(\sigma) \hat O^{\dagger -}_{-\alpha}(\sigma) \right)^n  \left(\hat O^-_{-\alpha}(0) \hat O^{\dagger -}_{-\alpha}(0) \right)^n \right> = \frac{(n!)^2}{|\sigma|^{n(d-2-2\alpha)}},
\eeq
while the three-point function reads
\beq
\left<  \left|\hat O^-_{-\alpha}(\sigma_1)\right|^{2n}   \left|\hat O^-_{-\alpha}(\sigma_2)\right|^{2n}  \left|\hat O^-_{-\alpha}(0)\right|^{2n} \right> = \frac{C_{nnn}}{|\sigma_1-\sigma_2|^{n\left(\frac{d}{2}-1-\alpha\right)}|\sigma_1|^{n\left(\frac{d}{2}-1-\alpha\right)}|\sigma_2|^{n\left(\frac{d}{2}-1-\alpha\right)}},
\eeq
where 
\beq\label{eq:C_nnn}
C_{nnn} = \sum_{k=0}^n \frac{n!}{(n-k)!k!}\frac{n!}{(n-k)!}\frac{n!}{k!}\frac{n!}{k!}\frac{n!}{(n-k)}\frac{n!}{k!}k! = 
(n!)^3\text{Fr}_n \, ,
\eeq
being Fr$_n$ the Franel numbers. Thus, the beta function is
\beq
\beta = - 2 n \,\epsilon \, \lambda_n + \frac{\pi ^{d/2-1}}{\Gamma \left(\frac{d}{2}-1\right)}\frac{C_{nnn}}{(n!)^2} \, \lambda_n^2 .
\eeq
This implies the existence of the following interacting fixed point
\beq
\lambda_{n *} =  \frac{2n}{(n!)\text{Fr}_n} \frac{\Gamma \left(\frac{d}{2}-1\right)}{\pi^{d/2-1}} \, \epsilon\, .
\eeq

\subsubsection{Correlators}
\label{sec:Monod_corr}

Now we study the correction to some correlators.

We can start with bulk correlators and then move to defect ones.
The first we consider is the bulk-bulk propagator, namely the two-point function of $\phi$. Its first correction is at the second order in $\lambda_n$, and it can be computed by the integral
\beq
\begin{split}
&\left< \phi(\sigma ; \theta_1, x_{\perp 1}) \phi^\dagger(0;\theta_2, x_{\perp 2})  \right>^{(2)}_{\lambda_n} =\mathcal{N} \,  \frac{\lambda_n^2}{2} \frac{e^{i\alpha(\theta_1-\theta_2)}}{(x_{\perp 1} x_{\perp 2})^\alpha} \times \\ 
&\times \int d^{d-2}\sigma_1 d^{d-2}\sigma_2 \, \frac{1}{|(\sigma-\sigma_1)^2+x_{\perp 1}^2|^{d/2-1-\alpha}}\frac{1}{|\sigma_2^2+x_{\perp 2}^2|^{d/2-1-\alpha}}\frac{1}{|\sigma_1-\sigma_2|^{(2n-1)(d-2-2\alpha)}} \, ,
\end{split}
\eeq
where the combinatorial factor $\mathcal{N} =2 n^2 n! (n-1)!\,$.
The above integral is better computed in momentum space. For this reason, we find it convenient to employ the representations in eq. \eqref{eq:fourier_xp} and eq. \eqref{eq:fourier_sigma}.
%
%
%
%
%
%
%
After performing all the trivial integrations and taking the leading order in $\epsilon$, we obtain 
%
%
\beq
\begin{split}
\left< \phi(\sigma ; \theta_1, x_{\perp 1}) \phi^\dagger(0;\theta_2, x_{\perp 2})  \right>^{(2)}_{\lambda_n} =&-\mathcal{N} \,  \lambda_n^2 e^{i\alpha(\theta_1-\theta_2)} \, \frac{\pi ^{d-3}}{2(n-1) \Gamma \left(\frac{d-2}{2 n}+1\right) \Gamma \left(\frac{(d-2) (2 n-1)}{2
   n}\right)} \times \\ 
&\times \int \frac{d^{d-2}k}{(2\pi)^{d-2}} K_\alpha (k x_{\perp 1})K_\alpha (k x_{\perp 2})  \, e^{i k\cdot \sigma} \,  \, .
\end{split}
\eeq
The integral over $k$ can be done, and we find
\begin{equation}
\begin{split}
 \int \frac{d^{d-2 }k}{(2\pi)^{d-2}}  \, K_{\a}(k x_{\perp 1})K_{\a}(k x_{\perp 1}) e^{i k(\sigma_1-\sigma_2)} = \frac{\pi^2}{\sin \pi \a} \left( \frac{1}{x_{\perp 1} x_{\perp 2}} \right)^{\frac{d-2}{2}}&\left[  c_{-\alpha}^- \,    F_{\hat{\Delta}^-, -\alpha}(\eta,\theta) \right.\\
 & \left.- c_{-\alpha}^+ \,    F_{\hat{\Delta}^+, -\alpha}(\eta,\theta)  \right] \, .
 \end{split}
\end{equation}
From the propagator we can read off the coefficient $\xi$ that now is different $1$, taking a value in $(0,1)$. We indeed find
%
%
\beq
\xi = 1- \frac{ 4 \pi \, n^3 \,   \Gamma^2 \left(\frac{d}{2}-1\right)  \csc \left(\pi \, \bar\alpha \right)}{\text{Fr}_n^2 \, (n-1) \, \Gamma \left(\frac{d-2}{2 n}+1\right) \Gamma
   \left(\bar\alpha+\frac{d}{2}-1\right)} \epsilon ^2 \, .
\eeq
Note that this value is less than $1$ in its range of validity, as required by unitarity. The same value can also be found by computing the one-point function $\left<|\phi|^2 \right>$ and employing the relation \eqref{oneptphi2full}. Another correlator we can study is the bulk one-point function of the field $|\phi|^{2m}$. 
At first order in the perturbation, only the cases $m \ge n$ get a non-vanishing contribution. This is easily computed to give
\beq
\left<|\phi|^{2m}(x_\perp) \right> =C_{\phi}^m \, m! \left[1 -\frac{2^{3-d} \pi ^{\frac{d-1}{2}}  m! }{\Gamma \left(\frac{d-1}{2}\right) ((m-n)!)^2} \left(\frac{c_{\phi \hat O_{-\alpha}^-}^2}{C_{\phi}}\right)^n  \, \lambda_{n *} \right] \frac{1}{x_\perp^{(d-2)m}} \, ,
\eeq
where $c_{\phi \hat O_{-\alpha}^-}$ is the coefficient reported in \eqref{bulkdefcoeff1} and $C_\phi=\Gamma(\frac{d-2}{2})/(4\pi^{\frac{d}{2}})$.

Now we can study two-point functions of some defect operators, such as the one of the operator $\left(\hat O^{-}_{-\alpha}(\sigma)\right)^m$. Also in this case, when $m\ge n$, then we already have a contribution at the first order. We find
\beq
\left<\left(\hat O^{-}_{-\alpha}(\sigma)\right)^m\left(\hat O^{- \dagger}_{-\alpha}(0)\right)^m \right> =  \frac{m!}{| \sigma | ^{m ( d-2-2 \alpha)}} \left(1-\frac{4 \pi^{\frac{d}{2}-1} m! \log (\mu |\sigma| )}{\Gamma \left(\frac{d}{2}-1\right) ((m-n)!)^2} \lambda_n\right)\, ,
\eeq
giving the anomalous dimension 
\beq
\gamma_{\hat O^m} = \frac{2 \pi^{\frac{d}{2}-1} m!}{\Gamma \left(\frac{d}{2}-1\right) ((m-n)!)^2} \lambda_{n *}  \, .
\eeq
Otherwise we need to push it to the second order. Let us just study the operator $\hat O^{-}_{-\alpha}$. The two-point function is 
%
%
\beq
\left<\hat O^{-}_{-\alpha}(\sigma)\hat O^{- \dagger}_{-\alpha}(0) \right> =\left[1- \frac{\pi ^{d-1} n \, \Gamma (n+1)^2 \csc \left(\pi \, \bar\alpha\right)}{(n-1) \Gamma \left(\frac{d}{2}-\bar\alpha\right)
   \Gamma \left(\bar\alpha+\frac{d}{2}-1\right)} \lambda_{n *}^2 \right]\frac{1}{|\sigma|^{d-2-2\alpha}}
\eeq
Thus, even at the second order the operator $\hat O^{-}_{-\alpha}$ does not acquire any anomalous dimension.

Now we compute the correction to the two-point function of the displacement operator, which has been defined for the unperturbed theory in eq. \eqref{eq:disp_mon}.
In order to compute the two-point function $\left< \hat D_z  \hat D_{\bar z} (0)\right>$ we observe that the operator $\hat O_{1+\alpha}^{+}$ does not couple to the perturbation, thus the computation reduces to the one of the two-point function of $\hat O_{-\alpha}^{-}$ done above. We get
\beq
\left<\hat D_z (\sigma) \hat D_{\bar z} (0)\right> = C_{\hat D}\left[1-\frac{\pi ^{d-1} n \, \Gamma (n+1)^2 \csc \left(\pi \, \bar\alpha\right)}{(n-1) \Gamma \left(\frac{d}{2}-\bar\alpha\right)
   \Gamma \left(\bar\alpha+\frac{d}{2}-1\right)} \lambda_{n *}^2  \right] \frac{1}{|\sigma|^{2(d-1)}} \, ,
\eeq
being $C_{\hat D}$ the unperturbed value reported in \eqref{eq:CD-h_relation} for $\xi =1$. At this point it is easy to check that the relation \eqref{eq:CD-h_relation} is broken at the second order in $\epsilon$. 

\subsubsection{Defect free energy}

The variation of the sphere-free energy can be computed by employing the general perturbative formula \eqref{eq:free_energ_gen}. We obtain
\beq
\Delta F =\frac{1}{\sin \left(\frac{\pi  d}{2}\right)}\frac{8 \pi  n^3  \Gamma
   \left(\frac{d}{2}-1\right)^2}{3 \,\text{Fr}_n^2 \,\Gamma (d-1)}\,\epsilon ^3  \, .
\eeq
From this last expression, we observe that when the $2<d<4$ the variation of the sphere-free energy is negative as expected. In addition, there is a simple pole when $d-2$ is even, signalling a variation in the  conformal anomaly. By specializing to $d=4$ and $d=6$ we obtain 
%
%
\begin{equation}
\Delta b =- \frac{8 n^3}{\text{Fr}_n^2} \epsilon ^3  \,, \qquad d =4\, ,
\end{equation}
and
\begin{equation}
\Delta a =- \frac{ n^3}{18\, \text{Fr}_n^2} \epsilon ^3  \,, \qquad d =6\, .
\end{equation}

\subsection{Monodromy defects coupled to Minimal Models}

The monodromy defect of a free complex scalar can also be coupled to a $(d-2)$-dimensional theory. In this section, as an example, we choose $d=4$ and a minimal model for the $p=2$ theory.
%
%
%
%
A first possibility is a coupling of the form
\begin{equation}
\label{eq:int_mon_min_sing}
S_{\text{\tiny int}} = g \int d^2 \sigma \, \left( \hat O^-_{-\a} \hat O^{- \dagger}_{-\a} \right)^n \hat\Phi \, ,
\end{equation}
where we may adopt $\alpha=\bar \alpha + \epsilon$ with $\bar \alpha = \frac{n-1}{n}+\frac{\Delta_{\hat{\Phi}}}{2n}$. It will also be more convenient to consider situations, where $C_{\hat\Phi \hat \Phi \hat \Phi}\neq0$, since the search of the fixed point happens to be easier: this, in particular, excludes the Ising model. The next simplest example will be the Tricritical Ising model,  and we can pick up ${\hat{\Phi}}=\varepsilon'$\footnote{In this section we consider both the analytic and anti-analytic part of the operators.} (in the convention of \cite{DiFrancesco:1997nk}),  with dimension $\Delta_{\varepsilon'}=6/5$, and we choose $n=1$. 
Another example is the Three-State Potts model, where we can take the spin operator $\sigma$, for which we have $\Delta_{\hat{\Phi}}=2/15$, and correspondingly $\bar \alpha = 1/15$  and $n=1$. Note that in both cases there are no other possible marginal deformations that can spoil our computation. At the end of this subsection we will consider a more general case where two marginal deformations can be turned on.

The computation of the $\beta$ function at one-loop is again done by employing the equation \eqref{BetaFuncGeneral_d}. Since the fields ${\hat{\Phi}}$ and $\hat O_{-\alpha}^{-}$ do not mix in the UV, we simply find
\beq
\beta_g = -2n \, \epsilon \, g + \pi \frac{C_{nnn}}{(n!)^2} C_{{\hat{\Phi}}{\hat{\Phi}}{\hat{\Phi}}} \, g^2 \, , 
\eeq
where we assumed $C_{{\hat{\Phi}}{\hat{\Phi}}}=1$ and $C_{nnn}$ is the same as in \eqref{eq:C_nnn}.
Thus, a non-trivial defect fixed point is again easily found at
\beq
g_* = \frac{2n}{\pi \, n!\, \text{Fr}_n\, C_{{\hat{\Phi}}{\hat{\Phi}}{\hat{\Phi}}}} \, \epsilon \, .
\eeq

The bulk-bulk propagator can be found exactly in the same way as  done in Sec. \ref{sec:Monod_corr} and we obtain the same functional expression \eqref{eq:two-point function_gaug_transf} with
%
%
%
%
%
%
\beq
\xi = 1-\frac{16 \, n^5  }{\text{Fr}_n^2 \, C_{{\hat{\Phi}}
   {\hat{\Phi}} {\hat{\Phi}}}^2 (\Delta_{\hat{\Phi}} +2 n-2)^2}\,\epsilon ^2 .
\eeq
We remind the reader that from this result one can extract the bulk one-point functions of $|\phi|^2$, of the stress tensor and the current. 

Similarly to the previous section, while the two-point function of the defect operator $\hat O^{-}_{-\alpha}$ does not acquire any anomalous dimension up to second order, we can consider the composite defect operator
%
%
%
$\left(\hat O^{-}_{-\alpha}\right)^m{\hat{\Phi}}$. We indeed find that if $m \ge n$ there is a non-trivial contribution already at the first order. We get
\beq
\left< \left(O^{-}_{-\alpha}\right)^m(\sigma){\hat{\Phi}}(\sigma) 
\left(O^{- \dagger}_{-\alpha}\right)^m(0){\hat{\Phi}}(0)\right> = \frac{m!}{|\sigma|^{2m(1-\alpha)+2 \Delta_{\hat{\Phi}}}} \left[1-\frac{4 \pi  C_{{\hat{\Phi}} {\hat{\Phi}} {\hat{\Phi}} }
    m! }{((m-n)!)^2} \log \left(\mu|\sigma|\right) \, g_* \right] \, ,
\eeq
which gives the anomalous dimension
\beq
\gamma_{\hat O^m{\hat{\Phi}}} = \frac{2 \pi  C_{{\hat{\Phi}} {\hat{\Phi}} {\hat{\Phi}} }
    m! }{((m-n)!)^2}\, g_* \, .
\eeq
If $m<n$ then we would need to go to second order, which in this case means to deal with the 4-point function of ${\hat{\Phi}}$. We will leave this computation to future work.

It may happen that the interaction in eq. \eqref{eq:int_mon_min_sing} is not the only marginal deformation given a value of $\alpha$ and a minimal model. For instance, one can consider a deformation made by two different defect operators as
\beq
S_{\text{\tiny int}} = g_1 \int d^2 \sigma \, \left( \hat O^-_{-\a} \hat O^{- \dagger}_{-\a} \right)^{n_1}{\hat{\Phi}}
 + g_2 \int d^2 \sigma \, \left( \hat O^-_{-\a} \hat O^{- \dagger}_{-\a} \right)^{n_2} \, .
\eeq
For the deformations to be both marginal one needs 
\begin{equation}
\Delta_{\hat \Phi}=2\frac{n_2-n_1}{n_2} \, ,\qquad \text{and} \qquad \bar \alpha=1-\frac{\Delta_{\hat \Phi}}{2(n_2-n_1)} \, .
\end{equation}
The one-loop $\beta$-functions can still be found by the general result \eqref{BetaFuncGeneral}, and they read
\begin{eqnarray}
&& \beta_{g_1} = -2 n_1 \epsilon \, g_1 + \frac{\pi}{C_{11}} \left(C_{111} \, C_{{\hat{\Phi}}{\hat{\Phi}}{\hat{\Phi}}} g_1^2 + 2 C_{112}g_1 g_2 \right)\,, \\ 
&&\beta_{g_2} =-2 n_2 \epsilon \, g_2 + \frac{\pi}{C_{22}} \left( C_{112} g_1^2 +  C_{222} g_2^2\right) \, .
\end{eqnarray}
We need the coefficient $C_{112}$ which is computed by
%
%
\beq
\begin{split}
C_{112}& =\sum_{k=n_2-n_1}^{n_1}\frac{(n_1!)^4 (n_2!)^2}{(k!)^2 (n_1-k)!
   ((n_2-k)!)^2 (k+n_1-n_2)!} \qquad 2 n_1 \ge n_2 \, ,
   \end{split}
\eeq
while it vanishes if $2n_1<n_2$.

To start, let us consider cases with $C_{{\hat{\Phi}}{\hat{\Phi}}{\hat{\Phi}}} = 0$. In this situation we may find fixed points with both $g_1$ and $g_2$ non-vanishing, which read
\begin{eqnarray}
 && g_1 = \pm \frac{ \sqrt{C_{11}} \sqrt{n_1} \sqrt{2 C_{112} C_{22} n_2-C_{11} C_{222} n_1}}{\pi \, 
    C_{112}^{3/2}}  \epsilon \, ,\\
&& g_2 =\frac{C_{11} n_1  }{\pi  C_{112}} \epsilon \, .
\end{eqnarray}
Note that we need to have $2 C_{112} C_{22} n_2-C_{11} C_{222} n_1> 0$ for the fixed points to be unitary. At this point it is simple to convince ourselves that there is no value with $n_2>n_1$ such that the condition is satisfied, and so the only interacting fixed point in this setup is the decoupled one with $g_{1 *}=0$ and $g_{2 *}=2 C_{22}n_2 \epsilon/(C_{222}\pi)$. 


The situation is different if we allow for non-vanishing $C_{{\hat{\Phi}}{\hat{\Phi}}{\hat{\Phi}}}$. In such cases, the condition becomes $8 C_{11} C_{112}^{2} C_{22} n_1 n_2+C_{111}^{2} C_{22}^2 C_{{\hat{\Phi}} {\hat{\Phi}} {\hat{\Phi}}}^2 n_2^2-4 C_{11}^2 C_{222}C_{112} n_1^2>0$, which in principle can be satisfied for judicious choices of $n_1$, $n_2$, and $C_{{\hat{\Phi}}{\hat{\Phi}}{\hat{\Phi}}}$. In practice however it happens to be a difficult task to find a working example. In fact, using the fact that $C^{(1,3)}_{(1,3), (1,3)}<\frac{4}{\sqrt{}3}$, one can rule out the $\Phi_{(1,3)}$ operators for any $\mathcal{M}(p+1, p)$. We postpone a conclusive analysis of this question for the future.

\section{Outlook}

This work explores scalar field theories with conformal defects, focusing on cases where the bulk remains free, and interactions are entirely localized on the defect submanifold $\Sigma_p$. These interactions are introduced by turning on slightly relevant defect deformations, which may take the form of self-interactions or actual couplings to lower-dimensional CFTs, such as Minimal Models in the case of surface defects. Along with identifying new perturbative conformal fixed points, we examine various interesting correlators, including bulk one-point functions, defect two-point functions, and defect Weyl anomalies. 

We conclude this paper by outlining several potential directions for future research.

A natural extension of this work is to explore conformal defects in free fermionic theories. While this has been studied extensively in the codimension-one setup \cite{MCAVITY1995522,Giombi:2021cnr,Herzog:2022jqv}, with applications to boundary conditions and defects in graphene \cite{PhysRevB.77.085423,PhysRevB.84.195434,Biswas:2022dkg,Semenoff:2022azt}, the higher-codimensional case poses additional challenges. Fermionic theories are more restrictive than scalar ones because the fundamental field's bare dimension is higher, limiting the number of relevant defect perturbations that can be constructed. In fact, quadratic couplings are already marginal in when $p=1$. Nonetheless, in the presence of a monodromy defect, it is possible to find relevant defect perturbations when $2<d<4$ \cite{Bianchi:2021snj}, and a similar analysis to the one performed in Sec.\ref{sec:monodromy} should be possible in such cases. In general, it would be interesting to extend the approach employed in \cite{Lauria:2020emq} to study the possible unitary defects in the theory of free scalar (see also \cite{Herzog:2022jqv} for the case of the free Maxwell theory) to the case of free Dirac fermions.   

The idea of putting interactions on the defects can also be explored in less symmetric setups. In particular, one can introduce wedges and composite defects, with the motivation coming \textit{e.g.} from the Bremmstrahlung function  computation \cite{Correa:2012at, Correa:2012hh,Fiol:2015spa} and ideas from $D$-brane physics \cite{Douglas:1995bn}, respectively. For recent results on defects with such geometries see \cite{Antunes:2021qpy,Bissi:2022bgu, Shimamori:2024yms, Ge:2024hei}

Throughout this manuscript, we have insisted on maintaining unitarity both in the bulk and on the defect. Another possibility would be to relax this assumption and consider non-unitary bulk theories. Such theories have been recently explored in the context of free higher-derivative theories with a boundary in \cite{Chalabi:2022qit}, and further extended to interacting theories in $d=6-\epsilon$ in \cite{Herzog:2024zxm}. However, higher-derivative theories in the presence of higher-codimensional defects remain an open area for future exploration.

Finally, we find it intriguing to extend our discussion to cases involving both bulk and defect interactions. Several works have explored defects in interacting scalar theories, including line defects \cite{Pannell:2023pwz}, surface defects \cite{Raviv-Moshe:2023yvq,Trepanier:2023tvb,Giombi:2023dqs}, monodromy defects \cite{Soderberg:2017oaa,Giombi:2021uae}, and wedges \cite{Antunes:2021qpy,Bissi:2022bgu}. It would be particularly interesting to study interacting bulk theories coupled to lower-dimensional field theories.

\section*{Acknowledgments}
The authors deeply thank Zohar Komargodski for insightful comments during the early stage of this work, and for reading and commenting the draft. The authors also thank Connor Behan, Lorenzo Di Pietro, Márk Mezei and Nikolay Slyunyaev for useful discussions.
JS is funded by the Knut and Alice Wallenberg Foundation under grant KAW 2021.0170, VR grant 2018-04438 and Olle Engkvists Stiftelse grant n. 2180108.

\appendix

\section{Bulk-bulk propagator for the Dirichlet coupling}
\label{app:BB_prop_DC}

In this appendix, we compute the leading modification of the bulk-bulk propagator in the case of the Dirichlet coupling employing perturbation theory.  This calculation serves as a consistency check of the general form \eqref{eq:prop_scal} in the presence of defect interactions. 

We take the perturbation
\begin{equation}
S=S_{\text{CFT}_p}+g_{0} \int\,d^p z\,\,\hat\phi \,\hat{\mathcal{O}} \, ,
\end{equation}
which induces the following modification to the propagator (at the first non-trivial order):
\begin{equation}
\begin{split}
\left<\phi(x)\phi(y) \right> & = \frac{C_\phi}{|x-y|^{d-2}} + \frac{g_0^2}{2} \int d^p \sigma_1 d^p \sigma_2 \left< \phi(x)\hat\phi(\sigma_1)\hat{\mathcal{O}}(\sigma_1)\hat\phi(\sigma_2)\hat{\mathcal{O}}(\sigma_2)\phi(y)\right> \\
& = \frac{C_\phi}{|x-y|^{d-2}} + g_0^2 \int d^p \sigma_1 d^p \sigma_2 \left< \phi(x)\hat \phi(\sigma_1)\right>\left<\hat{\mathcal{O}}(\sigma_1)\hat{\mathcal{O}}(\sigma_2)\right> \left<\hat \phi(\sigma_1)\phi(y)\right> \\
& =  \frac{C_\phi}{|x-y|^{d-2}} + g_0^2 \int d^p \sigma_1 \, d^p \sigma_2 \, 
\frac{C_\phi}{\left( x_\perp^2 + (\sigma_1-\sigma_x)^2 \right)^{\frac{d}{2}-1}}
\frac{C_{\hat{\mathcal{O}}}}{|\sigma_1-\sigma_2|^{2\hat{\Delta}}}
\frac{C_\phi}{\left( x_\perp^2 + (\sigma_2-\sigma_y)^2 \right)^{\frac{d}{2}-1}}\, .
\end{split}
\end{equation}
To solve the double integral we find it convenient to use the Fourier representations in \eqref{eq:fourier_xp} and \eqref{eq:fourier_sigma}. We get
\begin{equation}
\begin{split}
\left<\phi(x)\phi(y) \right> = \frac{C_\phi}{|x-y|^{d-2}} + & \frac{g_0^2 C_\phi^2 C_{\hat{\mathcal{O}}}}{(x_{\perp 1} x_{\perp 2})^{ \Delta_\phi -\frac{p}{2}}}\frac{\pi ^{3 p/2}\Gamma \left(\frac{p}{2}-\hat\Delta \right)}{4^{\hat\Delta +\Delta_\phi -p-1}\Gamma (\hat\Delta ) \Gamma^2
   (\Delta_\phi )}\times \\
   & \times \int\frac{ d^p k}{(2\pi)^p} \, 
  e^{i k (\sigma_x-\sigma_y)}    \frac{ K_{\Delta_\phi
   -\frac{p}{2}}(k\, x_{\perp 1}) K_{\Delta_\phi -\frac{p}{2}}(k\, x_{\perp 2})}{k^{2 (p-\hat\Delta -\Delta_\phi )}}\, .
\end{split}
\end{equation}
Now, since $p-\hat\Delta-\Delta_\phi \propto \epsilon$, at the leading order we can substitute $\hat\Delta  =p-\Delta_\phi$. The integral then simplifies to
\begin{equation}
\begin{split}
\left<\phi(x)\phi(y) \right> = \frac{C_\phi}{|x-y|^{d-2}} + & \frac{g_0^2 C_\phi^2 C_{\hat{\mathcal{O}}}}{(x_{\perp 1} x_{\perp 2})^{ \Delta_\phi -\frac{p}{2}}}\frac{4\pi ^{3 p/2}\Gamma \left(\Delta_\phi-\frac{p}{2} \right)}{\Gamma (p-\Delta_\phi ) \Gamma^2
   (\Delta_\phi )}\times \\
   & \times \int\frac{ d^p k}{(2\pi)^p} \, 
  e^{i k (\sigma_x-\sigma_y)}    K_{\Delta_\phi
   -\frac{p}{2}}(k\, x_{\perp 1}) K_{\Delta_\phi -\frac{p}{2}}(k\, x_{\perp 2})\, .
\end{split}
\end{equation}
By using the formula \eqref{eq:KK_int} one can easily see that we recover the propagator \eqref{eq:prop_scal} provided
\begin{equation}
    \xi = \frac{ \pi ^{\frac{1}{2} (p-q+2)} \csc \left(\frac{\pi  q}{2}\right)}{2 (2-q)
   \Gamma \left(1+\frac{p-q}{2}\right)}C_{\hat O} g_0^2 \, ,
\end{equation}
in agreement also with the one-point function in \eqref{eq:one-point_Dirich}.

\section{An alternative computation of Displacement Operators}
\label{app:displacement}
In this appendix we provide an alternative discussion of the displacement operator in the case of the quadratic perturbation $\hat \phi^2\equiv (O^{+}_{\{l=0\}})^2$. Specifically, we focus on the action
\begin{equation}
S = \frac{1}{2} \int d^d x \, \partial_\mu \phi \partial^\mu \phi + h_c \int d^p \sigma \hat\phi^2 \, .
\end{equation}
The displacement operator can be computed from the conservation of the stress-energy tensor together with the implementation of the equation of motion.
From the (bulk) stress-energy tensor
\begin{equation}
T_{\mu\nu} = \partial_\mu \phi \partial_\nu \phi -\frac{1}{2} \delta_{\mu\nu} \partial_\lambda \phi \partial^\lambda \phi - \frac{1}{4}\frac{d-2}{d-1} \left( \partial_\mu \partial_\nu -\delta_{\mu\nu}\partial^2 \right)\phi^2  \, ,
\end{equation}
we obtain its conservation equation, which reads
\begin{equation}
\partial^\mu T_{\mu \nu} = \partial_\nu\phi \, \partial^2 \phi \, .
\end{equation}
Then, combining the above equation with the equation of motion
\begin{equation}
-\partial^2 \phi + h_c \, \phi \, \delta^{d-p}(x_\perp)=0 \, ,
\end{equation}
we obtain 
\begin{equation}
\partial^\mu T_{\mu \nu} = \partial_\nu \phi \, \partial^2 \phi  = h_c \,  \hat\phi \, \partial_\nu \hat\phi \, \delta^{d-p} (x_\perp) \, .
\end{equation}
At this point, from the very definition of the displacement operator, we may write
\begin{equation}
\hat D^i = h_c \, \hat \phi \, \partial_\perp^i \hat \phi \, .
\end{equation}
Note that since the theory is quadratic, the operator dimensions simply add up. Thus, we expect that 
\begin{equation}
\Delta_{\hat D} = \Delta_{\hat \phi} + \Delta_{\partial_\perp \hat \phi} = \left( 1+p - \frac{d}{2} \right) + \frac{d}{2} = p+1 \, ,
\end{equation}
which is the correct result for the displacement operator.

A more rigorous derivation follows from the computation of the two-point function. Using Wick's theorem we have
\begin{equation}
\label{eq:DD_app}
\begin{split}
\left<\hat D^i(\sigma_1) \hat D^j(\sigma_2)\right>_{h_c} 
 =  h_c^2 \left<\hat \phi(\sigma_1) \hat \phi(\sigma_2)\right> \left< \partial_\perp^i \hat \phi(\sigma_1) \hat \partial_\perp^j \phi(\sigma_2)  \right>_{h_c} \, ,
\end{split}
\end{equation}
where we used $<\hat \phi\,\partial_\perp^i \hat \phi>=0$.

We need the correlators $\left< \partial^i_\perp \hat\phi \,\partial^j_\perp \hat\phi \right>$ that can be computed by taking the orthogonal derivatives of the propagator in eq. \eqref{eq:prop_h_scalar} and then sending to zero all the orthogonal components. 
%
%
%
We get
\begin{equation}
\label{eq:Dphi_Dphi_corr}
\left< \partial_\perp^i \hat\phi(\sigma_1)\partial_\perp^j\hat \phi(\sigma_2) \right>_{h_c} =\left< \partial_\perp^i \hat\phi(\sigma_1)\partial_\perp^j\hat \phi(\sigma_2) \right>_{0} = (d-2)C_{\phi} \, \frac{\delta^{ij}}{|\sigma|^{d}}\, .
\end{equation}
Notice that the conformal dimension of the defect operator $\partial_\perp \hat\phi$ remains unchanged along the flow and it reads
\begin{equation}
\Delta_{\partial_\perp\hat \phi} = \frac{d}{2} \, .
\end{equation}
This result is expected since $\partial_\perp^i\hat \phi$ and $\hat \phi$ are orthogonal defect operators, implying that the quadratic perturbation cannot modify such a correlator.

Finally, plugging the correlators \eqref{eq:Dphi_Dphi_corr} and \eqref{eq:OO_defect_scal} back into eq. \eqref{eq:DD_app} we obtain 
\begin{equation}
\begin{split}
\left<\hat D^i(\sigma_1) \hat D^j(\sigma_2)\right>_{h_c \rightarrow +\infty}  =& h_c^2 \times \left( \frac{2\left(p+2-d\right)  \sin \left( \frac{\pi}{2}  (d-p)\right) \Gamma \left(1+p-\frac{d}{2}\right)}{\pi^{1+p-\frac{d}{2}}} \frac{1}{ h^2_c}\frac{1}{|\sigma|^{2-d+2p}} \right) \times \\
&\times \left( (d-2) C_\phi \, \frac{\delta^{ij}}{|\sigma|^{d}}\right) =  C_{\hat D\hat D} \frac{\delta^{ij}}{|\sigma|^{2(p+1)}} \, ,
\end{split}
\end{equation}
with
\begin{equation}
    C_{\hat D\hat D} = \frac{\left(p+2-d\right)  \sin \left( \frac{\pi}{2}  (d-p)\right) \Gamma \left(1+p-\frac{d}{2}\right)\Gamma\left( \frac{d}{2} \right)}{\pi^{1+p}} \, .
\end{equation}
Notice that the scale $h_c$ cancels between the term coming from the definition of the displacement operator and the one arising from the large $h_c$ expansion in eq. \eqref{eq:OO_defect_scal}.
Additionally, as consistency check, $d=3$ and $p=2$ we find $C_{\hat D \hat D} = 1/(4\pi^2)$, which is twice the value of the coefficient of the displacement operator for the free scalar in the presence of a boundary with Dirichlet boundary conditions \cite{Herzog:2017kkj}. As noted in \cite{Raviv-Moshe:2023yvq}, this result is expected since, in this case, the defect reduces to an interface with Dirichlet boundary conditions.

\section{An useful integral}

In the main text we have done repetitive use of the following integral, which we prefers to report in this appendix rather than in the main text:
\beq
\begin{split}
\label{eq:KK_int}
\int \frac{d^p k}{(2\pi)^p} e^{i k \cdot \sigma}  K_{\alpha}(k \, x_{\perp 1} )K_{\alpha}(k \, x_{\perp 2} ) =  \frac{\pi}{4 \pi^{\frac{p}{2}}\sin \pi \a} \left( \frac{1}{x_{\perp 1} x_{\perp 2}} \right)^{\frac{p}{2}}\left[c^{(p)} (-\alpha) \,    F_{\frac{p}{2}-\a}(\eta)	 - c^{(p)} (\alpha) \,    F_{\frac{p}{2}+\a}(\eta)  \right]
\end{split}\,,
\eeq
being
\beq
c^{(p)}(\alpha) \equiv \frac{\Gamma\left( \frac{p}{2}+\a  \right)}{\Gamma\left(1+ \a \right)} \, ,
\eeq
and
\begin{equation}\label{defectblock}
F_{{\hat \Delta}}(\eta)=\left(\frac{\eta}{2} \right)^{\hat{\Delta}} \phantom{}_2 F_1\left( \frac{\hat{\Delta}}{2},\frac{\hat \Delta+1}{2}; \hat{\Delta}+1-\frac{p}{2}; \eta^2  \right) \,,
\end{equation}
where $\eta$ is the conformal ratio 
\beq
\eta \equiv  \frac{2 \, x_{\perp 1} x_{\perp 2}}{x_{\perp 1}^2 + x_{\perp 2}^2+\sigma^2} \, .
\eeq

\bibliographystyle{JHEP}
\bibliography{biblio}

\providecommand{\href}[2]{#2}\begingroup\raggedright\begin{thebibliography}{100}

\bibitem{DiFrancesco:1997nk}
P.~Di~Francesco, P.~Mathieu and D.~Senechal, \emph{{Conformal Field Theory}},
  Graduate Texts in Contemporary Physics. Springer-Verlag, New York, 1997,
  \href{https://doi.org/10.1007/978-1-4612-2256-9}{10.1007/978-1-4612-2256-9}.

\bibitem{Rychkov:2016iqz}
S.~Rychkov, \emph{{EPFL Lectures on Conformal Field Theory in D\ensuremath{>}=
  3 Dimensions}}, SpringerBriefs in Physics. 1, 2016,
  \href{https://doi.org/10.1007/978-3-319-43626-5}{10.1007/978-3-319-43626-5},
  [\href{https://arxiv.org/abs/1601.05000}{{\ttfamily 1601.05000}}].

\bibitem{Simmons-Duffin:2016gjk}
D.~Simmons-Duffin, \emph{{The Conformal Bootstrap}},  in \emph{{Theoretical
  Advanced Study Institute in Elementary Particle Physics}: {New Frontiers in
  Fields and Strings}}, pp.~1--74, 2017,
  \href{https://doi.org/10.1142/9789813149441_0001}{DOI}
  [\href{https://arxiv.org/abs/1602.07982}{{\ttfamily 1602.07982}}].

\bibitem{Poland:2018epd}
D.~Poland, S.~Rychkov and A.~Vichi, \emph{{The Conformal Bootstrap: Theory,
  Numerical Techniques, and Applications}},
  \href{https://doi.org/10.1103/RevModPhys.91.015002}{\emph{Rev. Mod. Phys.}
  {\bfseries 91} (2019) 015002}
  [\href{https://arxiv.org/abs/1805.04405}{{\ttfamily 1805.04405}}].

\bibitem{Billo:2016cpy}
M.~Bill\`o, V.~Gon\c{c}alves, E.~Lauria and M.~Meineri, \emph{{Defects in
  conformal field theory}},
  \href{https://doi.org/10.1007/JHEP04(2016)091}{\emph{JHEP} {\bfseries 04}
  (2016) 091} [\href{https://arxiv.org/abs/1601.02883}{{\ttfamily
  1601.02883}}].

\bibitem{Herzog:2020bqw}
C.~P. Herzog and A.~Shrestha, \emph{{Two point functions in defect CFTs}},
  \href{https://doi.org/10.1007/JHEP04(2021)226}{\emph{JHEP} {\bfseries 04}
  (2021) 226} [\href{https://arxiv.org/abs/2010.04995}{{\ttfamily
  2010.04995}}].

\bibitem{Bissi:2018mcq}
A.~Bissi, T.~Hansen and A.~S\"oderberg, \emph{{Analytic Bootstrap for Boundary
  CFT}}, \href{https://doi.org/10.1007/JHEP01(2019)010}{\emph{JHEP} {\bfseries
  01} (2019) 010} [\href{https://arxiv.org/abs/1808.08155}{{\ttfamily
  1808.08155}}].

\bibitem{Behan:2020nsf}
C.~Behan, L.~Di~Pietro, E.~Lauria and B.~C. Van~Rees, \emph{{Bootstrapping
  boundary-localized interactions}},
  \href{https://doi.org/10.1007/JHEP12(2020)182}{\emph{JHEP} {\bfseries 12}
  (2020) 182} [\href{https://arxiv.org/abs/2009.03336}{{\ttfamily
  2009.03336}}].

\bibitem{Behan:2021tcn}
C.~Behan, L.~Di~Pietro, E.~Lauria and B.~C. van Rees, \emph{{Bootstrapping
  boundary-localized interactions II. Minimal models at the boundary}},
  \href{https://doi.org/10.1007/JHEP03(2022)146}{\emph{JHEP} {\bfseries 03}
  (2022) 146} [\href{https://arxiv.org/abs/2111.04747}{{\ttfamily
  2111.04747}}].

\bibitem{Gimenez-Grau:2022czc}
A.~Gimenez-Grau, E.~Lauria, P.~Liendo and P.~van Vliet, \emph{{Bootstrapping
  line defects with O(2) global symmetry}},
  \href{https://doi.org/10.1007/JHEP11(2022)018}{\emph{JHEP} {\bfseries 11}
  (2022) 018} [\href{https://arxiv.org/abs/2208.11715}{{\ttfamily
  2208.11715}}].

\bibitem{Bianchi:2022sbz}
L.~Bianchi, D.~Bonomi and E.~de~Sabbata, \emph{{Analytic bootstrap for the
  localized magnetic field}},
  \href{https://doi.org/10.1007/JHEP04(2023)069}{\emph{JHEP} {\bfseries 04}
  (2023) 069} [\href{https://arxiv.org/abs/2212.02524}{{\ttfamily
  2212.02524}}].

\bibitem{Bianchi:2023gkk}
L.~Bianchi, D.~Bonomi, E.~de~Sabbata and A.~Gimenez-Grau, \emph{{Analytic
  bootstrap for magnetic impurities}},
  \href{https://doi.org/10.1007/JHEP05(2024)080}{\emph{JHEP} {\bfseries 05}
  (2024) 080} [\href{https://arxiv.org/abs/2312.05221}{{\ttfamily
  2312.05221}}].

\bibitem{Kapustin:2005py}
A.~Kapustin, \emph{{Wilson-'t Hooft operators in four-dimensional gauge
  theories and S-duality}},
  \href{https://doi.org/10.1103/PhysRevD.74.025005}{\emph{Phys. Rev. D}
  {\bfseries 74} (2006) 025005}
  [\href{https://arxiv.org/abs/hep-th/0501015}{{\ttfamily hep-th/0501015}}].

\bibitem{Deser:1993yx}
S.~Deser and A.~Schwimmer, \emph{{Geometric classification of conformal
  anomalies in arbitrary dimensions}},
  \href{https://doi.org/10.1016/0370-2693(93)90934-A}{\emph{Phys. Lett. B}
  {\bfseries 309} (1993) 279}
  [\href{https://arxiv.org/abs/hep-th/9302047}{{\ttfamily hep-th/9302047}}].

\bibitem{Polchinski:1998rq}
J.~Polchinski, \emph{{String theory. Vol. 1: An introduction to the bosonic
  string}}, Cambridge Monographs on Mathematical Physics. Cambridge University
  Press, 12, 2007,
  \href{https://doi.org/10.1017/CBO9780511816079}{10.1017/CBO9780511816079}.

\bibitem{Berenstein:1998ij}
D.~E. Berenstein, R.~Corrado, W.~Fischler and J.~M. Maldacena, \emph{{The
  Operator product expansion for Wilson loops and surfaces in the large N
  limit}}, \href{https://doi.org/10.1103/PhysRevD.59.105023}{\emph{Phys. Rev.
  D} {\bfseries 59} (1999) 105023}
  [\href{https://arxiv.org/abs/hep-th/9809188}{{\ttfamily hep-th/9809188}}].

\bibitem{Graham:1999pm}
C.~R. Graham and E.~Witten, \emph{{Conformal anomaly of submanifold observables
  in AdS / CFT correspondence}},
  \href{https://doi.org/10.1016/S0550-3213(99)00055-3}{\emph{Nucl. Phys. B}
  {\bfseries 546} (1999) 52}
  [\href{https://arxiv.org/abs/hep-th/9901021}{{\ttfamily hep-th/9901021}}].

\bibitem{Henningson:1999xi}
M.~Henningson and K.~Skenderis, \emph{{Weyl anomaly for Wilson surfaces}},
  \href{https://doi.org/10.1088/1126-6708/1999/06/012}{\emph{JHEP} {\bfseries
  06} (1999) 012} [\href{https://arxiv.org/abs/hep-th/9905163}{{\ttfamily
  hep-th/9905163}}].

\bibitem{Gustavsson:2003hn}
A.~Gustavsson, \emph{{On the Weyl anomaly of Wilson surfaces}},
  \href{https://doi.org/10.1088/1126-6708/2003/12/059}{\emph{JHEP} {\bfseries
  12} (2003) 059} [\href{https://arxiv.org/abs/hep-th/0310037}{{\ttfamily
  hep-th/0310037}}].

\bibitem{Asnin:2008ak}
V.~Asnin, \emph{{Analyticity Properties of Graham-Witten Anomalies}},
  \href{https://doi.org/10.1088/0264-9381/25/14/145013}{\emph{Class. Quant.
  Grav.} {\bfseries 25} (2008) 145013}
  [\href{https://arxiv.org/abs/0801.1469}{{\ttfamily 0801.1469}}].

\bibitem{Schwimmer:2008yh}
A.~Schwimmer and S.~Theisen, \emph{{Entanglement Entropy, Trace Anomalies and
  Holography}},
  \href{https://doi.org/10.1016/j.nuclphysb.2008.04.015}{\emph{Nucl. Phys. B}
  {\bfseries 801} (2008) 1} [\href{https://arxiv.org/abs/0802.1017}{{\ttfamily
  0802.1017}}].

\bibitem{Cvitan:2015ana}
M.~Cvitan, P.~Dominis~Prester, S.~Pallua, I.~Smoli\'c and T.~\v{S}temberga,
  \emph{{Parity-odd surface anomalies and correlation functions on conical
  defects}},  \href{https://arxiv.org/abs/1503.06196}{{\ttfamily 1503.06196}}.

\bibitem{Jensen:2018rxu}
K.~Jensen, A.~O'Bannon, B.~Robinson and R.~Rodgers, \emph{{From the Weyl
  Anomaly to Entropy of Two-Dimensional Boundaries and Defects}},
  \href{https://doi.org/10.1103/PhysRevLett.122.241602}{\emph{Phys. Rev. Lett.}
  {\bfseries 122} (2019) 241602}
  [\href{https://arxiv.org/abs/1812.08745}{{\ttfamily 1812.08745}}].

\bibitem{Herzog:2015ioa}
C.~P. Herzog, K.-W. Huang and K.~Jensen, \emph{{Universal Entanglement and
  Boundary Geometry in Conformal Field Theory}},
  \href{https://doi.org/10.1007/JHEP01(2016)162}{\emph{JHEP} {\bfseries 01}
  (2016) 162} [\href{https://arxiv.org/abs/1510.00021}{{\ttfamily
  1510.00021}}].

\bibitem{Herzog:2017kkj}
C.~Herzog, K.-W. Huang and K.~Jensen, \emph{{Displacement Operators and
  Constraints on Boundary Central Charges}},
  \href{https://doi.org/10.1103/PhysRevLett.120.021601}{\emph{Phys. Rev. Lett.}
  {\bfseries 120} (2018) 021601}
  [\href{https://arxiv.org/abs/1709.07431}{{\ttfamily 1709.07431}}].

\bibitem{FarajiAstaneh:2021foi}
A.~Faraji~Astaneh and S.~N. Solodukhin, \emph{{Boundary conformal invariants
  and the conformal anomaly in five dimensions}},
  \href{https://doi.org/10.1016/j.physletb.2021.136282}{\emph{Phys. Lett. B}
  {\bfseries 816} (2021) 136282}
  [\href{https://arxiv.org/abs/2102.07661}{{\ttfamily 2102.07661}}].

\bibitem{Chalabi:2021jud}
A.~Chalabi, C.~P. Herzog, A.~O'Bannon, B.~Robinson and J.~Sisti, \emph{{Weyl
  anomalies of four dimensional conformal boundaries and defects}},
  \href{https://doi.org/10.1007/JHEP02(2022)166}{\emph{JHEP} {\bfseries 02}
  (2022) 166} [\href{https://arxiv.org/abs/2111.14713}{{\ttfamily
  2111.14713}}].

\bibitem{Lewkowycz:2014jia}
A.~Lewkowycz and E.~Perlmutter, \emph{{Universality in the geometric dependence
  of Renyi entropy}},
  \href{https://doi.org/10.1007/JHEP01(2015)080}{\emph{JHEP} {\bfseries 01}
  (2015) 080} [\href{https://arxiv.org/abs/1407.8171}{{\ttfamily 1407.8171}}].

\bibitem{Bianchi:2015liz}
L.~Bianchi, M.~Meineri, R.~C. Myers and M.~Smolkin, \emph{{R\'enyi entropy and
  conformal defects}},
  \href{https://doi.org/10.1007/JHEP07(2016)076}{\emph{JHEP} {\bfseries 07}
  (2016) 076} [\href{https://arxiv.org/abs/1511.06713}{{\ttfamily
  1511.06713}}].

\bibitem{Prochazka:2018bpb}
V.~Prochazka, \emph{{The Conformal Anomaly in bCFT from Momentum Space
  Perspective}}, \href{https://doi.org/10.1007/JHEP10(2018)170}{\emph{JHEP}
  {\bfseries 10} (2018) 170}
  [\href{https://arxiv.org/abs/1804.01974}{{\ttfamily 1804.01974}}].

\bibitem{Bianchi:2019sxz}
L.~Bianchi and M.~Lemos, \emph{{Superconformal surfaces in four dimensions}},
  \href{https://doi.org/10.1007/JHEP06(2020)056}{\emph{JHEP} {\bfseries 06}
  (2020) 056} [\href{https://arxiv.org/abs/1911.05082}{{\ttfamily
  1911.05082}}].

\bibitem{Wang:2020xkc}
Y.~Wang, \emph{{Surface Defect, Anomalies and $b$-Extremization}},
  \href{https://arxiv.org/abs/2012.06574}{{\ttfamily 2012.06574}}.

\bibitem{Intriligator:2003jj}
K.~A. Intriligator and B.~Wecht, \emph{{The Exact superconformal R symmetry
  maximizes a}},
  \href{https://doi.org/10.1016/S0550-3213(03)00459-0}{\emph{Nucl. Phys. B}
  {\bfseries 667} (2003) 183}
  [\href{https://arxiv.org/abs/hep-th/0304128}{{\ttfamily hep-th/0304128}}].

\bibitem{Benini:2012cz}
F.~Benini and N.~Bobev, \emph{{Exact two-dimensional superconformal R-symmetry
  and c-extremization}},
  \href{https://doi.org/10.1103/PhysRevLett.110.061601}{\emph{Phys. Rev. Lett.}
  {\bfseries 110} (2013) 061601}
  [\href{https://arxiv.org/abs/1211.4030}{{\ttfamily 1211.4030}}].

\bibitem{Benini:2013cda}
F.~Benini and N.~Bobev, \emph{{Two-dimensional SCFTs from wrapped branes and
  c-extremization}}, \href{https://doi.org/10.1007/JHEP06(2013)005}{\emph{JHEP}
  {\bfseries 06} (2013) 005} [\href{https://arxiv.org/abs/1302.4451}{{\ttfamily
  1302.4451}}].

\bibitem{Kobayashi:2018lil}
N.~Kobayashi, T.~Nishioka, Y.~Sato and K.~Watanabe, \emph{{Towards a
  $C$-theorem in defect CFT}},
  \href{https://doi.org/10.1007/JHEP01(2019)039}{\emph{JHEP} {\bfseries 01}
  (2019) 039} [\href{https://arxiv.org/abs/1810.06995}{{\ttfamily
  1810.06995}}].

\bibitem{Klebanov:2011gs}
I.~R. Klebanov, S.~S. Pufu and B.~R. Safdi, \emph{{F-Theorem without
  Supersymmetry}}, \href{https://doi.org/10.1007/JHEP10(2011)038}{\emph{JHEP}
  {\bfseries 10} (2011) 038} [\href{https://arxiv.org/abs/1105.4598}{{\ttfamily
  1105.4598}}].

\bibitem{Zamolodchikov:1986gt}
A.~B. Zamolodchikov, \emph{{Irreversibility of the Flux of the Renormalization
  Group in a 2D Field Theory}}, {\emph{JETP Lett.} {\bfseries 43} (1986) 730}.

\bibitem{CARDY1988749}
J.~L. Cardy, \emph{Is there a c-theorem in four dimensions?},
  \href{https://doi.org/https://doi.org/10.1016/0370-2693(88)90054-8}{\emph{Physics
  Letters B} {\bfseries 215} (1988) 749}.

\bibitem{Komargodski:2011vj}
Z.~Komargodski and A.~Schwimmer, \emph{{On Renormalization Group Flows in Four
  Dimensions}}, \href{https://doi.org/10.1007/JHEP12(2011)099}{\emph{JHEP}
  {\bfseries 12} (2011) 099} [\href{https://arxiv.org/abs/1107.3987}{{\ttfamily
  1107.3987}}].

\bibitem{Casini:2012ei}
H.~Casini and M.~Huerta, \emph{{On the RG running of the entanglement entropy
  of a circle}}, \href{https://doi.org/10.1103/PhysRevD.85.125016}{\emph{Phys.
  Rev. D} {\bfseries 85} (2012) 125016}
  [\href{https://arxiv.org/abs/1202.5650}{{\ttfamily 1202.5650}}].

\bibitem{Casini:2015woa}
H.~Casini, M.~Huerta, R.~C. Myers and A.~Yale, \emph{{Mutual information and
  the F-theorem}}, \href{https://doi.org/10.1007/JHEP10(2015)003}{\emph{JHEP}
  {\bfseries 10} (2015) 003}
  [\href{https://arxiv.org/abs/1506.06195}{{\ttfamily 1506.06195}}].

\bibitem{Casini:2016udt}
H.~Casini, E.~Teste and G.~Torroba, \emph{{Relative entropy and the RG flow}},
  \href{https://doi.org/10.1007/JHEP03(2017)089}{\emph{JHEP} {\bfseries 03}
  (2017) 089} [\href{https://arxiv.org/abs/1611.00016}{{\ttfamily
  1611.00016}}].

\bibitem{Casini:2018cxg}
H.~Casini, R.~Medina, I.~Salazar~Landea and G.~Torroba, \emph{{Renyi relative
  entropies and renormalization group flows}},
  \href{https://doi.org/10.1007/JHEP09(2018)166}{\emph{JHEP} {\bfseries 09}
  (2018) 166} [\href{https://arxiv.org/abs/1807.03305}{{\ttfamily
  1807.03305}}].

\bibitem{Casini:2018nym}
H.~Casini, I.~Salazar~Landea and G.~Torroba, \emph{{Irreversibility in quantum
  field theories with boundaries}},
  \href{https://doi.org/10.1007/JHEP04(2019)166}{\emph{JHEP} {\bfseries 04}
  (2019) 166} [\href{https://arxiv.org/abs/1812.08183}{{\ttfamily
  1812.08183}}].

\bibitem{Casini:2022bsu}
H.~Casini, I.~Salazar~Landea and G.~Torroba, \emph{{Entropic g Theorem in
  General Spacetime Dimensions}},
  \href{https://doi.org/10.1103/PhysRevLett.130.111603}{\emph{Phys. Rev. Lett.}
  {\bfseries 130} (2023) 111603}
  [\href{https://arxiv.org/abs/2212.10575}{{\ttfamily 2212.10575}}].

\bibitem{Casini:2023kyj}
H.~Casini, I.~Salazar~Landea and G.~Torroba, \emph{{Irreversibility, QNEC, and
  defects}}, \href{https://doi.org/10.1007/JHEP07(2023)004}{\emph{JHEP}
  {\bfseries 07} (2023) 004}
  [\href{https://arxiv.org/abs/2303.16935}{{\ttfamily 2303.16935}}].

\bibitem{Friedan:2003yc}
D.~Friedan and A.~Konechny, \emph{{On the boundary entropy of one-dimensional
  quantum systems at low temperature}},
  \href{https://doi.org/10.1103/PhysRevLett.93.030402}{\emph{Phys. Rev. Lett.}
  {\bfseries 93} (2004) 030402}
  [\href{https://arxiv.org/abs/hep-th/0312197}{{\ttfamily hep-th/0312197}}].

\bibitem{Cuomo:2021rkm}
G.~Cuomo, Z.~Komargodski and A.~Raviv-Moshe, \emph{{Renormalization Group Flows
  on Line Defects}},
  \href{https://doi.org/10.1103/PhysRevLett.128.021603}{\emph{Phys. Rev. Lett.}
  {\bfseries 128} (2022) 021603}
  [\href{https://arxiv.org/abs/2108.01117}{{\ttfamily 2108.01117}}].

\bibitem{Nagar:2024mjz}
I.~Nagar, A.~Sever and D.-l. Zhong, \emph{{Planar RG flows on line defects}},
  \href{https://doi.org/10.1007/JHEP06(2024)110}{\emph{JHEP} {\bfseries 06}
  (2024) 110} [\href{https://arxiv.org/abs/2404.07290}{{\ttfamily
  2404.07290}}].

\bibitem{Castiglioni:2022yes}
L.~Castiglioni, S.~Penati, M.~Tenser and D.~Trancanelli, \emph{{Interpolating
  Wilson loops and enriched RG flows}},
  \href{https://doi.org/10.1007/JHEP08(2023)106}{\emph{JHEP} {\bfseries 08}
  (2023) 106} [\href{https://arxiv.org/abs/2211.16501}{{\ttfamily
  2211.16501}}].

\bibitem{Castiglioni:2023uus}
L.~Castiglioni, S.~Penati, M.~Tenser and D.~Trancanelli, \emph{{Wilson loops
  and defect RG flows in ABJM}},
  \href{https://doi.org/10.1007/JHEP06(2023)157}{\emph{JHEP} {\bfseries 06}
  (2023) 157} [\href{https://arxiv.org/abs/2305.01647}{{\ttfamily
  2305.01647}}].

\bibitem{Jensen:2015swa}
K.~Jensen and A.~O'Bannon, \emph{{Constraint on Defect and Boundary
  Renormalization Group Flows}},
  \href{https://doi.org/10.1103/PhysRevLett.116.091601}{\emph{Phys. Rev. Lett.}
  {\bfseries 116} (2016) 091601}
  [\href{https://arxiv.org/abs/1509.02160}{{\ttfamily 1509.02160}}].

\bibitem{Wang:2021mdq}
Y.~Wang, \emph{{Defect a-theorem and a-maximization}},
  \href{https://doi.org/10.1007/JHEP02(2022)061}{\emph{JHEP} {\bfseries 02}
  (2022) 061} [\href{https://arxiv.org/abs/2101.12648}{{\ttfamily
  2101.12648}}].

\bibitem{Shachar:2022fqk}
T.~Shachar, R.~Sinha and M.~Smolkin, \emph{{RG flows on two-dimensional
  spherical defects}},
  \href{https://doi.org/10.21468/SciPostPhys.15.6.240}{\emph{SciPost Phys.}
  {\bfseries 15} (2023) 240}
  [\href{https://arxiv.org/abs/2212.08081}{{\ttfamily 2212.08081}}].

\bibitem{Nozaki:2012qd}
M.~Nozaki, T.~Takayanagi and T.~Ugajin, \emph{{Central Charges for BCFTs and
  Holography}}, \href{https://doi.org/10.1007/JHEP06(2012)066}{\emph{JHEP}
  {\bfseries 06} (2012) 066} [\href{https://arxiv.org/abs/1205.1573}{{\ttfamily
  1205.1573}}].

\bibitem{Fursaev:2016inw}
D.~V. Fursaev and S.~N. Solodukhin, \emph{{Anomalies, entropy and boundaries}},
  \href{https://doi.org/10.1103/PhysRevD.93.084021}{\emph{Phys. Rev. D}
  {\bfseries 93} (2016) 084021}
  [\href{https://arxiv.org/abs/1601.06418}{{\ttfamily 1601.06418}}].

\bibitem{Cuomo:2022xgw}
G.~Cuomo, Z.~Komargodski, M.~Mezei and A.~Raviv-Moshe, \emph{{Spin impurities,
  Wilson lines and semiclassics}},
  \href{https://doi.org/10.1007/JHEP06(2022)112}{\emph{JHEP} {\bfseries 06}
  (2022) 112} [\href{https://arxiv.org/abs/2202.00040}{{\ttfamily
  2202.00040}}].

\bibitem{Gabai:2022vri}
B.~Gabai, A.~Sever and D.-l. Zhong, \emph{{Line Operators in
  Chern-Simons\textendash{}Matter Theories and Bosonization in Three
  Dimensions}},
  \href{https://doi.org/10.1103/PhysRevLett.129.121604}{\emph{Phys. Rev. Lett.}
  {\bfseries 129} (2022) 121604}
  [\href{https://arxiv.org/abs/2204.05262}{{\ttfamily 2204.05262}}].

\bibitem{Gabai:2022mya}
B.~Gabai, A.~Sever and D.-l. Zhong, \emph{{Line operators in
  Chern-Simons-Matter theories and Bosonization in Three Dimensions II:
  Perturbative analysis and all-loop resummation}},
  \href{https://doi.org/10.1007/JHEP04(2023)070}{\emph{JHEP} {\bfseries 04}
  (2023) 070} [\href{https://arxiv.org/abs/2212.02518}{{\ttfamily
  2212.02518}}].

\bibitem{Aharony:2023amq}
O.~Aharony, G.~Cuomo, Z.~Komargodski, M.~Mezei and A.~Raviv-Moshe,
  \emph{{Phases of Wilson lines: conformality and screening}},
  \href{https://doi.org/10.1007/JHEP12(2023)183}{\emph{JHEP} {\bfseries 12}
  (2023) 183} [\href{https://arxiv.org/abs/2310.00045}{{\ttfamily
  2310.00045}}].

\bibitem{Allais}
A.~Allais, \emph{{Magnetic defect line in a critical Ising model}},
  \href{https://arxiv.org/abs/1412.3449}{{\ttfamily 1412.3449}}.

\bibitem{ParisenToldin:2016szc}
F.~Parisen~Toldin, F.~F. Assaad and S.~Wessel, \emph{{Critical behavior in the
  presence of an order-parameter pinning field}},
  \href{https://doi.org/10.1103/PhysRevB.95.014401}{\emph{Phys. Rev. B}
  {\bfseries 95} (2017) 014401}
  [\href{https://arxiv.org/abs/1607.04270}{{\ttfamily 1607.04270}}].

\bibitem{Hanke}
A.~Hamke, \emph{{Critical absorbtion on defects in Ising magnets and binary
  alloys}}, {\emph{Phys. Rev. Lett.} {\bfseries 84} (2000) 2180}.

\bibitem{Allais:2014fqa}
A.~Allais and S.~Sachdev, \emph{{Spectral function of a localized fermion
  coupled to the Wilson-Fisher conformal field theory}},
  \href{https://doi.org/10.1103/PhysRevB.90.035131}{\emph{Phys. Rev. B}
  {\bfseries 90} (2014) 035131}
  [\href{https://arxiv.org/abs/1406.3022}{{\ttfamily 1406.3022}}].

\bibitem{Pannell:2024hbu}
W.~H. Pannell, \emph{{A note on defect stability in $d=4-\varepsilon$}},
  \href{https://arxiv.org/abs/2408.15315}{{\ttfamily 2408.15315}}.

\bibitem{Cuomo:2021kfm}
G.~Cuomo, Z.~Komargodski and M.~Mezei, \emph{{Localized magnetic field in the
  O(N) model}}, \href{https://doi.org/10.1007/JHEP02(2022)134}{\emph{JHEP}
  {\bfseries 02} (2022) 134}
  [\href{https://arxiv.org/abs/2112.10634}{{\ttfamily 2112.10634}}].

\bibitem{Giombi:2023dqs}
S.~Giombi and B.~Liu, \emph{{Notes on a Surface Defect in the $O(N)$ Model}},
  \href{https://arxiv.org/abs/2305.11402}{{\ttfamily 2305.11402}}.

\bibitem{Raviv-Moshe:2023yvq}
A.~Raviv-Moshe and S.~Zhong, \emph{{Phases of surface defects in Scalar Field
  Theories}}, \href{https://doi.org/10.1007/JHEP08(2023)143}{\emph{JHEP}
  {\bfseries 08} (2023) 143}
  [\href{https://arxiv.org/abs/2305.11370}{{\ttfamily 2305.11370}}].

\bibitem{Trepanier:2023tvb}
M.~Tr\'epanier, \emph{{Surface defects in the O(N) model}},
  \href{https://doi.org/10.1007/JHEP09(2023)074}{\emph{JHEP} {\bfseries 09}
  (2023) 074} [\href{https://arxiv.org/abs/2305.10486}{{\ttfamily
  2305.10486}}].

\bibitem{Giombi:2022vnz}
S.~Giombi, E.~Helfenberger and H.~Khanchandani, \emph{{Line defects in
  fermionic CFTs}}, \href{https://doi.org/10.1007/JHEP08(2023)224}{\emph{JHEP}
  {\bfseries 08} (2023) 224}
  [\href{https://arxiv.org/abs/2211.11073}{{\ttfamily 2211.11073}}].

\bibitem{Pannell:2023pwz}
W.~H. Pannell and A.~Stergiou, \emph{{Line defect RG flows in the
  \ensuremath{\varepsilon} expansion}},
  \href{https://doi.org/10.1007/JHEP06(2023)186}{\emph{JHEP} {\bfseries 06}
  (2023) 186} [\href{https://arxiv.org/abs/2302.14069}{{\ttfamily
  2302.14069}}].

\bibitem{Barrat:2023ivo}
J.~Barrat, P.~Liendo and P.~van Vliet, \emph{{Line defect correlators in
  fermionic CFTs}},  \href{https://arxiv.org/abs/2304.13588}{{\ttfamily
  2304.13588}}.

\bibitem{Harribey:2023xyv}
S.~Harribey, I.~R. Klebanov and Z.~Sun, \emph{{Boundaries and interfaces with
  localized cubic interactions in the O(N) model}},
  \href{https://doi.org/10.1007/JHEP10(2023)017}{\emph{JHEP} {\bfseries 10}
  (2023) 017} [\href{https://arxiv.org/abs/2307.00072}{{\ttfamily
  2307.00072}}].

\bibitem{Herzog:2017xha}
C.~P. Herzog and K.-W. Huang, \emph{{Boundary Conformal Field Theory and a
  Boundary Central Charge}},
  \href{https://doi.org/10.1007/JHEP10(2017)189}{\emph{JHEP} {\bfseries 10}
  (2017) 189} [\href{https://arxiv.org/abs/1707.06224}{{\ttfamily
  1707.06224}}].

\bibitem{Lauria:2020emq}
E.~Lauria, P.~Liendo, B.~C. Van~Rees and X.~Zhao, \emph{{Line and surface
  defects for the free scalar field}},
  \href{https://doi.org/10.1007/JHEP01(2021)060}{\emph{JHEP} {\bfseries 01}
  (2021) 060} [\href{https://arxiv.org/abs/2005.02413}{{\ttfamily
  2005.02413}}].

\bibitem{Zamolodchikov:1987ti}
A.~B. Zamolodchikov, \emph{{Renormalization Group and Perturbation Theory Near
  Fixed Points in Two-Dimensional Field Theory}}, {\emph{Sov. J. Nucl. Phys.}
  {\bfseries 46} (1987) 1090}.

\bibitem{Bianchi:2021snj}
L.~Bianchi, A.~Chalabi, V.~Proch\'azka, B.~Robinson and J.~Sisti,
  \emph{{Monodromy defects in free field theories}},
  \href{https://doi.org/10.1007/JHEP08(2021)013}{\emph{JHEP} {\bfseries 08}
  (2021) 013} [\href{https://arxiv.org/abs/2104.01220}{{\ttfamily
  2104.01220}}].

\bibitem{Nishioka:2021uef}
T.~Nishioka and Y.~Sato, \emph{{Free energy and defect $C$-theorem in free
  scalar theory}}, \href{https://doi.org/10.1007/JHEP05(2021)074}{\emph{JHEP}
  {\bfseries 05} (2021) 074}
  [\href{https://arxiv.org/abs/2101.02399}{{\ttfamily 2101.02399}}].

\bibitem{Rychkov:2015naa}
S.~Rychkov and Z.~M. Tan, \emph{{The $\epsilon$-expansion from conformal field
  theory}}, \href{https://doi.org/10.1088/1751-8113/48/29/29FT01}{\emph{J.
  Phys. A} {\bfseries 48} (2015) 29FT01}
  [\href{https://arxiv.org/abs/1505.00963}{{\ttfamily 1505.00963}}].

\bibitem{Bashmakov:2016pcg}
V.~Bashmakov, M.~Bertolini, L.~Di~Pietro and H.~Raj, \emph{{Scalar Multiplet
  Recombination at Large N and Holography}},
  \href{https://doi.org/10.1007/JHEP05(2016)183}{\emph{JHEP} {\bfseries 05}
  (2016) 183} [\href{https://arxiv.org/abs/1603.00387}{{\ttfamily
  1603.00387}}].

\bibitem{Brax:2023goj}
P.~Brax and S.~Fichet, \emph{{Casimir Forces in CFT with Defects and
  Boundaries}}, \href{https://doi.org/10.3390/physics6020036}{\emph{Physics}
  {\bfseries 6} (2024) 544} [\href{https://arxiv.org/abs/2312.02281}{{\ttfamily
  2312.02281}}].

\bibitem{Gubser:2002vv}
S.~S. Gubser and I.~R. Klebanov, \emph{{A Universal result on central charges
  in the presence of double trace deformations}},
  \href{https://doi.org/10.1016/S0550-3213(03)00056-7}{\emph{Nucl. Phys. B}
  {\bfseries 656} (2003) 23}
  [\href{https://arxiv.org/abs/hep-th/0212138}{{\ttfamily hep-th/0212138}}].

\bibitem{Brust:2016gjy}
C.~Brust and K.~Hinterbichler, \emph{{Free \ensuremath{\square}$^{k}$ scalar
  conformal field theory}},
  \href{https://doi.org/10.1007/JHEP02(2017)066}{\emph{JHEP} {\bfseries 02}
  (2017) 066} [\href{https://arxiv.org/abs/1607.07439}{{\ttfamily
  1607.07439}}].

\bibitem{Komargodski:2016auf}
Z.~Komargodski and D.~Simmons-Duffin, \emph{{The Random-Bond Ising Model in
  2.01 and 3 Dimensions}},
  \href{https://doi.org/10.1088/1751-8121/aa6087}{\emph{J. Phys. A} {\bfseries
  50} (2017) 154001} [\href{https://arxiv.org/abs/1603.04444}{{\ttfamily
  1603.04444}}].

\bibitem{Cardy:1988cwa}
J.~L. Cardy, \emph{{Is There a c Theorem in Four-Dimensions?}},
  \href{https://doi.org/10.1016/0370-2693(88)90054-8}{\emph{Phys. Lett. B}
  {\bfseries 215} (1988) 749}.

\bibitem{Dotsenko:1985hi}
V.~S. Dotsenko and V.~A. Fateev, \emph{{Operator Algebra of Two-Dimensional
  Conformal Theories with Central Charge C \ensuremath{<}= 1}},
  \href{https://doi.org/10.1016/0370-2693(85)90366-1}{\emph{Phys. Lett. B}
  {\bfseries 154} (1985) 291}.

\bibitem{Giombi:2014xxa}
S.~Giombi and I.~R. Klebanov, \emph{{Interpolating between $a$ and $F$}},
  \href{https://doi.org/10.1007/JHEP03(2015)117}{\emph{JHEP} {\bfseries 03}
  (2015) 117} [\href{https://arxiv.org/abs/1409.1937}{{\ttfamily 1409.1937}}].

\bibitem{Fei:2015oha}
L.~Fei, S.~Giombi, I.~R. Klebanov and G.~Tarnopolsky, \emph{{Generalized
  $F$-Theorem and the $\epsilon$ Expansion}},
  \href{https://doi.org/10.1007/JHEP12(2015)155}{\emph{JHEP} {\bfseries 12}
  (2015) 155} [\href{https://arxiv.org/abs/1507.01960}{{\ttfamily
  1507.01960}}].

\bibitem{Giombi:2021uae}
S.~Giombi, E.~Helfenberger, Z.~Ji and H.~Khanchandani, \emph{{Monodromy defects
  from hyperbolic space}},
  \href{https://doi.org/10.1007/JHEP02(2022)041}{\emph{JHEP} {\bfseries 02}
  (2022) 041} [\href{https://arxiv.org/abs/2102.11815}{{\ttfamily
  2102.11815}}].

\bibitem{Dowker:1987mn}
J.~S. Dowker, \emph{{Casimir Effect Around a Cone}},
  \href{https://doi.org/10.1103/PhysRevD.36.3095}{\emph{Phys. Rev. D}
  {\bfseries 36} (1987) 3095}.

\bibitem{MCAVITY1995522}
D.~McAvity and H.~Osborn, \emph{Conformal field theories near a boundary in
  general dimensions},
  \href{https://doi.org/https://doi.org/10.1016/0550-3213(95)00476-9}{\emph{Nuclear
  Physics B} {\bfseries 455} (1995) 522}.

\bibitem{Giombi:2021cnr}
S.~Giombi, E.~Helfenberger and H.~Khanchandani, \emph{{Fermions in AdS and
  Gross-Neveu BCFT}},
  \href{https://doi.org/10.1007/JHEP07(2022)018}{\emph{JHEP} {\bfseries 07}
  (2022) 018} [\href{https://arxiv.org/abs/2110.04268}{{\ttfamily
  2110.04268}}].

\bibitem{Herzog:2022jqv}
C.~P. Herzog and A.~Shrestha, \emph{{Conformal surface defects in Maxwell
  theory are trivial}},
  \href{https://doi.org/10.1007/JHEP08(2022)282}{\emph{JHEP} {\bfseries 08}
  (2022) 282} [\href{https://arxiv.org/abs/2202.09180}{{\ttfamily
  2202.09180}}].

\bibitem{PhysRevB.77.085423}
A.~R. Akhmerov and C.~W.~J. Beenakker, \emph{Boundary conditions for dirac
  fermions on a terminated honeycomb lattice},
  \href{https://doi.org/10.1103/PhysRevB.77.085423}{\emph{Phys. Rev. B}
  {\bfseries 77} (2008) 085423}.

\bibitem{PhysRevB.84.195434}
J.~A.~M. van Ostaay, A.~R. Akhmerov, C.~W.~J. Beenakker and M.~Wimmer,
  \emph{Dirac boundary condition at the reconstructed zigzag edge of graphene},
  \href{https://doi.org/10.1103/PhysRevB.84.195434}{\emph{Phys. Rev. B}
  {\bfseries 84} (2011) 195434}.

\bibitem{Biswas:2022dkg}
S.~Biswas and G.~W. Semenoff, \emph{{Massless fermions on a half-space: the
  curious case of 2+1-dimensions}},
  \href{https://doi.org/10.1007/JHEP10(2022)045}{\emph{JHEP} {\bfseries 10}
  (2022) 045} [\href{https://arxiv.org/abs/2208.06374}{{\ttfamily
  2208.06374}}].

\bibitem{Semenoff:2022azt}
G.~W. Semenoff, \emph{{Boundary ferromagnetism in zigzag edged graphene}},
  \href{https://doi.org/10.1063/5.0135165}{\emph{J. Math. Phys.} {\bfseries 64}
  (2023) 071902} [\href{https://arxiv.org/abs/2211.09282}{{\ttfamily
  2211.09282}}].

\bibitem{Correa:2012at}
D.~Correa, J.~Henn, J.~Maldacena and A.~Sever, \emph{{An exact formula for the
  radiation of a moving quark in N=4 super Yang Mills}},
  \href{https://doi.org/10.1007/JHEP06(2012)048}{\emph{JHEP} {\bfseries 06}
  (2012) 048} [\href{https://arxiv.org/abs/1202.4455}{{\ttfamily 1202.4455}}].

\bibitem{Correa:2012hh}
D.~Correa, J.~Maldacena and A.~Sever, \emph{{The quark anti-quark potential and
  the cusp anomalous dimension from a TBA equation}},
  \href{https://doi.org/10.1007/JHEP08(2012)134}{\emph{JHEP} {\bfseries 08}
  (2012) 134} [\href{https://arxiv.org/abs/1203.1913}{{\ttfamily 1203.1913}}].

\bibitem{Fiol:2015spa}
B.~Fiol, E.~Gerchkovitz and Z.~Komargodski, \emph{{Exact Bremsstrahlung
  Function in $N=2$ Superconformal Field Theories}},
  \href{https://doi.org/10.1103/PhysRevLett.116.081601}{\emph{Phys. Rev. Lett.}
  {\bfseries 116} (2016) 081601}
  [\href{https://arxiv.org/abs/1510.01332}{{\ttfamily 1510.01332}}].

\bibitem{Douglas:1995bn}
M.~R. Douglas, \emph{{Branes within branes}}, {\emph{NATO Sci. Ser. C}
  {\bfseries 520} (1999) 267}
  [\href{https://arxiv.org/abs/hep-th/9512077}{{\ttfamily hep-th/9512077}}].

\bibitem{Antunes:2021qpy}
A.~Antunes, \emph{{Conformal bootstrap near the edge}},
  \href{https://doi.org/10.1007/JHEP10(2021)057}{\emph{JHEP} {\bfseries 10}
  (2021) 057} [\href{https://arxiv.org/abs/2103.03132}{{\ttfamily
  2103.03132}}].

\bibitem{Bissi:2022bgu}
A.~Bissi, P.~Dey, J.~Sisti and A.~S\"oderberg, \emph{{Interacting conformal
  scalar in a wedge}},
  \href{https://doi.org/10.1007/JHEP10(2022)060}{\emph{JHEP} {\bfseries 10}
  (2022) 060} [\href{https://arxiv.org/abs/2206.06326}{{\ttfamily
  2206.06326}}].

\bibitem{Shimamori:2024yms}
S.~Shimamori, \emph{{Conformal field theory with composite defect}},
  \href{https://doi.org/10.1007/JHEP08(2024)131}{\emph{JHEP} {\bfseries 08}
  (2024) 131} [\href{https://arxiv.org/abs/2404.08411}{{\ttfamily
  2404.08411}}].

\bibitem{Ge:2024hei}
D.~Ge, T.~Nishioka and S.~Shimamori, \emph{{Localized RG flows on composite
  defects and $\mathcal{C}$-theorem}},
  \href{https://arxiv.org/abs/2408.04428}{{\ttfamily 2408.04428}}.

\bibitem{Chalabi:2022qit}
A.~Chalabi, C.~P. Herzog, K.~Ray, B.~Robinson, J.~Sisti and A.~Stergiou,
  \emph{{Boundaries in free higher derivative conformal field theories}},
  \href{https://doi.org/10.1007/JHEP04(2023)098}{\emph{JHEP} {\bfseries 04}
  (2023) 098} [\href{https://arxiv.org/abs/2211.14335}{{\ttfamily
  2211.14335}}].

\bibitem{Herzog:2024zxm}
C.~P. Herzog and Y.~Zhou, \emph{{An Interacting, Higher Derivative, Boundary
  Conformal Field Theory}},  \href{https://arxiv.org/abs/2409.11072}{{\ttfamily
  2409.11072}}.

\bibitem{Soderberg:2017oaa}
A.~S\"oderberg, \emph{{Anomalous Dimensions in the WF O($N$) Model with a
  Monodromy Line Defect}},
  \href{https://doi.org/10.1007/JHEP03(2018)058}{\emph{JHEP} {\bfseries 03}
  (2018) 058} [\href{https://arxiv.org/abs/1706.02414}{{\ttfamily
  1706.02414}}].

\end{thebibliography}\endgroup

\end{document}